\documentclass[%
 reprint,
 twocolumn,
 amsmath,amssymb,
 aps,
prb,
nofootinbib
]{revtex4-2}

\usepackage{graphicx}
\usepackage{dcolumn}
\usepackage{bm}
\usepackage{color}
\usepackage{braket}
\usepackage{lipsum}
\usepackage{mathdots}
\usepackage[version=3]{mhchem} 
\usepackage{ulem}
\usepackage{color, soul}
\usepackage{xspace}
\usepackage{url}
\usepackage{appendix}
\usepackage{hyperref}

\newcommand{\green}{\textcolor{green}}

\renewcommand{\[}     {\left[}

\renewcommand{\_}[1]  {_\textrm{#1}} 

\preprint{APS/123-QED}

\begin{document}

\title{Topological spin...}

\date{\today}

\title{
Quantum-Geometric Theory of Current Generation Driven by Lattice Vibration
%
}

\author{Tomonari Meguro$^1$}
 \email{meguro.tomonari@mbp.phys.kyushu-u.ac.jp} 
%
 %
\author{Hiroaki Ishizuka$^2$}
 \email{ishizuka@hiroishizuka.com}
\author{Kentaro Nomura$^{1,}$$^3$}
 \email{nomura.kentaro@mbp.phys.kyushu-u.ac.jp}

 \affiliation{%
$^1$Department of Physics, Kyushu University, Fukuoka 819-0395, Japan \\
$^2$Department of Physics, Institute of Science Tokyo, Meguro, Tokyo 152-8551, Japan \\
$^3$Quantum and Spacetime Research Institute, Kyushu University, Fukuoka, 819-0395, Japan
}%

\begin{abstract}
Phonons couple to electrons, and their time-dependent dynamics can continuously deform electronic Bloch wave functions. 
{\it Electron-phonon quantum geometry} arises from the resulting dependence on electron crystal momentum $\bm k$ and phonon coordinates ${{\bm Q}}$ and affects physical phenomena, e.g., transport phenomena.
Here, we present the theory of the nonlinear electric current generation directly driven by the phonon dynamics and elucidate its quantum geometric aspects.
Developing a nonadiabatic semiclassical wave-packet theory, 
we derive the phonon-induced current rectification and second-harmonic generation to second order in the phonon coordinates.
We identify the two distinct contributions, which are characterized by the geometrical quantities defined in the two parameter spaces; 
${{\bm Q}}$-space and the $(\bm k, {{\bm Q}})$-mixed space.
The adiabatic contribution is linear in phonon frequency and governed by Berry-curvature-related structures, 
whereas the leading nonadiabatic contribution is quadratic in the frequency and originates from quantum metrics and Christoffel symbols. 
Analytical results for a massive Dirac electron reveal finite in-gap current plateaus and characteristic hump peaks near band edges. 
Numerical calculations for the Haldane and Kane-Mele models confirm these features through valley- and spin-resolved Dirac electrons. 
These results provide a systematic geometric description of phonon-driven nonlinear transports beyond the adiabatic theory and open a new viewpoint for analyzing phonon-driven phenomena.

\end{abstract}

 \maketitle

\section{Introduction}\label{sec:intro}

Phonons --- the collective excitations of lattice vibrations --- play central roles in many solid-state phenomena. 
Through electron-phonon coupling, they contribute to superconductivity, carrier relaxation \cite{Mahan_text, Ziman_text}, etc.
%
%
More recently, coherent phonons~\cite{Zeiger1992,Merlin1997}, phase-synchronized lattice oscillations spanning many unit cells, have become control tools. 
Coherent driving can transiently reshape electronic and magnetic states \cite{Rini2007,Forst2011,Mankowsky2014,Cavalleri2015,Mankowsky2016RPP,Nova2017,Afanasiev2021,JFransson2023}.
%
%
Among them, transport phenomena provide a direct manifestation of this dynamical role of phonons. 
They also strongly influence electronic mobility and optical properties in solids \cite{Giustino2017,Ziman_text}.
Electron-phonon scattering is a basic ingredient of conventional charge transport from the perspective of a relaxation mechanism~\cite{Mahan_text}. 
Besides, 
driven lattice motion itself can also generate or control electric currents. 
%
Specifically, coherent phonons can modify photocurrents generated in optically driven systems \cite{Shi2021,Luo2021,Okamura2022,JHu_PRL2025}. 
Chiral phonons~\cite{ZhangNiu2014,Zhu2018Science,Ishito2023a} 
can also generate electric currents in the adiabatic regime \cite{DYao2022}.
Various studies have up to now established lattice dynamics as an active driving way of electronic transport.

A useful framework for understanding nontrivial charge transport has been developed by focusing on geometric properties of electronic wave functions~\cite{Xiao2010,Nagaosa2010, Bandyopadhyay2024,Liu2025, AGaoNN2025}. 
These properties describe how quantum states vary across a parameter space, for example crystal momentum for Bloch electrons of the Bloch Hamiltonian $\hat{\mathcal H}_{\bm k}$ .
%
Such geometric quantities contain two distinct ingredients: the Berry curvature and quantum metric. 
Berry curvature measures how the quantum states are bent by varying 
the momentum
and governs anomalous Hall effect~\cite{Sundaram1999,Xiao2010,Nagaosa2010} and adiabatic charge pumping \cite{Thouless1983,QNiu1984}. 
In contrast, the quantum metric measures the distance between quantum states with different momentum and enters nonlinear transport and optical responses \cite{YGao2014,Sodemann2015,CXiao2019NLHE,Bandyopadhyay2024,Liu2025, AGaoNN2025}, phenomena beyond the adiabatic regime. 
%
%
The concepts of Berry curvature and the quantum metric are particularly useful for understanding intrinsic electric responses, and are collectively termed quantum geometry.

\begin{figure}[t]
   \includegraphics[width=1.0\hsize]{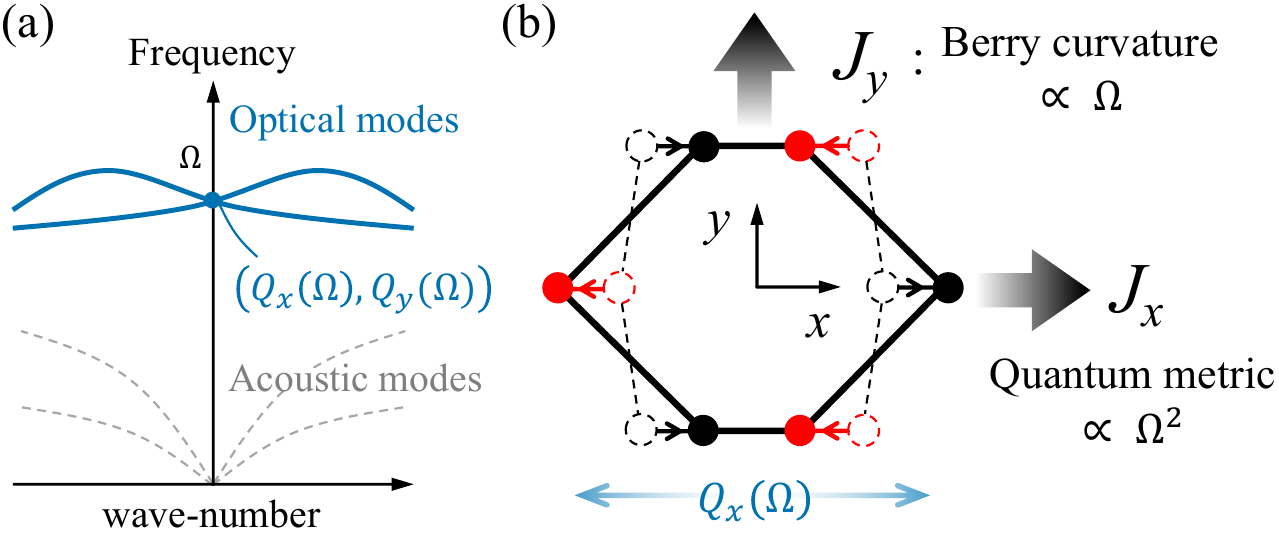}
\caption{
{
Schematic illustrations of optical phonon modes and electric currents generated directly by lattice vibrations in, e.g., a honeycomb lattice. 
(a)~Schematic of the phonon dispersion showing two in-plane optical modes (blue lines). 
Their Fourier amplitudes are denoted by $Q_x(\Omega)$ and $Q_y(\Omega)$. 
We omit the out-of-plane phonon modes.
(b)~Second-order currents induced by, e.g., the $x$-polarized optical phonon.
%
The $x$- and $y$-direction currents originate from quantum-metric and Berry-curvature, scaling as $\Omega^2$ and $\Omega$, respectively. 
%
}
}
\label{fig:current_schematic_honeycomb}
\end{figure}

Crystal momentum need not be the only relevant parameter when considering electron quantum geometry coupled with other external parameters.
In electron-phonon-coupled systems, coherently excited phonons can periodically drive electrons 
{by temporal deforming the crystal potential.} 
%
Consequently, the instantaneous Bloch states depend not only on crystal momentum 
{but also on the direction and amplitude of lattice motion, giving rise to a mixed parameter space.
}
%
%
%
The corresponding curvature and metric structures
describe how electronic states deform under lattice motion and changes in momentum direction. 
We call these geometric structures of eigenstates in this mixed space as {\it electron-phonon quantum geometry}. 
%
%

Understanding phonon-driven transport beyond the adiabatic regime still remains an unexplored area.
In the adiabatic limit, electrons follow instantaneous Bloch states during lattice motion,~{as described within the Born-Oppenheimer framework \cite{Longuet1958,Mead_RMP1992}}. 
The Berry curvature defined in the mixed parameter space has been studied to provide a geometric description of the resulting transport~\cite{DYao2022} and other phenomena~\cite{YRen2021, DSaparov2022}. 
Meanwhile, Hu et al. connected the electron-phonon quantum geometry to phonon-mediated optical responses \cite{JHu_PRL2025}, and addresses the deformation of electron wavefunction by phonon dynamics.
At finite phonon frequencies, lattice dynamics induces interband mixing, deviating from adiabatic dynamics. 
However, until now, 
the resulting corrections to transport, which is directly driven by the phonon, remain insufficiently understood, and the challenge is to construct a framework and derive these corrections systematically while preserving their geometric interpretation.

In this work, we develop a theory of phonon-driven electron dynamics and current generation based on a semiclassical wave-packet formalism. 
The wave-packet formalism provides an intuitive description of electron dynamics that incorporates quantum-geometric effects and phonon-induced interband mixing. 
Using this formulation, we derive the second-order rectified and second-harmonic currents induced by phonon dynamics. 
The current consists of adiabatic and leading nonadiabatic contributions. 
The former is linear in the phonon frequency $\Omega$ and involves Berry-curvature-like structures, 
whereas 
the latter is quadratic in $\Omega$ and governed by quantum metrics and the associated Christoffel symbols.
We then apply the theory to massive Dirac electrons, obtaining analytical conductivities that exhibit finite in-gap plateaus and characteristic band-edge humps reflecting electron-phonon quantum geometry. 
{
Fig.~\ref{fig:current_schematic_honeycomb} schematically illustrates how optical phonon dynamics generate electric currents in a honeycomb lattice.
} 
We further investigate the Haldane and Kane-Mele lattice models, whose conductivity behaviors can be qualitatively understood in terms of valley- and spin-resolved massive Dirac electrons. 
These results establish a systematic geometric description of phonon-driven nonlinear transport beyond adiabatic theory.


\section{Nonadiabatic wave-packet theory}\label{sec:general}

\iftrue

\subsection{Wave-packet}

To investigate how phonon dynamics affects electron dynamics, we consider the following electronic Hamiltonian
\begin{align}\label{eq:totalHam}
    \hat{H}^{\rm tot}[\hat{\bm x},\hat{\bm p};{\bm Q}(t)]
    &=
    \hat{H}[\hat{\bm x},\hat{\bm p};{\bm Q}(t)]
    +\hat{H}'(\hat{\bm x}).
\end{align}
Here, $\hat H$ represents the lattice-periodic electronic Hamiltonian, whose {$\hat{\bm p}$ and $\hat{\bm x}$ dependences originate from the electronic kinetic energy and the periodic crystal potential, respectively}, while $\hat H'$ denotes a spatially inhomogeneous term.
{
The dependence of the position operator in $\hat{H}'(\hat{\bm x})$ enters as potentials slowly vary in position, such as electromagnetic potential, inhomogeneous lattice distortion.
In the following formulation, we assume that this potential enters the Hamiltonian only as an electrostatic potential $\phi(\hat{\bm{x}})$
\footnote{
In addition to the electrostatic potential, one may also incorporate static lattice distortions, finite-momentum lattice vibrations, and spin textures. 
In the present work, however, we do not address effects originating from these spatial dependences, as they are outside the scope of this study. 
Related works can be found in Refs.~\cite{Sundaram1999,YOnishi_PRB2026,HYoshida_PRL2026,YRen_PRB2026}
}.
}







{
The time dependence of the Hamiltonian $\hat{H}$ in Eq.~(\ref{eq:totalHam}) arises from lattice vibrations. 
To describe these vibrations, we introduce the displacement vector {$\Delta\bm R_\kappa(t)$} for the $\kappa$-th sublattice~(the number of the sublattice is arbitrary). 
Since we consider spatially uniform phonons such as optical modes at the $\Gamma$-point in this formalism, 
we do not consider the spatial dependence of lattice displacement. 
%
%
%
Following the standard normal-mode analysis based on the dynamical matrix~\cite{Grosso2000}, 
we expand the lattice displacement vector in terms of the phonon eigenmodes as 
\begin{align}
    {\Delta{\bm R}_\kappa(t)}
    &=
    \sum_{\alpha\in{\rm opt}}
    {\bm e_{\kappa\alpha}}Q_\alpha(t),
\end{align}
with $\alpha$ being the label of the optical modes.
Here, 
$\bm e_{\kappa\alpha}$ is the dimensionless polarization vector specifying the relative displacement pattern of the sublattices in the $\alpha$-th optical mode, 
and $Q_\alpha(t)$ describes its time-dependent amplitude. 
We refer to this amplitude as ``phonon coordinate'' in this paper and collectively denote them by ${\bm Q}(t)=\{Q_\alpha(t)\}_{\alpha\in{\rm opt}}$;  
each component represents an independent optical vibration
\footnote{
{We note that the bold symbol ${\bm Q}$ represents a set of phonon normal coordinates, rather than the Cartesian components of a real-space vector.}}.
We assume that the phonon coordinate is much smaller than the lattice constant, $|{\bm Q}(t)| \ll a$, and is periodic in time with period $T$.
The lattice vibrations couple to electrons through the deformation of the periodic crystal potential. 
Specifically, the displacement pattern associated with each phonon coordinate $Q_\alpha(t)$ modifies the crystal potential experienced by electrons, thereby introducing the ${\bm Q}(t)$ dependence into $\hat H$. 
Consequently, $\partial_{Q_\alpha}\hat H$ describes the change in the electronic Hamiltonian induced by an infinitesimal displacement along the $\alpha$-th optical mode, characterizing the corresponding electron-phonon coupling.
}

\else

\subsection{Setup}

To investigate how phonon dynamics affects electron dynamics, we consider the following electronic Hamiltonian
{
\begin{align}\label{eq:totalHam}
    \hat{H}^{\rm tot}[\hat{\bm x},\hat{\bm p}; \{\bm u_\kappa(t)\}]
    &=
    \hat{H}[\hat{\bm x},\hat{\bm p};\{\bm u_\kappa(t)\}]
    +\hat{H}'(\hat{\bm x}).
\end{align}
}
Here, $\hat H$ represents the lattice-periodic electronic Hamiltonian, whose {$\hat{\bm p}$ and $\hat{\bm x}$ dependences originate from the electronic kinetic energy and the periodic crystal potential, respectively}, while $\hat H'$ denotes a spatially inhomogeneous term.
{
The dependence of the position operator in $\hat{H}'(\hat{\bm x})$ enters as potentials slowly vary in position, such as electromagnetic potential, inhomogeneous lattice distortion.
In the following formulation, we assume that this potential enters the Hamiltonian only as an electrostatic potential $\phi(\hat{\bm{x}})$
\footnote{
In addition to the electrostatic potential, one may also incorporate static lattice distortions, finite-momentum lattice vibrations, and spin textures. 
In the present work, however, we do not address effects originating from these spatial dependences, as they are outside the scope of this study. 
Related works can be found in Refs.~\cite{Sundaram1999,YOnishi_PRB2026,YRen_PRB2026}
}.
}

{
%
$\{ \bm u_\kappa(t) \}$ in Eq.~(\ref{eq:totalHam}) collectively denotes the lattice displacement vectors for the $\kappa$-th sublattice~($\kappa=1,\ldots,l_s$). 
%
%
Since we consider spatially uniform phonons such as optical modes at the $\Gamma$-point in this formalism, 
we do not consider the spatial dependence of lattice displacement. 
%
Following the standard normal-mode analysis based on the dynamical matrix~\cite{Grosso2000}, 
we expand the lattice displacement vector in terms of the phonon eigenmodes as 
\begin{align}
    {\bm u}_\kappa(t)
    &=
    \sum_{\alpha\in{\rm opt}}
    {\bm e_{\kappa\alpha}}Q_\alpha(t),
\end{align}
with $\alpha$ being the label of the optical modes.
Here, 
$\bm e_{\kappa\alpha}$ is the dimensionless polarization vector specifying the relative displacement pattern of the sublattices in the $\alpha$-th optical mode, 
and $Q_\alpha(t)$ describes its time-dependent amplitude. 
We refer to this amplitude as ``phonon coordinate'' in this paper and collectively denote them by ${\bm Q}(t)=\{Q_\alpha(t)\}_{\alpha\in{\rm opt}}$;  
each component represents an independent optical vibration.
We assume that the phonon coordinate is much smaller than the lattice constant, $|{\bm Q}(t)| \ll a$, and is periodic in time with period $T$.
}

{
The electron-phonon coupling is introduced through the dependence of the electronic Hamiltonian on the phonon coordinates $Q_\alpha(t)$. 
To make this dependence explicit, we expand the electronic Hamiltonian $\hat H$ in Eq.~(1) in terms of the lattice displacement vectors $\bm u_\kappa(t)$, assuming $|\bm u_\kappa(t)|\ll a$. 
Using the normal-mode expansion in Eq.~(3), we obtain
\begin{align}
    \hat{H}[\hat{\bm x},\hat{\bm p} ; \{\bm u_\kappa\}]
    &=
     \hat{H}^{(0)}[\hat{\bm x},\hat{\bm p}] 
     +
     \sum_{\kappa=1}^{l_s} 
     \frac{ \partial \hat{H} }{ \partial \bm u_\kappa }
     \cdot
     \bm u_\kappa(t) 
     +
     \mathcal{O}(u^2)
     \nonumber\\
     &=
     \hat{H}^{(0)}[\hat{\bm x},\hat{\bm p}] 
     +
     \sum_{\alpha\in{\rm opt}}
     Q_\alpha(t)  \hat{G}_\alpha
     +
     \mathcal{O}(Q^2)
\end{align}
where $\hat{H}^{(0)}[\hat{\bm x},\hat{\bm p}] $ is original electronic Hamiltonian, and 
\begin{align}
    \hat{G}_\alpha
     &=
     \frac{1}{\sqrt{N}}
     \sum_{\kappa=1}^{l_s} \bm e_{\kappa\alpha} \cdot
     \frac{ \partial \hat{H} }{ \partial \bm u_\kappa }
\end{align}
characterizes the electron-phonon coupling associated with the $\alpha$-th optical mode. In the following, we express the lattice-displacement dependence of the electronic Hamiltonian in terms of the phonon coordinates ${\bm Q}(t)$, so that the total Hamiltonian reads
\begin{align}
    \hat{H}^{\rm tot}[\hat{\bm x},\hat{\bm p};{\bm Q}(t)]
    &=
    \hat{H}[\hat{\bm x},\hat{\bm p};{\bm Q}(t)]
    +\hat{H}'(\hat{\bm x}).
\end{align}
}

\subsection{Wave-packet}

\fi

%
%

%
Defining the Bloch Hamiltonian for $\hat{H}[\hat{\bm x},\hat{\bm p},  {\bm Q}(t)]$ as 
{$\hat{\mathcal{H}}({\bm{k}};{\bm Q}) = e^{-i\bm{k}\cdot\hat{\bm{x}}}\hat{H} e^{i\bm{k}\cdot\hat{\bm{x}}}$}, 
one can obtain the instantaneous eigenstates and eigenvalues as follows,
      \begin{align}\label{eq:inst_eigeneq}
          \hat{\mathcal{H}}(\bm k;{{\bm Q}}) \ket{ u_{n}(\bm k;{{\bm Q}}) } 
           = 
          \varepsilon_{n}(\bm k;{{\bm Q}})  \ket{ u_{n}(\bm k;{{\bm Q}}) }.
      \end{align}
where $n$ is the band index and $\bm k$ is the crystal momentum.
The phonon coordinate $Q_\alpha(t)$ serves as a parameter that deforms the electronic Bloch wavefunctions.
{
The construction of Eq.~(\ref{eq:inst_eigeneq}) is based on the spirit of the Born-Oppenheimer approximation, in which the lattice displacements are treated as fixed parameters when determining the electronic eigenstates~\cite{Longuet1958,Mead_RMP1992}. 
However, we do not impose the adiabatic approximation on the subsequent electron dynamics, allowing for nonadiabatic transitions between instantaneous eigenstates induced by the time evolution of the phonon coordinates ${\bm Q}(t)$.}

Following the semiclassical theory of electron dynamics \cite{Sundaram1999,Xiao2010}, 
we construct a wave-packet type wavefunction that is localized around $\bm{x}_c$ and $\bm{k}_c$ in phase space. Here, $\bm{x}_c$ and $\bm{k}_c$ denote the wave-packet centers in position and momentum space, respectively. Its time evolution is obtained from a variational principle.
Supposing the instantaneous Bloch eigenstates as the basis function, we construct the wave-packet wavefunction for the $n$-th band as, 
      \begin{align}
        \ket{\Psi_{n,\bm x_c, \bm k_c}}
        &=
        \int_{\bm k}\sum_{l} C_{n\bm k}(l;t) \; e^{i\bm k\cdot \hat{\bm x}} \ket{u_{l}(\bm k;{{\bm Q}})},\label{eq:psi_wave-packet}
      \end{align}
where 
the momentum integration measure is defined as 
$\int_{\bm{k}} \equiv \int_{\rm BZ} d^d\bm{k}/(2\pi)^d$ over the Brillouin zone (BZ).
The expansion coefficient $C_{n\bm k}(l;t)$ is referred to as the wave-packet coefficient and depends on $\bm k$ in general.
The subscript $n$ labels the band of interest, while the index $l$ in parentheses labels the band component mixed with band $n$ during the time evolution, including $l=n$.
%
This effect is referred to as the nonadiabatic correction \cite{HYoshida_PRL2026,YRen_PRB2026} and plays a crucial role in our study. 
%
%
We impose the normalization condition on the wave-packet wave function 
$\braket{\Psi_{n,\bm x_c, \bm k_c} | \Psi_{n,\bm x_c, \bm k_c}}=1$, i.e., 
      \begin{align}
        \int_{\bm k} \sum_{l} |C_{n\bm k}(l;t)|^2 = 1.
      \end{align}

%
In the present formulation, we determine $C_{n\bm k}$ by imposing the following two conditions. 
First, they are chosen such that the wave-packet wavefunction is sharply peaked around $(\bm{x}_c,\bm{k}_c)$ in phase space, which leads to the condition 
$ \bm x_c = \bra{\Psi_{n,\bm x_c, \bm k_c}} \hat{\bm x} \ket{\Psi_{n,\bm x_c, \bm k_c}} $ 
and
$ \bm k_c = \bra{\Psi_{n,\bm x_c, \bm k_c}} \hat{\bm k} \ket{\Psi_{n,\bm x_c, \bm k_c}} $.
Second, the time evolution of the wave-packet coefficients is determined by requiring the wave-packet to satisfy the 
following time-dependent Schr{\"o}dinger equation,
      \begin{align}\label{eq:Sch_eq_time}
        i\hbar \frac{\partial}{\partial t}\ket{\Psi_{n,\bm x_c, \bm k_c}} 
        &=
        {\hat H}^{\rm tot}\ket{\Psi_{n,\bm x_c, \bm k_c}}.
      \end{align}
Here, the time dependence of the wave-packet coefficients enters through the center coordinates, $\bm x_c, \bm k_c$, and through the dynamics of the phonon coordinate ${{\bm Q}}$.
%
For the position dependence of $\hat{H}'$ in Eq.(\ref{eq:Sch_eq_time}), we perform the gradient expansion around $\bm{x}_c$.
Since the higher order gradient terms are outside the scope of the present work, we retain only the zeroth-order term, i.e., $\phi(\bm x_c)$.
By imposing the initial condition that the electrons are in the focused band $n$ at $t_0$,
we obtain
\begin{align}\label{eq:Cn_solution}
    %
    %
    %
    %
    C_{n\bm k}(l\ne n; t )
              &=
              -i  \int_{t_0}^t d\tau~
                e^{-i\gamma_{l\bm k}(t,\tau)-i\gamma_{n\bm k} (\tau,t)}  
                \nonumber \\ 
                &\ \ \ \ \ \ \ \ \ \ \ \ \ \ \ \times
                a^{{{\bm Q}}}_{ln,\alpha} 
                \dot{Q}_\alpha (\tau) C_{n\bm k}(n;t ).
\end{align}
For the intra-band coefficient, from the first condition mentioned above, we impose that its squared modulus is narrow distribution function around $\bm{k}_c$ in momentum space
\footnote{
Strictly speaking, we first assume that the effect of inter-band mixing is small,
$ 1 \gg \sum_{l(\ne n)} |M_{ln,\bm k}(t)|^2 $.
We then assume that the squared modulus of the intra-band coefficient is sharply localized around $\bm{k}_c$ and takes a form such that 
$ |C_{n\bm k}(n;t)|^2 \approx \delta( \bm k - \bm k_c(t) ) $.
These assumptions are adequate for the perturbative treatment of nonadiabatic effects considered in this work.}
.

In Eq. (\ref{eq:Cn_solution}),
we introduce the inter-band Berry connection with respect to the phonon coordinate as 
$a^{{{\bm Q}}}_{ln,\alpha} = -i \bra{u_{l}(\bm k;{{\bm Q}})} \partial_{Q_\alpha} \ket{u_{n}(\bm k;{{\bm Q}})}$.
This quantity represents the inter-band transitions driven by the dynamics of the phonon coordinate. 
While there are second- and higher order contributions of such transitions on the actual calculation of Eq. (\ref{eq:Cn_solution}), one does not need to incorporate them since they eventually do not appear in following calculations of electric current.
The explicit expression for the second-order wave-packet coefficient is presented in Appendix.~A.
We also introduce 
\begin{align}
        \gamma_{n\bm k}(t_2,t_1) 
        = 
        \frac{1}{\hbar}\int_{t_1}^{t_2} d\tau
        \Bigl(
             \varepsilon^{\rm tot}_{n\bm k}(\bm x_c (\tau), {{\bm Q}}(\tau))+\hbar \dot{Q}_{\alpha}(\tau) A^{{{\bm Q}}}_{n,\alpha}
        \Bigr),
\end{align}
which denotes the dynamical and geometrical phases of electrons in the $n$-th band, accumulated by the dynamics of the phonon coordinate between times $t_1$ and $t_2$.
Here, the $\bm{x}_c$-dependence of the eigenenergy enters only through the electrostatic potential $\phi(\bm{x}_c)$, which gives an overall energy shift, 
$\varepsilon^{\rm tot}_{n\bm k}(\bm x_c, {{\bm Q}}) = \varepsilon_{n\bm k}({{\bm Q}}) - e \phi(\bm x_c)$.
$A^{{{\bm Q}}}_{n,\alpha} = -i \bra{u_{n}(\bm k;{{\bm Q}})} \partial_{Q_\alpha} \ket{u_{n}(\bm k;{{\bm Q}})}$ is the intra-band Berry connection for the $n$-th band. 
%

We close this subsection by commenting on the gauge choice for the basis function used in constructing the wave-packet wavefunction.
In constructing a wave-packet wave function, there is freedom in the choice of the basis function used for the expansion. 
When dealing with electronic states in solids, a natural choice for such a basis function is the Bloch eigenfunctions. 
In this case, there are two possible choices: 
one may use the Bloch eigenfunctions at the equilibrium positions, say $\ket{u_{n}(\bm k;\bm 0)}$, obtained from the static Hamiltonian that does not include the electron-phonon interaction, 
or the instantaneous Bloch eigenfunctions $\ket{u_{n}(\bm{k};{\bm Q})}$ obtained from Eq.(\ref{eq:inst_eigeneq}).
%

When choosing the $\ket{u_{n}(\bm k;\bm 0)}$-basis,  
the resulting equation of motion for the wave-packet coefficients does not include the effect of the Berry connection.
This corresponds to fixing the gauge of the Bloch wavefunction along the phonon-coordinate direction ${{\bm Q}}$, while the gauge in momentum space can still be chosen arbitrarily. 
This choice is valid when a global gauge fixing is possible in the ${{\bm Q}}$-space overall, irrespective of the situation in momentum space \cite{Meguro_PRB2026}. 
By contrast, choosing the $\ket{u_{n}(\bm{k};{\bm Q})}$-basis corresponds to not imposing such a gauge fixing. This formulation is also applicable even when there exists a nontrivial region in which a global gauge fixing cannot be performed in the ${{\bm Q}}$-space.
In the following, we aim to formulate the wave-packet dynamics while properly incorporating the nontrivial geometric structure associated with ${\bm Q}$. 
We therefore use the instantaneous eigenfunction basis $\ket{u_{n}(\bm{k};{\bm Q})}$ in the following formulation.
%

\subsection{Equations of motion}


The equation of motion for the wave packet in Eq.~\eqref{eq:psi_wave-packet} is given by 
the stationary condition of the action,
\begin{align}
    S 
    &=
    \int dt L[ \bm x_c(t), \bm k_c(t) ],
\end{align}
where $L[ \bm x_c(t), \bm k_c(t) ]$ is the Lagrangian, 
\begin{align}\label{eq:Lag_def}
    L 
    &= 
    \bra{\Psi_{n, \bm x_c, \bm k_c}}
    \Bigl( i\hbar\frac{d}{dt} - {\hat H}^{\rm tot}  \Bigr)
    \ket{\Psi_{n, \bm x_c, \bm k_c}}.
\end{align}
%
%
By substituting Eq.~\eqref{eq:psi_wave-packet} into $\ket{\Psi_{n, \bm x_c, \bm k_c}}$, the Lagrangian reads (See {Appendix} for the details of derivation), 
\begin{align}\label{eq:Lag_form}
        L 
        &=
        \hbar \dot{\bm x}_c \cdot \bm k_c 
       -\hbar \dot{\bm k}_c \cdot \tilde{\bm A}^{\bm k_c}_n
       -\hbar \dot{{{\bm Q}}} \cdot {\bm{\widetilde A}^{{\bm Q}}_n}
       -\mathcal{E}_n.
\end{align}
Here,
\begin{align}\label{eq:dynA_k}
        \tilde{\bm A}^{\bm k_c}_{n}
        &=
        \sum_{m,l} 
        M^*_{mn,\bm k_c}(t) \;
        \bm a^{\bm k_c}_{ml}\; 
        M_{ln,\bm k_c}(t), 
\end{align}
is the dynamical Berry connection in the momentum space, where 
$\bm a^{\bm k_c}_{ml} = -i \bra{u_{m}(\bm k_c;{{\bm Q}})} \partial_{\bm k_c} \ket{u_{l}(\bm k_c;{{\bm Q}})}$ 
is the inter-band Berry connection, and $M_{ln,\bm k}$ is defined by $ C_{n\bm k}(l;t) = M_{ln,\bm k}(t) C_{n\bm k}(n;t)$ with $M_{nn,\bm k}=1$.
The dynamical Berry connection for phonon coordinate, ${{\widetilde A}^{{{\bm Q}}}_{n,\alpha}}$, is defined in the same manner as Eq.~\eqref{eq:dynA_k} but with the derivative by {$Q_{\alpha}$}.
%
The time-dependence of $C_{n\bm k}$ and $M_{ln,\bm k}$ manifest as the time dependence of $\bm x_c(t), \bm k_c(t)$, and ${{\bm Q}}(t)$ in Eq.~\eqref{eq:Lag_form}. 
%
In Eq.~\eqref{eq:Lag_form}, we ignored the Berry connection with respect to the position $\bm x_c$, assuming that the effects of the second-order gradient in $\partial_{\bm x_c}$ are negligible.

{
We also introduce the $n$-th wave-packet energy which consists of two parts
$\mathcal{E}_n = \mathcal{E}_n^{\rm WP} + \mathcal{E}_n^{\rm dyn}$. 
Here, $\mathcal{E}_n^{\rm WP}$ is given by 
\begin{align}\label{eq:WP_energy}
    \mathcal{E}_n^{\rm WP}
    &=
    \varepsilon^{\rm tot}_{n \bm k_c}(\bm x_c,{{\bm Q}}), 
\end{align}
and corresponds to the wave-packet energy in the conventional adiabatic wave-packet theory. 
On the other hand, $\mathcal{E}_n^{\rm dyn}$ is termed dynamical energy; 
it describes a nonadiabatic contribution arising from the mixing with other bands induced by the time evolution.
It reads~(see detailed derivation in Appendix.~B), 
\begin{align}
    \label{eq:dynamical_WPen}
    \mathcal{E}_n^{\rm dyn}
    &=
    \sum_{l(\ne n)}
    M^*_{ln,\bm k_c}(t) 
    \left( -i\hbar \frac{d}{dt} + \varepsilon^{\rm tot}_{l \bm k_c}(\bm x_c,{{\bm Q}}) \right)
    M_{ln,\bm k_c}(t),
\end{align}
}
where total time-derivative means
\begin{align}
    &\frac{d}{dt}M_{ln,\bm k_c}(t)
    =
    \nonumber\\
    &\ \ \ \  
      \dot{\bm x}_c \cdot \partial_{\bm x_c} M_{ln,\bm k_c}(t)
     +\dot{\bm k}_c \cdot \partial_{\bm k_c} M_{ln,\bm k_c}(t)
     +{\dot{{\bm Q}}} \cdot \partial_{{{\bm Q}}} M_{ln,\bm k_c}(t).
\end{align}

By applying the principle of least action to Eq.~(\ref{eq:Lag_form}),
we obtain the equations of motion for $\bm x_c$ and $\bm k_c$,
\begin{align}
        \label{eq:EOM_k}
        \hbar \dot{k}_{c,i} 
        &=
        -e E_i,
        \\
        \label{eq:EOM_x}
        \dot{x}_{c,i}
        &=
        \frac{1}{\hbar}\frac{\partial {\mathcal E}_n}{\partial k_{c,i}}
       -
        \tilde{\Omega}^{\bm k_c \bm k_c}_{n,ij} \dot{k}_{c,j}
       -
        \tilde{\Omega}^{\bm k_c {{\bm Q}}}_{n,i\alpha} \dot{Q}_{\alpha}
              \nonumber\\
       &\ \ \ 
       +
        \frac{d}{dt} {\rm Im}
        \Biggl[ \sum_{l(\ne n)} M^*_{ln, \bm k_c}(t) \frac{\partial M_{ln, \bm k_c}(t)}{\partial k_{c,i}} \Biggr]
\end{align}
where $\bm E = - \partial_{\bm x_c}\phi(\bm x_c)$ is the electric field and
%
\begin{align}\label{eq:dyn_BC_ku}
    \tilde{\Omega}^{\bm k_c {{\bm Q}}}_{n,i\alpha}
    &=
    \partial_{k_{c,i}} \tilde{A}^{{{\bm Q}}}_{n,\alpha}
    -
    \partial_{Q_\alpha} \tilde{A}^{\bm k_c}_{n,i}.
\end{align}
is the dynamical Berry curvature for the mixed parameter space $(\bm k_c,{{\bm Q}})$; 
$\tilde{\Omega}^{\bm k \bm k}_{n,ij}$ is the dynamical Berry curvature for the momentum, which is defined in the same manner as Eq.~\eqref{eq:dyn_BC_ku}.
In Eq.~(\ref{eq:EOM_x}), the first term corresponds to the conventional group velocity of an electron in the $n$-th band, and 
the second and third terms describe the anomalous velocities associated with the momentum change and the phonon dynamics, respectively. 
The fourth term originates from the gradient of the additional energy induced by the mixing with other bands, as mentioned in Eq.~(\ref{eq:dynamical_WPen}).

As discussed in the previous subsection, the wave-packet coefficients used in this work are obtained within the approximation up to first order in the inter-band transitions. Nevertheless, the above formulation of the equations of motion is independent of this approximation and holds for general wave-packet coefficients.

\section{Electric current }

\subsection{Distribution function}\label{subsec:notes_distribution}


In the Boltzmann theory, the electric current is given by
\begin{align}\label{eq:current_general}
    J_i &= -e \sum_n \int_{\bm k} f_{n\bm k}\; \dot{\bm x}_{c},
\end{align}
where $f_{n\bm k}$ is the electron distribution function for the $n$-th band and $\dot{\bm x}$ represents the wave-packet velocity given by Eq.(\ref{eq:EOM_x}).
The explicit form of the electron distribution function should be determined from the Boltzmann equation and depends on both the phonon dynamics {${\bm Q}(t)$} and the relevant scattering mechanisms. 
%
%
In this work, the effect of scattering is taken into account by the relaxation-time approximation 
%
%
focusing on the $\Omega \tau \gg 1$ limit, where $\tau$ is the relaxation time and $\Omega$ denotes the characteristic angular frequency of the phonon dynamics.

Within the constant relaxation-time approximation, and for a periodically driven external field, the formal solution for the distribution function reads~\cite{Harada2023},
      \begin{align}\label{eq:el_Distribution}
        f_{n\bm k}(t)
        &=
        \frac{1}{1-e^{-\frac{T}{\tau}}}
        \int_0^T \frac{dt'}{\tau} e^{-\frac{t'}{\tau}}
        f_{\rm FD}(\varepsilon_{n\bm k} (t-t'))
        %
        %
        %
        \\ \notag
        &=
        \frac{1}{T} \int_0^T dt'
        f_{\rm FD}(\varepsilon_{n\bm k} (t')),
      \end{align}
where $ f_{\rm FD}(E)=1/(e^{\beta(E-\mu)}+1)$ is the Fermi-Dirac distribution function where $\beta=1/k_B T$ and $\mu$ is the chemical potential. 
In the second line, we transform the expression by using the fact that $\varepsilon_{n\bm k}(t)$ is periodic in time with the period $T$.
Equation (\ref{eq:el_Distribution}) means we should employ the periodically averaged Fermi-Dirac distribution function as the electron's distribution function.
When one expands $f_{\rm FD}(\varepsilon_{n\bm k} (t'))$ around $Q_\alpha=0$, it takes the following form;  
$f_{\rm FD}(\varepsilon_{n\bm k} (t')) = f_{\rm FD}(\varepsilon_{n\bm k}(\bm 0) ) + Q_\alpha (t') \partial_{Q_\alpha} f^{\rm FD}(\varepsilon_{n\bm k} ({{\bm Q}}))|_{{{\bm Q}}=\bm 0} + {\mathcal O}(Q^2)$, where $\varepsilon_{n\bm k}(\bm 0)$ denotes the eigenenergy for the Bloch Hamiltonian without the electron-phonon coupling.
Since $Q_\alpha(t')$ is periodic in time with $T$, the time-average of $f_{\rm FD}(\varepsilon_{n\bm k} (t'))$ in Eq. (\ref{eq:el_Distribution}) eventually reduces to $f_{\rm FD}(\varepsilon_{n\bm k}(\bm 0) )$.
Therefore, in the current calculation below, we may simply use the equilibrium distribution function $f_{\rm FD}(\varepsilon_{n\bm k} (\bm 0) )$.

\subsection{Adiabatic and nonadiabatic velocity}\label{sec:velocity}
To study the second-order nonlinear current generation induced by the dynamics of the phonon coordinate, we expand the wave-packet velocity in Eq. (\ref{eq:current_general}) up to the second order in ${{\bm Q}}$, assuming a small phonon amplitude. 
%
%
The wave-packet velocity consists of two terms, 
\begin{align}
    \label{eq:xad&xna}
    \dot{x}_{c,i} = \dot{x}^{\rm ad}_{c,i} + \dot{x}^{\rm na}_{c,i}.
\end{align}
The former one in Eq.~(\ref{eq:xad&xna}) is the adiabatic velocity, which contains terms proportional to $Q_\alpha Q_\beta$ or $\dot{Q}_\alpha Q_\beta$, 
      \begin{align}\label{eq:WPvel_ad}
        \dot{x}^{\rm ad}_{c,i}
        &=
        \frac{1}{\hbar}\frac{ \partial \varepsilon^{(2)}_{n\bm k_c} }{\partial k_{c,i}}
       +
        {\mathcal T}_{n,i\alpha\beta} \; \dot{Q}_\alpha Q_\beta,
      \end{align}
where 
\begin{align}
    \varepsilon^{(2)}_{n\bm k_c}
    &=
    \frac{1}{2} Q_{\alpha} Q_{\beta} 
         \left.
         {\rm Re}
        \Bigl[ 
              \varepsilon_{n\bm k_c}({{\bm Q}})
              \braket{ 
               u_{n\bm k_c}({{\bm Q}}) |
              \partial_{Q_\alpha}  \partial_{Q_\beta} u_{n\bm k_c}({{\bm Q}}) }
        \Bigr]  \right|_{{{\bm Q}}= \bm 0}
        \nonumber \\
        &\ \ \ \ 
        +
        Q_\alpha Q_\beta 
         \left.
         {\rm Re}
        \Bigl[ 
              \bra{ u_{n\bm k_c}({{\bm Q}}) }
              \partial_{ Q_\beta }\hat{\mathcal H}_{\bm k}({{\bm Q}})
              \ket{ \partial_{Q_\alpha}u_{n\bm k_c}({{\bm Q}}) } 
        \Bigr] \right|_{{{\bm Q}}=\bm 0},
\end{align}
and ${\mathcal T}_{n,i\alpha\beta}$ consists of two parts, ${\mathcal T}_{n,i\alpha\beta} = {\mathcal T}^{\partial^2}_{n,i\alpha\beta} + {\mathcal T}^{\Omega}_{n,i\alpha\beta}$, 
\begin{widetext}
\begin{align}\label{eq:WPvel_ad_T}
    {\mathcal T}^{\partial^2}_{n,i\alpha\beta}
    &=
        2{\rm Im}
        \sum_{l(\ne n)}
        \left.
        \frac{ v^{i;\beta}_{nl} \lambda^\alpha_{ln} + v^{i}_{nl} \lambda^{\alpha\beta}_{ln} }
             { \varepsilon^2_{nl,\bm k_c} }
        \right|_{{{\bm Q}}=\bm 0},
        \\
    {\mathcal T}^{\Omega}_{n,i\alpha\beta}
    &=
       -2{\rm Im}\sum_{l(\ne n)}
       \left.
       \Biggl\{
                \frac{\partial_{k_{c,i}} \varepsilon_{nl,\bm k_c}}
                     {\varepsilon_{nl,\bm k_c}} 
                     a^{{{\bm Q}}}_{nl,\beta} a^{{{\bm Q}}}_{ln,\alpha}
              + \frac{\partial_{Q_\alpha} \varepsilon_{nl,\bm k_c}}
                     {\varepsilon_{nl,\bm k_c}} 
                     a^{\bm k_c}_{nl,i} a^{{{\bm Q}}}_{ln,\beta}
              +2\frac{\partial_{Q_\beta} \varepsilon_{nl,\bm k_c}}
                     {\varepsilon_{nl,\bm k_c}} 
                     a^{\bm k_c}_{nl,i} a^{{{\bm Q}}}_{ln,\alpha}
       \Biggr\}
       \right|_{{{\bm Q}}=\bm 0},
\end{align}
\end{widetext}
with 
$\varepsilon_{nl,\bm k_c} = \varepsilon_{n\bm k_c} - \varepsilon_{l\bm k_c}$
being the difference of the eigenenergy, 
$ v^{i;\beta}_{nl} = \bra{u_{n\bm k_c}({{\bm Q}})} \partial_{Q_\beta}\partial_{k_{c,i}}\hat{\mathcal H}_{\bm k_c}({{\bm Q}}) \ket{u_{l\bm k_c}({{\bm Q}})} $, and 
$\lambda^{\alpha\beta}_{ln} = \bra{u_{l\bm k_c}({{\bm Q}})} \partial_{Q_\beta}\partial_{Q_\alpha}\hat{\mathcal H}_{\bm k_c}({{\bm Q}}) \ket{u_{n\bm k_c}({{\bm Q}})}$.
In Eq. (\ref{eq:WPvel_ad_T}), 
${\mathcal T}^{\partial^2}_{n,i\alpha\beta}$ involves gradients of the Hamiltonian in both the momentum and the phonon-coordinate directions, as well as a second derivative with respect to the phonon coordinate.
Compared with the conventional group velocity, which arises from the single-band expectation value of the velocity operator, 
${\mathcal T}^{\partial^2}_{n,i\alpha\beta}$ may be viewed as an additional contribution for the group velocity. 
The first term in ${\mathcal T}^{\partial^2}_{n,i\alpha\beta}$ originates from phonon corrections to both the velocity operator and the electronic state in the expectation value, whereas the second term originates from the phonon correction to the electronic state alone.
On the other hand, in ${\mathcal T}^{\Omega}_{n,i\alpha\beta}$, the imaginary part of the product of the inter-band Berry connections appears and is weighted by the difference of the eigenenergies, which can be viewed as the nonadiabatic effect in our formalism. 
${\mathcal T}^{\Omega}_{n,i\alpha\beta}$ imprints the structure of the Berry curvature for the ${{\bm Q}}$-space and the mixed $(\bm k_c, {{\bm Q}})$-space.
%
%
%
%
%
%
%
%

$\dot{x}^{\rm na}_{c,i}$ in Eq.~(\ref{eq:xad&xna}) is the nonadiabatic velocity, which reads
%
      \begin{align}\label{eq:WPvel_na}
        \dot{x}^{\rm na}_{c,i}
        &=
         -\hbar \Gamma^{\bm k {{\bm Q}} {{\bm Q}}}_{[i, \alpha \beta],n} 
                \dot{Q}_\alpha \dot{Q}_\beta
         +\hbar\frac{\partial \varphi_{n,\alpha\beta}}{\partial k_{c,i}}\dot{Q}_\alpha \dot{Q}_\beta,
      \end{align}
where we introduce the Christoffel symbol for mixed parameter $(\bm k_c, {{\bm Q}})$,
\begin{align}
    \Gamma^{\bm k {{\bm Q}} {{\bm Q}}}_{[i, \alpha \beta],n} 
                &=
                \frac{1}{2} \sum_{l(\ne n)}
                \left.
                \frac{
                     \partial_{Q_\alpha} g^{\bm k {{\bm Q}}}_{i\beta,nl}
                     + \partial_{Q_\beta} g^{\bm k {{\bm Q}}}_{i\alpha,nl}
                     - \partial_{k_{c,i}}g^{{{\bm Q}} {{\bm Q}}}_{\alpha\beta,nl}
                }
                {\varepsilon_{nl,\bm k_c}}
                \right|_{{{\bm Q}}=\bm 0},
\end{align}
with $g^{\bm k {{\bm Q}}}_{i\alpha,nl} = {\rm Re}[ a^{\bm k_c}_{nl,i} a^{{{\bm Q}}}_{ln,\alpha} ]$ being the inter-band quantum metric,  
and the scalar potential induced by the phonon as
\begin{align}
    \varphi_{n,\alpha\beta}
    &= 
    \left.
                \sum_{l(\ne n)} 
                \frac{ g^{{{\bm Q}}{{\bm Q}}}_{\alpha\beta,nl}}{\varepsilon_{nl,\bm k_c}}
                \frac{\varepsilon_{l\bm k_c}-2\varepsilon_{nl,\bm k_c}}{\varepsilon_{nl,\bm k_c}}
    \right|_{{{\bm Q}}=\bm 0}
\end{align}
for electrons of the $n$-th band.
The nonadiabatic velocity $\dot{x}^{\rm na}_{c,i}$ is proportional to $\dot{Q}_\alpha \dot{Q}_\beta$. 
We note that the adiabatic velocity is in the order $\mathcal{O}(\hbar^0)$ and contains the Berry-curvature like structure, 
whereas 
the nonadiabatic velocity is $\mathcal{O}(\hbar^1)$ and contains the quantum metric and the corresponding Christoffel symbol.
%
%
%

Eqs.~(\ref{eq:WPvel_ad}) and (\ref{eq:WPvel_na}) can be viewed as a geodesic-like equation projected onto the momentum direction of an extended phase space $\xi = (\bm k, \bm x, {{\bm Q}})$. 
The Christoffel-symbol term represents a geometric inertial force generated by the motion of the phonon coordinate ${{\bm Q}}$, while $\partial_{k_i}\varphi_{n,\alpha\beta}$, i.e., the potential term acts as a conservative force in momentum space. 
The term $\dot{x}^{\rm na}_{c,i}$ reflects the underlying symplectic structure of the phase-space dynamics rather than a conventional second-order acceleration.
Note that this equation does not represent a purely Riemannian geodesic equation, but also the structure such that the Hamilton kinetics on symplectic manifold and the Christoffel geometry arising from the quantum geometry.


In what follows, since no bare variables $\bm{x}$ and $\bm{k}$ appear other than the coordinates of the wave-packet center, we omit the subscript $c$ for notational simplicity.

\subsection{ {Second-order current response to phonon coordinate} }

Based on the equations of motion in Eqs.~(\ref{eq:WPvel_ad}) and (\ref{eq:WPvel_na}), 
we formulate the second-order nonlinear current generated by the dynamics of the phonon coordinate. 
To this end, 
{we consider monochromatic optical phonons at the $\Gamma$-point,}
%
      \begin{align}
        {{\bm Q}} (t)
        &=
         {{\bm Q}}(\Omega) e^{-i\Omega t} + {{\bm Q}}(-\Omega) e^{i\Omega t}.
      \end{align}
%
Here, $\Omega$ is the eigen-frequency of the optical mode, 
and 
${{\bm Q}}(\omega)$ [$={{\bm Q}}^*(-\omega)$] is the amplitude. 


Combining the wave-packet velocity in Eqs. (\ref{eq:WPvel_ad}) and (\ref{eq:WPvel_na}) and the formula of the electric current in Eq. (\ref{eq:current_general}) with the distribution function in Eq. (\ref{eq:el_Distribution}), 
we find that the phonon induces 
DC and SHG responses,
\begin{align}\label{eq:J_DC_SHG}
    J_i (t) &= J_i (\omega=0) + 2{\rm Re} \Bigl[ J_i (\omega=2\Omega) e^{-2i\Omega t} \Bigr],
\end{align}
where 
\begin{align}\label{eq:J_Fourier}
        J_i (\omega) &= \frac{1}{T} \int_0^T dt \; J_i (t) e^{i\omega t}
\end{align}
is the Fourier component of current with frequency $\omega$.
The first term of Eq.(\ref{eq:J_DC_SHG}), the nonlinear DC current, is given by 
\begin{align}
   \label{eq:rectified_current}
    J_i (\omega=0)
    &=
    {\rm Re} \Bigl[ \sigma_{i;\alpha\beta} (0; \Omega, -\Omega) Q_\alpha(\Omega) Q_\beta(-\Omega)\Bigr],
\end{align}
where $\sigma_{i;\alpha\beta} (0; \Omega, -\Omega)$ is the nonlinear DC conductivity. 
The second term of Eq.(\ref{eq:J_DC_SHG}), the nonlinear SHG current, is given by 
\begin{align}
    \label{eq:SHGcurrent}
    J_i (\omega=2\Omega)
    &=
    \sigma_{i;\alpha\beta} (2\Omega; \Omega, \Omega) Q_\alpha(\Omega) Q_\beta(\Omega), 
\end{align}
with $\sigma_{i;\alpha\beta} (2\Omega; \Omega, \Omega)$ being the nonlinear SHG conductivity.

\subsubsection{DC conductivity }

{
We find that the DC conductivity consists of four contributions, 
\begin{align}
    \sigma_{i;\alpha\beta} (0; \Omega, -\Omega) 
&=
\sigma^{\rm ad, \Omega}_{i;\alpha\beta} (0; \Omega, -\Omega)
+\sigma^{\rm ad, \partial^2}_{i;\alpha\beta} (0; \Omega, -\Omega)\notag\\
&+\sigma^{\rm na}_{i;\alpha\beta} (0; \Omega, -\Omega)
+\sigma^{\rm surf}_{i;\alpha\beta} (0; \Omega, -\Omega),\label{eq:sigma-full}
\end{align}
where
\begin{align}
          \label{eq:sigma_ad_Omega_DC}
          &\sigma^{\rm ad, \Omega}_{i;\alpha\beta} (0; \Omega, -\Omega)
          =
          2ie \Omega \sum_n \int_{\bm k} f_{\rm FD}(\varepsilon_{n\bm k} (\bm 0) )
          {\mathcal T}^\Omega_{n,i\alpha\beta},
          \\
          \label{eq:sigma_ad_partial_DC}
          &\sigma^{\rm ad, \partial^2}_{i;\alpha\beta} (0; \Omega, -\Omega)
          =
          2i e \Omega \sum_n \int_{\bm k} f_{\rm FD}(\varepsilon_{n\bm k} (\bm 0) )
          {\mathcal T}^{\partial^2}_{n,i\alpha\beta},\\
          \label{eq:sigma_na_DC}
        &\sigma^{\rm na}_{i;\alpha\beta} (0; \Omega, -\Omega)
        \nonumber\\
        &\ \ 
        =
        2e\hbar \Omega^2\sum_n \int_{\bm k} f_{\rm FD}(\varepsilon_{n\bm k} (\bm 0) )
        \;
        \Bigl(
              \Gamma^{\bm k {{\bm Q}} {{\bm Q}}}_{[i,\alpha\beta],n}
              - 
              \varphi_{n,\alpha\beta}
        \Bigr),\\
  &\sigma^{\rm surf}_{i;\alpha\beta}(0; \Omega, -\Omega)
  =
  \frac{2e}{\hbar} \sum_n \int_{\bm k} f'_{\rm FD} (\varepsilon_{n\bm k}(\bm 0))
                  \frac{\partial \varepsilon_{n\bm k}(\bm 0)}{\partial k_i} 
                  \varepsilon^{(2)}_{n\bm k}.
\end{align}
}

{
The first two terms of $\sigma_{i;\alpha\beta} (0; \Omega, -\Omega)$, Eqs.~(\ref{eq:sigma_ad_Omega_DC}) and (\ref{eq:sigma_ad_partial_DC}), arise from the ${\mathcal T}_{n,i\alpha\beta}$-term in Eq. (\ref{eq:WPvel_ad}), which is related to the anomalous velocity. 
These terms are linear in the phonon frequency, indicating that they are adiabatic contributions.
The conductivity in Eq.~(\ref{eq:sigma_ad_Omega_DC}) can be understood as originating from the Berry curvature, as discussed in Sec.~\ref{sec:velocity}. 
To be precise, the contribution is given by a derivative of the interband Berry connection between the $n$-th and $l$-th bands weighted by their energy difference, not by the Berry curvature of $n$-th band. 
Here, the transitions to the other bands are driven by the dynamics of the phonon coordinate. 
Nevertheless, as we will see in the next section, 
the expression reduces to one involving the Berry curvature itself in two-band models.
We note that both ${\mathcal T}^\Omega_{n,i\alpha\beta}$ and ${\mathcal T}^{\partial^2}_{n,i\alpha\beta}$ involve derivatives with respect to the phonon coordinate ${{\bm Q}}$, in addition to momentum derivatives. Therefore, the derivatives cannot simply be transferred by integration by parts. 
Correspondingly, these conductivities cannot be reduced to Fermi-surface contributions, but are intrinsic conductivities involving all electrons in the Fermi sea.
}

The third term of $\sigma_{i;\alpha\beta} (0; \Omega, -\Omega)$ is a contribution quadratic in the phonon frequency, which arises from the nonadiabatic wave-packet velocity in Eq.(\ref{eq:WPvel_na}).
This conductivity represents the leading deviation from the adiabatic regime induced by the phonon dynamics with finite-frequency. 
Its expression involves the Christoffel symbols and the scalar potential. 
Rather than the quantum metric of the $n$-th band itself, the contribution is governed by the quantum metric weighted by the band energy differences and by its derivatives. 
As shown below, this contribution is expressed by the quantum metric in two-band models.

The fourth term is the Fermi surface contribution, which is inferred from the $\bm k$-derivative of $\varepsilon^{(2)}_{n\bm k}$ in Eq. (\ref{eq:WPvel_ad}).
Naively, in the presence of a Fermi surface, 
Eq.~\eqref{eq:sigma-full} implies a finite current in the zero-frequency limit of the phonon dynamics, i.e., a current generated by a static lattice displacement. 
This unphysical behavior is a consequence of our assumption that the relaxation time is much longer than the period of phonon dynamics, which is mentioned in Sec. III. A 
\footnote{See also Ref.~\cite{Meguro_PRB2026} for the detailed explanation and treatment of the surface term $\sigma^{\rm surf}_{i;\alpha\beta}$.}. 
%
A proper treatment of this contribution requires explicitly incorporating a finite relaxation time~\cite{Harada2023}, which lies outside the scope of the present work.
Hence, in this work, we focus on the contribution proportional to $\Omega^{n}$ ($n\ge1$).

For the case of the linearly polarized phonons, i.e., $\sigma_{i;\alpha\alpha}(0,\Omega,-\Omega)$, 
the adiabatic contributions $\sigma_{i;\alpha\alpha}^{\rm ad, \Omega/\partial^2} (0; \Omega,-\Omega)$ vanish as they are purely imaginary. 
%
This is also because the adiabatic velocity is proportional to $\dot Q_\alpha Q_\alpha$, whose time average vanishes over one oscillation period 
%
Physically, the phonon coordinates retrace the same path during each cycle, resulting in no net charge transport. 
Consequently, linearly polarized phonons generate the rectification current only from the nonadiabatic and Fermi-surface contributions.

\subsubsection{SHG conductivity}
Similar to the DC conductivity, the SHG conductivity consists of four terms, 
\begin{align}
\sigma_{i;\alpha\beta} (2\Omega; \Omega, \Omega) 
&= \sigma^{\rm ad, \Omega}_{i;\alpha\beta} (2\Omega; \Omega, \Omega)
+\sigma^{\rm ad, \partial^2}_{i;\alpha\beta} (2\Omega; \Omega, \Omega)
\nonumber\\
&
+\sigma^{\rm na}_{i;\alpha\beta} (2\Omega; \Omega, \Omega)
+\sigma^{\rm surf}_{i;\alpha\beta} (2\Omega; \Omega, \Omega), 
\end{align}
where
\begin{align}
          \label{eq:sigma_ad_Omega_SHG}
          &\sigma^{\rm ad, \Omega}_{i;\alpha\beta} (2\Omega; \Omega, \Omega)
          =
          ie \Omega \sum_n \int_{\bm k} f_{\rm FD}(\varepsilon_{n\bm k} (\bm 0) )\; 
          {\mathcal T}^\Omega_{n,i\alpha\beta},
          \\
          \label{eq:sigma_ad_partial_SHG}
          &\sigma^{\rm ad, \partial^2}_{i;\alpha\beta} (2\Omega; \Omega, \Omega)
          =
          i e \Omega \sum_n \int_{\bm k} f_{\rm FD}(\varepsilon_{n\bm k} (\bm 0) )\; 
          {\mathcal T}^{\partial^2}_{n,i\alpha\beta},
\\
\label{eq:sigma_na_SHG}
        &\sigma^{\rm na}_{i;\alpha\beta} (2\Omega; \Omega, \Omega)
        \nonumber\\
        &
        =
        -e\hbar \Omega^2 \sum_n \int_{\bm k} f_{\rm FD}(\varepsilon_{n\bm k} (\bm 0) )
        \; \left( \Gamma^{\bm k {{\bm Q}} {{\bm Q}}}_{[i,\alpha\beta],n} - \varphi_{n,\alpha\beta} \right),
\\
\label{eq:sigma_surf_SHG}
  &\sigma^{\rm surf}_{i;\alpha\beta}(2\Omega; \Omega, \Omega)
  =
  \frac{e}{\hbar} \sum_n \int_{\bm k} f'_{\rm FD} (\varepsilon_{n\bm k}(\bm 0))\; 
                  \frac{\partial \varepsilon_{n\bm k}(\bm 0)}{\partial k_i} 
                  \varepsilon^{(2)}_{n\bm k}.
\end{align}
Similar to the DC current, the first two terms arise from the anomalous velocity, the third term is related to the Christoffel symbol and the quantum metric, and the last term is the Fermi-surface contribution.
%
%
As can be seen from Eq.~\eqref{eq:J_DC_SHG}, 
the real and imaginary parts of the complex SHG amplitude, ${\rm Re}[J_i(\omega=2\Omega)]$ and ${\rm Im}[J_i(\omega=2\Omega)]$, correspond to the $\cos 2\Omega t$ and $\sin 2\Omega t$ components of the current, respectively.

\subsection{Inversion and time-reversal symmetries}\label{sec:III.D}

Next, we discuss the symmetry constraints on the nonlinear current. 
The electric current is a polar vector and odd under time-reversal operation.
The phonon coordinate is polar vector and time-reversal-even.
Consequently, {$\dot{{\bm Q}}$} is a time-reversal-odd polar vector.
These properties imply that $\sigma^{\rm ad}_{i;\alpha\beta}$ is odd under ${\mathcal P}$, even under ${\mathcal T}$, and odd under $\mathcal{PT}$ operations. 
On the other hand, $\sigma^{\rm na}_{i;\alpha\beta}$ is odd under ${\mathcal P}$, odd under ${\mathcal T}$, and even under $\mathcal{PT}$ operations.
Therefore, $\sigma^{\rm ad}_{i;\alpha\beta}$ is forbidden in ${\mathcal P}$-symmetric or ${\mathcal PT}$-symmetric materials, whereas $\sigma^{\rm na}_{i;\alpha\beta}$ is forbidden when either ${\mathcal P}$ or ${\mathcal T}$ exists.



\section{Dirac model}\label{sec:conDirac}


{Dirac Hamiltonian provides effective low-energy descriptions of various two-dimensional materials, including graphene-based systems and transition-metal dichalcogenides. 
Motivated by these {materials}, we first consider a Dirac model to analytically investigate the phonon-induced current and its quantum-geometric origin.}

\subsection{Hamiltonian}

{
Considering a graphene-like honeycomb lattice, we focus on two degenerate in-plane optical modes at the $\Gamma$-point, which is schematically depicted in Fig.~\ref{fig:current_schematic_honeycomb}(a). 
These modes describe relative displacements of the two sublattices along two orthogonal in-plane directions. 
Choosing these directions as the $x$ and $y$ axes, 
we denote the corresponding phonon coordinates by $Q_x$ and $Q_y$, respectively, and write ${\bm Q}=(Q_x,Q_y)$.
Then, Hamiltonian reads,}
\begin{align}
    \label{eq:Ham_conDirac}
    \hat{\mathcal H}_{\xi}(\bm k,{{\bm Q}})
     &=
     \bm R_\xi(\bm k, {{\bm Q}}) \cdot \hat{\bm \tau},
     \\
     %
     %
     \bm R_\xi(\bm k, {{\bm Q}})&=
     \begin{pmatrix}
         \hbar v\xi k_x -g Q_y
         - \xi\lambda(Q_x k_y 
         + Q_y k_x) \\
         -\hbar vk_y -g\xi Q_x
         - \lambda(-Q_x k_x 
         + Q_y k_y) \\
         \Delta - \xi\Delta_{\phi} \notag
     \end{pmatrix}.
\end{align}
%
%
Here, $\xi=+(-)$ is the helicity of the Dirac electron,
$\hat{\bm \tau}$ is the vector of Pauli matrices, 
$v$ is the velocity of the Dirac electron, 
$g$ and $\lambda$ are the coupling constant between electrons and phonons,
%
and $\Delta$ is the sublattice-dependent mass term (Semenoff mass) while $\Delta_\phi$ is the Haldane mass.
%
%
%
Here, in addition to the pseudo-gauge-field coupling proportional to $g$, 
we also consider a term that couples the momentum and $Q_{x,y}$ (the ones proportional to $\lambda$). 
The $\lambda$ couplings naturally appear in solid-state realizations of two-dimensional Dirac electrons, as discussed in Appendix.~C.

\begin{figure*}[t]
   \includegraphics[width=1.0\hsize]{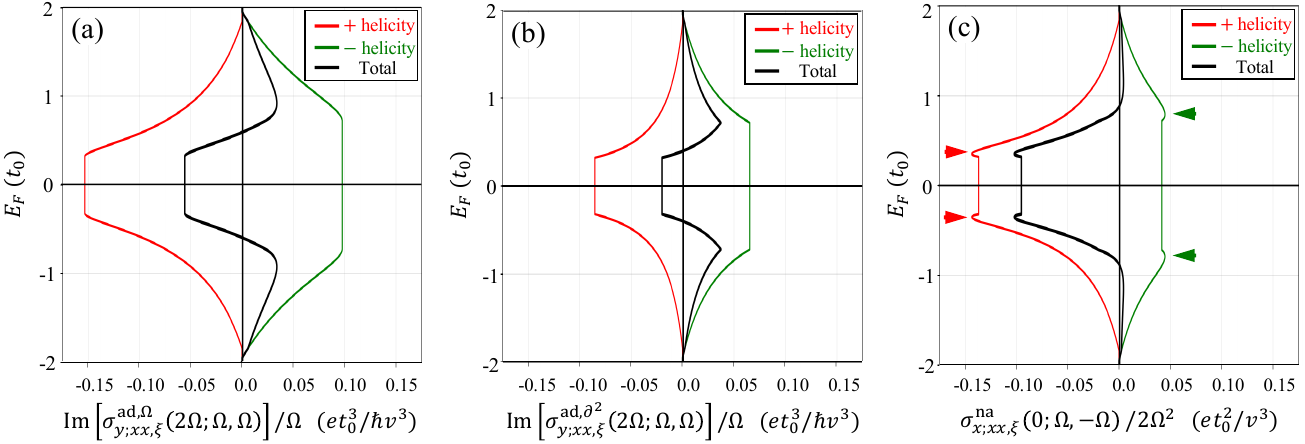}
\caption{
%
{
Fermi-level ($E_F$) dependences of the adiabatic SHG conductivities [(a) and (b)] and the nonadiabatic rectification conductivity (c) for the continuum Dirac Hamiltonian in Eq.~\eqref{eq:Ham_conDirac}.
The corresponding SHG~[(a) and (b)] and rectification~(c) currents are given by Eqs.~\eqref{eq:SHGcurrent} and \eqref{eq:rectified_current}, respectively. 
}
The nonadiabatic SHG conductivity is related to the rectification conductivity by $\sigma^{\rm na}_{x;xx,\xi}(0;\Omega,-\Omega)=-2\sigma^{\rm na}_{x;xx,\xi}(2\Omega;\Omega,\Omega)$.
In each panel, the red (green) line depicts the contribution from $\xi=+1$ ($\xi=-1$) Dirac electron, and the black line is the sum of the two.
The arrows in (c) depict the maxima of the nonadiabatic rectification conductivity near the band edges.
In the numerical evaluations of these conductivities,  
the model parameters are set as $\frac{\hbar v}{t_0}=\frac{\sqrt{3}}{2}, g=\frac{3\sqrt{3}t_0}{2}, \lambda=\frac{3t_0}{4}, \Delta=0.2t_0, $ and $\Delta_\phi = 0.3\sqrt{3} t_0$ with $t_0$ being an energy unit which is set as $t_0=1$. We also non-dimensionalize length unit.
On this parameter setting, $m_+ \approx -0.32t_0$ and $m_- \approx 0.72t_0$.
The cutoff energy is also set as $E_{\rm cut}=\hbar v\frac{2\pi}{3} \approx 1.81 t_0$.
On the lattice realization discussed in Sec.~\ref{sec:Haldane}, $\Lambda_{\rm cut}$ corresponding to this cutoff energy is momentum between $K$ and $M$ point in the Brillouin zone for the honeycomb lattice with lattice constant $a=1$.
}
\label{fig:sigma_conDiracmodel}
\end{figure*}


\subsection{${\mathcal T}^\Omega_{n,i\alpha\beta}$-part in adiabatic current}


%
For the two-band model, the adiabatic SHG conductivity arising from the Berry-curvature-structure, Eq.~\eqref{eq:sigma_ad_Omega_SHG}, reduces to
{
\begin{align}
 \sigma^{\rm ad, \Omega}_{i;\alpha\beta,\xi} (2\Omega; \Omega, \Omega)=i\Omega \; S^{\rm ad, \Omega}_{i;\alpha\beta,\xi},
 \end{align}
}
where
\begin{align}
  \label{eq:ad_conductivity_Omega}
  S^{\rm ad, \Omega}_{i;\alpha\beta,\xi}
  &=
  -e   \int_{\bm k} 
  W_{\bm k \xi} \; {\mathcal I}^{\rm ad, \Omega}_{i;\alpha\beta,\xi}(\bm k),
  \\
  W_{\bm k\xi} 
  &= f_{\rm FD}( R_\xi(\bm k, \bm 0) )-f_{\rm FD}(-R_\xi(\bm k, \bm 0)),
  \\
  \label{eq:ad_conductivity_Omega_I}
  {\mathcal I}^{\rm ad, \Omega}_{i;\alpha\beta,\xi}(\bm k)
  &=
  \left.
  \Biggl\{
        \frac{\partial R_\xi}{\partial {k_i}}
        \frac{\Omega^{{{\bm Q}} {{\bm Q}}}_{\beta\alpha, \xi}}{R_\xi}
       +\frac{\partial R_\xi}{\partial {Q_\alpha}}
       \frac{\Omega^{\bm k {{\bm Q}}}_{i\beta, \xi}}{R_\xi}
       +2\frac{\partial R_\xi}{\partial {Q_\beta}}
       \frac{\Omega^{\bm k {{\bm Q}}}_{i\alpha, \xi}}{R_\xi}
  \Biggr\}\right|_{{{\bm Q}}=0}, 
\end{align}
$R_\xi = |\bm R_\xi(\bm k, {{\bm Q}})|$ and
\begin{align}
        \Omega^{{{\bm Q}} {{\bm Q}}}_{\beta\alpha, \xi}
        &=
        -\frac{m_\xi}{2R^3_\xi} \Bigl(\partial_{Q_\beta} \bm R_\xi \times \partial_{Q_\alpha} \bm R_\xi \Bigr)_z,
        \\
        \Omega^{\bm k {{\bm Q}}}_{i\alpha, \xi}
        &=
        -\frac{m_\xi}{2R^3_\xi} \Bigl(\partial_{k_i} \bm R_\xi \times \partial_{Q_\alpha} \bm R_\xi \Bigr)_z,
\end{align}
are the Berry curvatures of the conduction band with ${m_\xi}=\Delta-\xi\Delta_\phi$ being the mass term for the helicity $\xi$.
The Berry curvature for the valence bands is the same as those for the conduction bands except for the sign, i.e., they are $-\Omega^{{{\bm Q}} {{\bm Q}}}_{ \beta\alpha,\xi}$ and $-\Omega^{\bm k {{\bm Q}}}_{ i\alpha,\xi}$.
The momentum integral $\int_{\bm{k}}$ is over a sphere $|\bm k|\le\Lambda_{\rm cut}$.
At $T=0$, $W_{\bm k\xi}$ serves as a weight function, restricting the integral to the exterior of the isotropic Fermi surface. %
%
%
%

In the following, we focus on 
the ${(y;xx)}$-component for $\sigma^{{\rm ad},\Omega}_{i;\alpha\beta, \xi}$.
{
This component describes a nonlinear current along the $y$ direction induced by lattice vibrations along the $x$ direction, as illustrated in Fig.~\ref{fig:current_schematic_honeycomb}(b).
}
At zero temperature, {$S^{{\rm ad},\Omega}_{y;xx,\xi}$} for the Hamiltonian in Eq.~\eqref{eq:Ham_conDirac} reads 
\begin{align}\label{eq:ad_conductivity_Omega_an}
{S^{{\rm ad},\Omega}_{y;xx,\xi}}
&=
\frac{3e g\lambda}{8\pi v^2}
\begin{cases}
\dfrac{2}{3}{\rm sign}(m_\xi) - F_\xi(E_{\rm cut}),
&
|E_F|<|m_\xi|,
\\[10pt]
F_\xi(|E_F|)- F_\xi(E_{\rm cut}),
&
|E_F|>|m_\xi|,
\end{cases}
\end{align}
where 
\begin{align}\label{eq:ad_conductivity_Omega_Fxi}
    F_\xi(x)
&=
\frac{m_\xi}{x}
-\frac{m_\xi^3}{3x^3},
\end{align}
$E_F$ is the Fermi energy, 
and $E_{\rm cut}=\hbar v \Lambda_{\rm cut}$ is the cutoff energy. 
%

In Fig.~\ref{fig:sigma_conDiracmodel}(a), the solid red~(green) line shows the $E_F$ dependence of the adiabatic SHG conductivity, Eq. (\ref{eq:ad_conductivity_Omega_an}), for the $\xi=+$ ($\xi=-$) Dirac electron; it corresponds to the electric current generated by that Dirac electron.
Here, we set the cutoff energy to $E_{\rm cut}=\hbar v\frac{2\pi}{3} \approx 1.81 t_0$.
The current direction of each Dirac electron depends on the sign of $\xi$, reflecting the sign of $m_\xi$ ($m_+ 
\approx -0.32t_0$ and $m_- \approx 0.72t_0$), and increases as $E_F$ approaches the gap centered at $E_F=0$.
Notably, the conductivity remains finite when $E_F$ lies within the gap, forming a plateau similar to that found for the non-adiabatic spin-motive force~\cite{Meguro_PRB2026}.
The height of the plateau depends on $E_\text{cut}$, as
evident from Eq.~(\ref{eq:ad_conductivity_Omega_an}).  
In the limit $E_{\rm cut}\rightarrow\infty$,  
the in-gap conductivity reads ${ S^{{\rm ad},\Omega}_{y;xx,\xi} }=\frac{e g\lambda}{4\pi v^2}\text{sgn}(m_\xi)$, which is independent of the magnitude of $m_\xi$ and gives the same conductivity-magnitude for $\xi=\pm$. 
Outside the gap, the magnitude of {$S^{{\rm ad},\Omega}_{y;xx,\xi}$} decreases monotonically with increasing $|E_F|$. 
%

The total conductivity induced by the phonon dynamics is obtained by the sum of the two contributions, which is shown by the solid black line in Fig.~\ref{fig:sigma_conDiracmodel}(a).
The total conductivity remains non-zero inside the gap $|E_F|\le\min(|m_+|,|m_-|)$ for a finite $E_\text{cut}$, if $m_+\ne m_-$; the two contributions cancel exactly only at $E_\text{cut}\to\infty$.
%
%
%
Outside the gap, the total conductivity shows a non-monotonic $|E_F|$ dependence with a broad peak at $|E_F|\sim\max(|m_+|,|m_-|)$ due to partial cancellation of the conductivity by the two nodes.

%
%


\subsection{${\mathcal T}^{\partial^2}_{n,i\alpha\beta}$-part in adiabatic current}

We next look into the adiabatic SHG conductivity arising from ${\mathcal T}^{\partial^2}_{n,i\alpha\beta}$, Eq.~(\ref{eq:sigma_ad_partial_SHG}).
For the helicity $\xi$, it is given by 
%
\begin{align}
\sigma^{\rm ad, \partial^2}_{i;\alpha\beta,\xi} (2\Omega; \Omega, \Omega)
=
i\Omega \; S^{\rm ad, \partial^2}_{i;\alpha\beta,\xi},
\end{align}
%
where
\begin{align}
    \label{eq:ad_conductivity_partial}
    S^{\rm ad, \partial^2}_{i;\alpha\beta,\xi}
    &=
    -e  \int_{\bm k} 
        W_{\bm k \xi}\; \mathcal{I}^{\rm ad, \partial^2}_{i;\alpha\beta,\xi}(\bm k),
    \\
    \label{eq:ad_conductivity_partial_I}
    \mathcal{I}^{\rm ad, \partial^2}_{i;\alpha\beta,\xi}(\bm k)
    &=
        \left.
        \frac{\bm R_\xi \cdot \Bigl( \partial_{k_i} \partial_{Q_\beta} \bm R_\xi \times \partial_{Q_\alpha} \bm R_\xi \Bigr)}
             { 4 |R_\xi|^3 }\right|_{{{\bm Q}}=0}.
\end{align}
%
%
In the zero-temperature limit, it reads, 
          \begin{align}\label{eq:ad_conductivity_partial_an}
            {S^{\rm ad, \partial^2}_{y;xx,\xi}}
            &=\frac{e g \lambda }{8\pi v^2 }
              \begin{cases}
                {\rm sign }(m_\xi) 
                - 
                \frac{m_\xi }{E_{\rm cut}},
                & \text{( $|E_F| < |m_\xi|$ )}, \\[10pt]
                \frac{m_\xi}{|E_F|} 
                - 
                \frac{m_\xi}{E_{\rm cut}} ,
                & \text{( $|E_F| > |m_\xi|$ )},
              \end{cases}
          \end{align}
%
%
%

The $E_F$ dependence of ${S^{\rm ad, \partial^2}_{y;xx,\xi}}$ is shown in Fig.~\ref{fig:sigma_conDiracmodel}(b) where the solid red~(green) line is the contribution from the $\xi=+(-)$ Dirac electron.
The overall behavior of ${S^{\rm ad, \partial^2}_{y;xx,\xi}}$ is similar to that of ${S^{\rm ad, \partial^2}_{y;xx,\xi}}$;
with a finite $E_\text{cut}$, a finite conductivity occurs in the insulating state, and a non-monotonic $E_F$ dependence appears as a consequence of the partial cancellation between the two nodes with different gap size. 
A slight difference, however, appears at $|E_F|=|m_\pm|$.
Since ${S^{\rm ad, \partial^2}_{y;xx,\xi}}$ 
contains only a factor proportional to $1/|E_F|$, as in Eq.~\eqref{eq:ad_conductivity_partial_an}, the red and green curves 
show a sharp edge at 
the ends of the plateau, resulting in a cusp of total conductivity (black line) at $|E_F|\sim\max(|m_+|,|m_-|)$.

%
%

\subsection{Nonadiabatic current}

The nonadiabatic velocity induce both rectification conductivity and SHG conductivity. 
For the two-band model, the corresponding formula,  Eqs.~\eqref{eq:sigma_na_DC} and \eqref{eq:sigma_na_SHG}, read, 
\begin{align}
\sigma^{\rm na}_{i;\alpha\beta,\xi} (0; \Omega, -\Omega)
=   2\Omega^2 S^{\rm na}_{ i;\alpha\beta,\xi}
= - 2\sigma^{\rm na}_{i;\alpha\beta,\xi} (2\Omega; \Omega, \Omega),
\end{align} 
where
\begin{align}
    \label{eq:na_conductivity}
    S^{\rm na}_{ i;\alpha\beta,\xi}
    &=
    -e\hbar \int_{\bm k} 
            W_{\bm k\xi}\; \mathcal{I}^{\rm na}_{ i;\alpha\beta,\xi}(\bm k)
    \\
    \label{eq:na_conductivity_I}
    \mathcal{I}^{\rm na}_{ i;\alpha\beta,\xi}(\bm k)
    &=
          \left.
          \left(
                \frac{ 1}{4}
                \frac{ \partial_{k_{c,i}}  g^{{{\bm Q}} {{\bm Q}}}_{ \alpha \beta} }
                     { R_\xi }
               -
                \frac{ \partial_{Q_{\alpha}}  g^{\bm k {{\bm Q}}}_{ i\beta}  }
                     { R_\xi }
               +
                \frac{5}{4}
                \frac{\partial R_\xi }{\partial k_{c,i}}
                \frac{ g^{{{\bm Q}} {{\bm Q}}}_{ \alpha\beta} }
                     { R_\xi^2  }
          \right)
          \right|_{{{\bm Q}}=0}.
\end{align}
Here, 
\footnotesize
\begin{align}\label{eq:2band_qm_QQ}
    g_{\alpha\beta,\xi}^{{{\bm Q}} {{\bm Q}}}
    &=
    \frac{ 
           m_\xi^2 (\partial_{Q_\alpha} \bm R_\xi) \cdot (\partial_{Q_\beta} \bm R_\xi)
         + (\bm R_\xi \times \partial_{Q_\alpha} \bm R_\xi)_z 
           (\bm R_\xi \times \partial_{Q_\beta} \bm R_\xi)_z
          }
    {4R_\xi^4},
\end{align}
\normalsize
\footnotesize
\begin{align}\label{eq:2band_qm_kQ}
    g_{i\alpha,\xi}^{\bm k {{\bm Q}}}
    &=
    \frac{ 
           m_\xi^2 (\partial_{k_i} \bm R_\xi) \cdot (\partial_{Q_\alpha} \bm R_\xi)
         + (\bm R_\xi \times \partial_{k_i} \bm R_\xi)_z 
           (\bm R_\xi \times \partial_{Q_\alpha} \bm R_\xi)_z
          }
    {4R_\xi^4},
\end{align}
\normalsize
are the quantum metrics for the conduction band 
in the ${{\bm Q}}$-space and the $(\bm k, {{\bm Q}})$-mixed space.
%
%
%
%
%
For instance, the {$(x;xx)$}-component of $S^{\rm na}_{ i;\alpha\beta,\xi}$
at zero temperature reads 
\begin{align}
\label{eq:na_conductivity_an}
{ S^{\rm na}_{x;xx,\xi} }
=
\frac{\xi e \hbar g\lambda}{64\pi v^2}
\begin{cases}
-\dfrac{64}{15|m_\xi|}
+G_\xi(E_{\rm cut}),
&
|E_F|<|m_\xi|,
\\[15pt]
-G_\xi(|E_F|)
+G_\xi(E_{\rm cut}),
&
|E_F|>|m_\xi|,
\end{cases}
\end{align}
with 
\begin{align}
\label{eq:sigma_na_an_Gxi}
G_\xi(x)
=
\frac{5}{x}
+\frac{8m_\xi^2}{3x^3}
-\frac{17m_\xi^4}{5x^5}.    
\end{align}
{
This component describes a nonlinear current along the $x$ direction induced by lattice vibrations along the same direction, as illustrated in Fig.~\ref{fig:current_schematic_honeycomb}(b).
}
The $E_F$ dependence of the nonadiabatic rectification conductivity is shown in Fig.~\ref{fig:sigma_conDiracmodel}(c), which shows an $E_F$ dependence similar to Figs.~\ref{fig:sigma_conDiracmodel}(a) and \ref{fig:sigma_conDiracmodel}(b). 

Compared with the adiabatic conductivities in Eqs. (\ref{eq:ad_conductivity_Omega_an}) and (\ref{eq:ad_conductivity_partial_an}), however, 
the nonadiabatic conductivity of each Dirac electron remains finite and shows a plateau in the gap, and its leading order is in $|m_\xi|^{-1}$. 
Therefore, unlike the adiabatic conductivity, the contributions from the two nodes do not generally cancel.
For the case in Fig.~\ref{fig:sigma_conDiracmodel}, 
the $\xi=+$ Dirac electron has the narrower gap than for the $\xi=-$ Dirac electron.
Hence, as shown in Fig.~\ref{fig:sigma_conDiracmodel}(c) by the red solid curve, the in-gap conductivity of the $\xi=+$ Dirac electron is greater than that of the $\xi=-$ Dirac electron.
%
%
%
Furthermore, the two contributions from $\xi=\pm$ Dirac electrons do not cancel even in the limit $E_{\rm cut} \rightarrow \infty$ ($G_\xi(E_F)\rightarrow 0$).
In fact, in this limit, the in-gap conductivity for the $\xi=\pm$ Dirac electron reads
$\frac{-\xi e \hbar g\lambda}{15\pi v^2 |m_\xi|}$ and explicitly depends on the magnitude of $|m_\xi|$, contrasting to the adiabatic conductivity.

Outside the gap, the nonadiabatic conductivity exhibits an $|E_F|$ dependence that is distinct from those of the adiabatic conductivities.
Near the band edge, the nonadiabatic conductivity of each Dirac electron exhibits a maximum, which is indicated by the arrow in Fig. \ref{fig:sigma_conDiracmodel} (c). 
This hump reflects the non-monotonicity of $G_\xi(E_F)$ in Eq.~(\ref{eq:sigma_na_an_Gxi}), i.e., the opposite signs between the third term and the first and second terms lead to the non-monotonicity.
%
%

%
%
%


The total nonadiabatic conductivity is shown in Fig. \ref{fig:sigma_conDiracmodel} (c) as the solid black line.
Its $E_F$ dependence is broadly similar to that of the adiabatic conductivity; the total conductivity forms a plateau with finite value if $|E_F|\le\min(|m_+|,|m_-|)$ and decreases monotonically with increasing $|E_F|$.
However, it also exhibits distinct features that are absent in the adiabatic conductivity.
Inside the gap $|E_F|\le\min(|m_+|,|m_-|)$, for the parameter set in Fig.~\ref{fig:sigma_conDiracmodel}, the total conductivity has the negative value, which reflects the fact that the $\xi=+$ Dirac electron gives the stronger negative contribution than $\xi=-$ mentioned above.
In addition, since the two contribution do not cancel exactly even in the limit $E_{\rm cut} \rightarrow \infty$, 
the in-gap total conductivity generally remains non-zero consequently.
Once $E_F$ moves outside the gap, it exhibits a hump near the band edges.
%
%
We emphasize that 
this maximum of the total nonadiabatic conductivity arises from the intrinsic maxima already present in the helicity-resolved conductivities, 
unlike the maximum in the adiabatic conductivity.

\section{Lattice models}\label{sec:modelcalc}
We next look into the impact of the nonadiabatic correction on the nonlinear electric conductivity in lattice models.
Unlike the Dirac model studied in Sec.~\ref{sec:conDirac}, whose result depends on the cutoff, lattice models are free of cutoffs.
Therefore, it provides a complementary viewpoint to the continuum model. 
\subsection{Haldane model} \label{sec:Haldane}

\subsubsection{Hamiltonian}

We first consider the Haldane model~\cite{Haldane1988} with the sublattice-dependent potential $\Delta$, the Hamiltonian of which reads 
\begin{align}\label{eq:Haldane_ham0}
        {\hat H}^{(0)}
        &=
         -t_0\sum_{\bm R}\sum_{l=1}^3
         \left(
               \; {\hat c}^{\dagger}_{A, \bm R + \bm \tau_l}{\hat c}_{B,\bm R}
               +
               h.c.
         \right)
         \nonumber \\
        &\ \ \ 
        +\Delta\sum_{\bm R}
        \left(
              {\hat c}^{\dagger}_{A, \bm R }{\hat c}_{A,\bm R}
             -
              {\hat c}^{\dagger}_{B, \bm R }{\hat c}_{B,\bm R}
        \right)
         \nonumber\\
         &\ \ \ 
        + 
        t_{\phi}\sum_{\braket{\braket{ij}}} e^{i\nu_{ij}\phi}
          {\hat c}^{\dagger}_{i} {\hat c}_{j},
      \end{align}
Here, ${\hat c}^{\dagger}_{A(B), \bm R }$ is the creation operator of spinless electrons on $A(B)$-sublattice of unit cell $\bm R$, $\nu_{ij}=1$ ($\nu_{ij}=-1$) for the clockwise (anti-clockwise) NNN hopping paths, and
$\bm \tau_l$ are the vectors that connect the two neighboring sites: 
$\bm \tau_1 = a(0,1/\sqrt{3})^{\rm T}$, 
$\bm \tau_2 = a(-1/2,-1/2\sqrt{3})^{\rm T}$, 
$\bm \tau_3 = a(1/2,-1/2\sqrt{3})^{\rm T}$.
%
%
%
%
%
%
The pristine Haldane model ($\Delta=0$) is centrosymmetric. Hence, both adiabatic and nonadiabatic conductivities are prohibited, as argued in Sec.~\ref{sec:III.D}.
The Semenoff mass breaks inversion symmetry, and hence, enables a non-zero phonon-induced current.
%

To calculate the nonlinear electric current induced by phonons, we consider the coupling of electrons to the linear order in phonon coordinate ${{\bm Q}}(t)$. 
{
As in the Dirac model, we consider two optical phonon modes at the $\Gamma$-point, ${\bm Q}=(Q_x,Q_y)$, as illustrated in Fig.~\ref{fig:current_schematic_honeycomb}(a). 
}
%
%
The electrons couple to the phonon coordinate through modulation of the hopping integrals by a lattice displacement.
%
For the nearest-neighbor hopping, 
one can find that the following electron-phonon coupling Hamiltonian should be added to ${\hat H}^{(0)}$ in Eq. (\ref{eq:Haldane_ham0}) (the detailed calculations for the derivation are presented in Appendix.~C), 
\begin{align}\label{eq:Haldane_HamQ}
        \delta {\hat H}({{\bm Q}}(t))
        &=
        -t_0 \sum_{\bm R}\sum_{l=1}^3 
             \left( \frac{\beta}{a^2}  ({{\bm Q}}(t) \cdot \bm\tau_l) \; 
                    {\hat c}^\dagger_{A,\bm R+ \bm \tau_l} {\hat c}_{B,\bm R} + h.c. 
             \right),
\end{align}
where $\beta$ is a dimensionless parameter with $\beta\simeq3$ for graphene, 
{
and ${{\bm Q}}(t) \cdot \bm\tau_l = Q_x(t)\tau_{l,x} + Q_y(t)\tau_{l,y}$.
}
As mentioned above, we focus on the optical phonon at the $\Gamma$-point; the NNN hopping associated with Haldane mass does not change up to the first order of the phonon coordinate.
{
Acoustic phonons also do not modify the hopping integrals and hence do not couple to electrons through hopping modulation, 
since they at $\bm q=\bm 0$ correspond to simply rigid translations of the lattice. 
}
By performing the Fourier transformation, 
${\hat c}_{A/B, \bm R} = \frac{1}{\sqrt{N}} \sum_{\bm k} {\hat c}_{A/B, \bm k} e^{i\bm k\cdot \bm R}$, 
to the full Hamiltonian ${\hat H} = {\hat H}^{(0)} + \delta {\hat H}$, 
the Bloch Hamiltonian reads,
      \begin{align}
        \label{eq:Haldane_fullHam_Bloch}
        \hat{\mathcal H}(\bm k, {{\bm Q}})
        &=
        \sum_{\mu=0,1,2,3} R_{\mu}(\bm k, {{\bm Q}}) \hat{\tau}_{\mu}
        \\
        R_1 &= -t_0  \sum_{l=1}^3 
              \left( 1 + \frac{\beta }{a^2} ( {{\bm Q}}(t) \cdot \bm \tau_l ) \right)
              {\rm cos}(\bm k\cdot\bm \tau_l)
        \notag\\
        R_2 &= -t_0  \sum_{l=1}^3 
              \left( 1 + \frac{\beta }{a^2} ( {{\bm Q}}(t) \cdot \bm \tau_l ) \right)
              {\rm sin}(\bm k\cdot\bm \tau_l)
        \notag\\
        R_3 &= \Delta -2t_\phi{\rm sin}\phi \sum_{n=1,2,3}{\rm sin}(\bm k\cdot \bm a_n) \notag
        \\
        R_0 &= 2t_\phi{\rm cos}\phi \sum_{n=1,2,3}{\rm cos}(\bm k\cdot \bm a_n) \notag
      \end{align}
      where $\bm a_1 = \bm \tau_1 -\bm \tau_2, \bm a_2 = \bm \tau_2-\bm \tau_3, \bm a_3 = \bm \tau_3-\bm \tau_1$, and $\hat{\tau}_0$ is the $2\times 2$ identity matrix.


\begin{figure*}[t]
   \includegraphics[width=1.0\hsize]{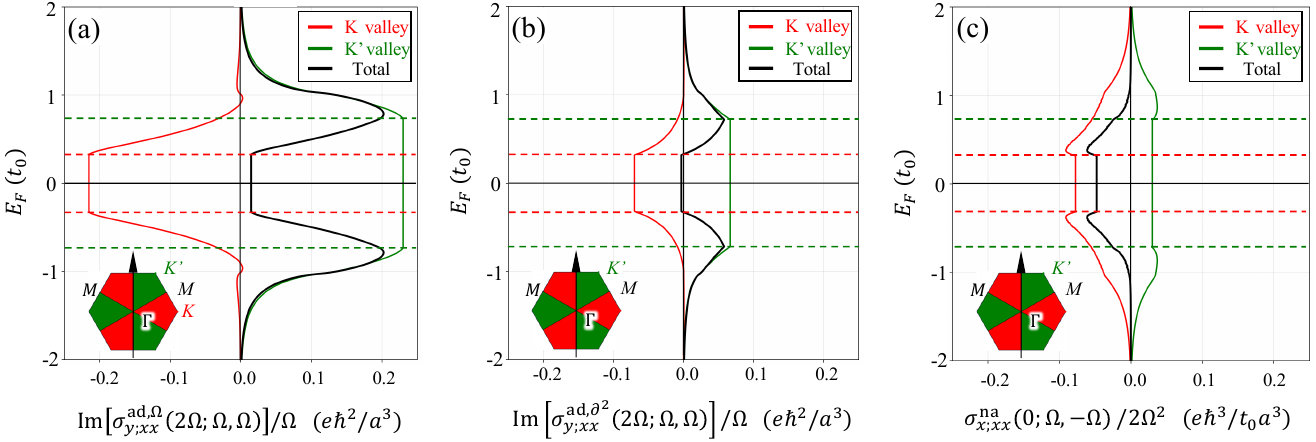}
\caption{
{Fermi-level ($E_F$) dependences of 
the adiabatic SHG conductivities [(a) and (b)] 
and 
the nonadiabatic rectification conductivity (c) 
for the Haldane model in Eq.~(\ref{eq:Haldane_ham0}) are depicted by the black solid line in each panel.
The nonadiabatic SHG conductivity is related to the rectification conductivity by $\sigma^{\rm na}_{x;xx,\xi}(0;\Omega,-\Omega)=-2\sigma^{\rm na}_{x;xx,\xi}(2\Omega;\Omega,\Omega)$.}
%
The red (green) line represents the contribution from electrons around $K$ ($K'$) valley in BZ.
The decomposition of the black solid line is depicted as inset in each panel;
we define the $\overline{M}$ point as the mirror image of the
$M$ point with respect to the $k_y$ axis. 
The Brillouin zone is then divided into six regions by the $\Gamma$--$M$ line, the $\Gamma$--$\overline{M}$ line, and the $k_y$ axis. 
The integral over each region containing the $K$ ($K'$) valley is assigned to the contribution from the $K$ ($K'$) valley.
The model parameters are set as $t_0=1, a=1, \Delta=0.2, t_\phi=0.1, $ and $\phi=\frac{\pi}{2}$.
On this parameter setting, the band edge for $K$-valley is approximately $0.32$ whereas that for $K'$-valley is $ 0.72$.
}
\label{fig:sigma_Haldanemodel}
\end{figure*}

The Haldane model in Eq.~\eqref{eq:Haldane_ham0} consists of two bands: One valence and one conduction band. 
Typically, the bottom of the conduction band and the top of the valence band are at the $K$ and $K'$ points, whose effective low energy model is given by massive Dirac electrons.
Indeed, expanding the Bloch Hamiltonian in Eq.~(\ref{eq:Haldane_fullHam_Bloch}) around these points yields the continuum Dirac Hamiltonian in Eq.~(\ref{eq:Ham_conDirac}), where the $\xi=+$ and $\xi=-$ models correspond to the $K$ and $K'$ valleys, respectively. 
The parameters in the Dirac Hamiltonian read,
\begin{align}\label{eq:lattice_conDirac_corr}
v=\frac{\sqrt{3}at_0}{2\hbar},
\qquad
g=\frac{\sqrt{3}\beta t_0}{2a},
\qquad
\lambda=\frac{\beta t_0}{4}. 
 \end{align}
For the numerical evaluation of conductivities, 
we set the parameters as 
$a=1, \Delta=0.2 t_0, t_\phi= 0.1 t_0, $ and $ \phi = \frac{\pi}{2}$. 
%

\subsubsection{Conductivity}

The symmetry of the Hamiltonian often poses constraints on the conductivities
$\sigma^{\rm ad,\Omega}_{i;\alpha\beta}$, $\sigma^{\rm ad,\partial^2}_{i;\alpha\beta}$, and $\sigma^{\rm na}_{i;\alpha\beta}$. 
For the Haldane model, the $C_{3z}$ symmetry implies 
$\sigma_{x;xx}=-\sigma_{x;yy}=-\sigma_{y;xy}=-\sigma_{y;yx}$ and 
$\sigma_{y;xx}=-\sigma_{y;yy}=\sigma_{x;xy}=\sigma_{x;yx}$
.
Hence, there are only two independent matrix elements in the conductivity tensor, {$\sigma_{x;xx}$ and $\sigma_{y;xx}$}.
This constraint holds for all contributions, $\sigma^{\rm ad,\Omega}_{i;\alpha\beta}$, $\sigma^{\rm ad,\partial^2}_{i;\alpha\beta}$, and $\sigma^{\rm na}_{i;\alpha\beta}$.

In addition, the magnetic mirror symmetry ${\mathcal M}_x {\mathcal T}$ further poses constraints on the conductivity tensor. 
The anti-unitary operator corresponding to the ${\mathcal M}_x {\mathcal T}$ operation is defined by $\hat{\Theta}_{{\mathcal M}_x {\mathcal T}} =\hat{\mathcal{M}}_x \mathcal{K}$. 
Here, $\mathcal{K}$ is the complex conjugation operator.
Meanwhile, 
$\hat{\mathcal{M}}_x$ is the mirror operator acting on the electron-creation/annihilation operators such that $ \hat{{\mathcal M}}_x \hat{c}_{A/B, \bm k} \hat{\mathcal M}_x^{-1} = \hat{c}_{A/B, \bar{\bm k}}$, where $\bar{\bm k}= (s^{\bm k}_{x}k_x, s^{\bm k}_{y}k_y) $ with $s^{\bm k}_{x}=+1,\; s^{\bm k}_{y}=-1$.
%
%
%
%
%
%
%
%
Under the symmetry, the Bloch Hamiltonian satisfies the relation $\hat{\mathcal H}(\bm k,{{\bm Q}}) = \hat{\mathcal H}^*(\bar{\bm k}, {\bar{{\bm Q}}})$.
%
%
Here, ${\bar{{\bm Q}}}$ is defined as ${\bar{{\bm Q}}}= (s^{{{\bm Q}}}_{x}Q_x, s^{{{\bm Q}}}_{y}Q_y) $ with $s^{{{\bm Q}}}_{x}=-1,\; s^{{{\bm Q}}}_{y}=+1$.
Consequently, the vector $R_\mu$ satisfies $R_{0,1,3}(\bm k, {{\bm Q}}) = R_{0,1,3}(\bar{\bm k}, {\bar{{\bm Q}}})$ and $R_{2}(\bm k, {{\bm Q}}) = -R_{2}(\bar{\bm k}, {\bar{{\bm Q}}})$.
Meanwhile, the following relations of geometric quantities are satisfied: 
$g^{\bm k{{\bm Q}}}_{i\alpha}(\bm k,{{\bm Q}}) 
= 
s^{\bm k}_i 
s^{{{\bm Q}}}_\alpha 
g^{\bm k{{\bm Q}}}_{i\alpha}(\bar{\bm k}, {\bar{{\bm Q}}})$ 
and 
$\Omega^{\bm k{{\bm Q}}}_{i\alpha}(\bm k,{{\bm Q}}) 
= 
-s^{\bm k}_i 
s^{{{\bm Q}}}_\alpha 
\Omega^{\bm k{{\bm Q}}}_{i\alpha}(\bar{\bm k}, {\bar{{\bm Q}}})$.
The relations for $g^{{{\bm Q}}{{\bm Q}}}_{\alpha\beta}$ and $\Omega^{{{\bm Q}}{{\bm Q}}}_{\alpha\beta}$ are supposed in the same way.
We summarize the derivations of the above relations in Appendix.~D.
%
%
%
%
%
These relations lead to the constraints for the conductivities as
$\sigma^{\rm ad,\Omega/\partial^2}_{i;\alpha\beta} = -s^{\bm k}_{i}s^{{{\bm Q}}}_{\alpha}s^{{{\bm Q}}}_{\beta} \sigma^{\rm ad,\Omega/\partial^2}_{i;\alpha\beta}$
and 
$\sigma^{\rm na}_{i;\alpha\beta} = s^{\bm k}_{i}s^{{{\bm Q}}}_{\alpha}s^{{{\bm Q}}}_{\beta} \sigma^{\rm na}_{i;\alpha\beta}$.
Therefore, $\sigma^{\rm ad,\Omega/\partial^2}_{x;xx}$ is odd under ${\mathcal M}_x{\mathcal T}$, whereas $\sigma^{\rm ad,\Omega/\partial^2}_{y;xx}$ is even;
it implies
$
\sigma^{\rm ad}_{x;xx}=
\sigma^{\rm ad}_{x;yy}=
\sigma^{\rm ad}_{y;xy}=
\sigma^{\rm ad}_{y;yx}
=0.
$
Similarly, for the nonadiabatic conductivity, $\sigma^{\rm na}_{x;xx}$ is even under ${\mathcal M}_x{\mathcal T}$, whereas $\sigma^{\rm na}_{y;xx}$ is odd. 
%
Therefore,
$
\sigma^{\rm na}_{y;xx}=
\sigma^{\rm na}_{y;yy}=
\sigma^{\rm na}_{x;xy}=
\sigma^{\rm na}_{x;yx}
=0.
$
Hence, we focus on $\sigma^{\rm ad,\Omega/\partial^2}_{y;xx}$ and $\sigma^{\rm na}_{x;xx}$ in the following.
{
These representative conductivity components describe the current responses to the $x$-polarized optical phonon, as discussed for the Dirac model and schematically illustrated in Fig.~\ref{fig:current_schematic_honeycomb}(b).
}

Since Eq.~(\ref{eq:Haldane_fullHam_Bloch}) is a two-band model, the conductivity formulae in Eqs. (\ref{eq:ad_conductivity_Omega}), (\ref{eq:ad_conductivity_partial}), and (\ref{eq:na_conductivity}) can be used to compute the conductivity. 
We replace $R_{\xi,\mu}$ of Eq.~(\ref{eq:Ham_conDirac}) to $R_\mu$ of Eq.~(\ref{eq:Haldane_fullHam_Bloch}) and change the range of integral to the 1st BZ of the honeycomb lattice. 
We denote the conductivities obtained through these replacements as $\sigma^{\rm ad, \Omega}_{i;\alpha\beta}~(S^{\rm ad, \Omega}_{i;\alpha\beta})$, 
$\sigma^{\rm ad, \partial^2}_{i;\alpha\beta}~(S^{\rm ad, {\partial^2}}_{i;\alpha\beta})$, 
and 
$\sigma^{\rm na}_{i;\alpha\beta}~(S^{\rm na}_{i;\alpha\beta})$.
%
%
%

The black solid lines in Fig.~\ref{fig:sigma_Haldanemodel} show the $E_F$ dependence of the adiabatic and nonadiabatic conductivities. 
%
%
%
In Fig.~\ref{fig:sigma_Haldanemodel}(a), 
$S^{\rm ad, \Omega}_{y;xx}$ forms a plateau inside the gap (between the two red dashed lines). 
Outside the gap, $S^{\rm ad, \Omega}_{y;xx}$ shows a strong $E_F$ dependence with a maximum slightly below the $K'$-valley band edge. 
One can also observe a discontinuous $E_F$ dependence near $E_F \approx 1.02$ which corresponds to the energy of the high-symmetry $M$ point in BZ.
This discontinuity is also observed in Figs. (b) and (c).
The behavior of $S^{\rm ad, {\partial^2}}_{y;xx}$ shown in Fig.~\ref{fig:sigma_Haldanemodel} (b) is similar to that of $S^{\rm ad, {\Omega}}_{y;xx}$ in Fig.~\ref{fig:sigma_Haldanemodel}(a). 
Its $E_F$ dependence is sharper than that for $S^{\rm ad, {\Omega}}_{y;xx}$.
The $E_F$ dependence of $S^{\rm na}_{x;xx}$, shown in Fig.~\ref{fig:sigma_Haldanemodel}(c), also resembles that of the adiabatic conductivity; $S^{\rm na}_{x;xx}$ forms a plateau inside the gap and decays to zero at high energies, away from the gap. 
In contrast to the adiabatic conductivity, the magnitude of the in-gap nonadiabatic conductivity increases as the gap width decreases. 
%
Immediately outside the band edges of the $K$ valley, we find that the $E_F$ dependence of $S^{\rm na}_{x;xx}$ exhibits a characteristic hump maximum, which is absent in the adiabatic conductivity.
%
%

%
%
%
To identify the origin of the $E_F$ dependence described above, we decompose the momentum integral into contributions of the $K$ and $K'$ valleys. 
To this end, we group the Brillouin zone into two subspaces
as illustrated in the inset of Fig.~\ref{fig:sigma_Haldanemodel}; The electrons closer to the $K$ point and those around $K'$ point.
The contribution from the $K$ ($K'$) valley is shown by the red~(green) solid line in Fig.~\ref{fig:sigma_Haldanemodel}; by definition, their sum gives the total conductivity.
%
The two contributions and the total conductivity resemble those of the continuum model in Sec.~\ref{fig:sigma_conDiracmodel}.
This coincidence is reasonable considering that the low-energy expansion of the Haldane model about each valley reduces to the Dirac Hamiltonian in Sec.~\ref{sec:conDirac}, as mentioned in Eq.~(\ref{eq:lattice_conDirac_corr}).
%
%
%
Indeed, the $E_F$ dependences of the valley-resolved conductivities in Fig.~\ref{fig:sigma_Haldanemodel} are qualitatively reproduced by the continuum results in Fig.~\ref{fig:sigma_conDiracmodel} 
when $\Lambda_{\rm cut}$ is chosen to the distance between the $K$ and $M$ points in the BZ.
Both the in-gap plateaus of the conductivity for each valley contribution and their decay outside the corresponding gaps 
are the consequences of the Dirac electrons. 
In addition, the valley decomposition in Fig.~\ref{fig:sigma_Haldanemodel}(c) can trace the hump-like $E_F$ dependence of the nonadiabatic conductivity.
%
%
%
From the above observations, 
we conclude that the main features of the conductivity in the Haldane model can be largely understood in terms of the Dirac electrons near the $K$ and $K'$ valleys.

We note that there is noticeable difference between the lattice and continuum models regarding the Berry curvature contribution inside the gap.
The in-gap total conductivity in Fig.~\ref{fig:sigma_Haldanemodel}(a) deviates from its continuum counterpart in Fig.~\ref{fig:sigma_conDiracmodel}(a). 
Although the continuum Dirac model correctly captures the Berry-curvature structure at low energies, 
it cannot correctly describe the contributions at high energies, i.e., the states away from the $K$ and $K'$ points. 
This limitation is relevant to Fermi-sea contributions, whose values depend on the ultraviolet completion supplied by the lattice model. 
Hence, the discrepancy suggests that, at least for $S^{\rm ad, \Omega}_{y;xx}$, electronic states away from the Dirac cones also contribute appreciably to the in-gap current. 
%
%
%
%

Outside the gap, the conductivities on the lattice model show differences from that on the continuum model.
$S^{\rm ad, \Omega/\partial^2}_{y;xx}$ on the lattice model shown in Figs. \ref{fig:sigma_Haldanemodel}(a) and (b) decrease more rapidly with increasing $|E_F|$, compared with that of the continuum model. 
Regarding $S^{\rm na}_{x;xx}$, its hump structure in Fig. \ref{fig:sigma_Haldanemodel}(c) is broader than that in Fig. \ref{fig:sigma_conDiracmodel} (c).
In addition, both $S^{\rm ad, \Omega/\partial^2}_{y;xx}$ and $S^{\rm na}_{x;xx}$ exhibit weak non-monotonic $E_F$ dependence near the energy corresponding to the $M$ point. 
These differences have two main origins. 
Firstly, once $|E_F|$ reaches the energy of the $M$-point in BZ, the electronic states in the Fermi sea can no longer be described by the continuum Dirac model of Eq.(\ref{eq:Ham_conDirac}). 
Secondly, the continuum model in Sec.~\ref{sec:conDirac} retains only the isotropic terms linear in momentum, whereas the full lattice model contains higher-order momentum corrections, including trigonal warping.
%
Thus, we conclude the Haldane-model conductivities are largely governed by Dirac electrons near the $K$ and $K'$ points, with quantitative deviations arising from states away from the Dirac points and higher-order lattice corrections.

We now examine the validity of interpreting the results on the Haldane lattice model in terms of the continuum Dirac model.
For this purpose, we introduce the polar coordinates, $\bm{k}=(k\cos\theta,k\sin\theta)$, and decompose the integrands of the conductivities in Eqs.~(\ref{eq:ad_conductivity_Omega_I}), (\ref{eq:ad_conductivity_partial_I}), and (\ref{eq:na_conductivity_I}) into angular multipoles, 
\begin{align}  
  \mathcal{I}^{\rm ad, \Omega/\partial^2}_{y;xx,\xi}(k,\theta)
  &=
  \sum_{l=0}^{3}
  \mathcal{C}^{\rm ad,\Omega/\partial^2}_{\xi,l}(k)\; {\rm cos}(l\theta),
  \\
  \mathcal{I}^{\rm na}_{x;xx,\xi}(k,\theta)
  &=
  \sum_{l=0}^{5}
  \mathcal{C}^{\rm na}_{\xi,l}(k)\; {\rm cos}(l\theta).
\end{align}
The full multipole expansions are given in Appendix.~E. 
We then expand the lattice Hamiltonian of Eq.~(\ref{eq:Haldane_fullHam_Bloch}) up to the second order of the momentum $k$ around the $K$ and $K'$ points. 
The $\mathcal O(k^2)$ terms modify both the Fermi surface and the conductivity integrands. 
However, the latter corrections are either quantitatively small or subleading in $k$ compared with the terms already present in the linear Dirac model.
We therefore assume the dominant effect of the $\mathcal O(k^2)$ terms is the Fermi surface deformation.

For the isotropic Fermi surface of the linear Dirac model, only the monopole component ($l=0$) survives angular integration.
The $\mathcal O(k^2)$ terms induce a trigonal deformation, allowing the octupole component ($l=3$) to survive as well.
The relevant monopole and octupole components are given by
\begin{align}
  \mathcal{C}^{\rm ad,\Omega}_{\xi,0}(k) &= \frac{3m_\xi g\lambda \hbar^2v^2 k^2}{4 R^5_\xi}, \ \  
  \mathcal{C}^{\rm ad,\Omega}_{\xi,3}(k) = \frac{3\xi m_\xi \lambda^2 \hbar^2v^2 k^3}{4 R^5_\xi},
  \\
  &\ \ \ \ \ \ \ \ \ 
  \mathcal{C}^{\rm ad,\partial^2}_{\xi,0}(k) = \frac{m_\xi g\lambda }{4 R^3_\xi},
  \\
  \label{eq:na_monopole}
  \mathcal{C}^{\rm na}_{\xi,0}(k)
    &=
    \frac{\xi g\lambda}{16 R^7_\xi}
    \left( 2m_\xi^4 -9m_\xi^2 \hbar^2 v^2k^2 -\frac{5}{2}\hbar^4v^4k^4 \right),
  \\
  \mathcal{C}^{\rm na}_{\xi,3}(k)
    &=
    \Bigl( \frac{17 }{4}g^2 - 6m_\xi^2 \lambda^2  \Bigr) \frac{\hbar^2 v^2 k^3}{16 R^7_\xi} 
    -\frac{7\lambda^2 \hbar^4 v^4 k^5}{64 R^7_\xi}.
\end{align}
We find that the octupole correction leaves the in-gap behavior unchanged, but slightly modifies the response outside the gaps. 
In particular, the adiabatic conductivity $S^{\rm ad,\Omega}_{y;xx,\xi}$ decays more rapidly, whereas the hump structure of the nonadiabatic conductivity $S^{\rm na}_{x;xx,\xi}$ near the gap edge becomes broader. 
These changes bring the continuum-model results into closer agreement with those of the full lattice model.

Finally, we discuss the hump-like peak of the nonadiabatic conductivity and clarify its quantum-geometric origin.
As mentioned above, this peak mainly originates from the monopole component, Eq.~(\ref{eq:na_monopole}), in the integrand.
At zero temperature and outside the gap, the $E_F$ dependence of $S^{\rm na}_{x;xx,\xi}$ is determined by the derivative regarding $E_F$, and it reads 
$ d S^{\rm na}_{x;xx,\xi}/dE_F \propto {\mathcal{C}}^{\rm na}_{\xi,0}(k_F)$ where $k_F$ is the Fermi wavenumber;
the extremum of $S^{\rm na}_{x;xx,\xi}$ directly corresponds to a sign change of the monopole component ${\mathcal{C}}^{\rm na}_{\xi,0}$ at the Fermi surface.
This sign change directly results from a competition among the three quantum-metric-derived terms entering $\mathcal{I}^{\rm na}_{x;xx,\xi}$, i.e.,  
$\partial_{k_i} g^{{{\bm Q}}{{\bm Q}}}_{\alpha\beta}$, 
$\partial_{Q_\alpha} g^{\bm k{{\bm Q}}}_{i\beta}$, and
$g^{{{\bm Q}}{{\bm Q}}}_{\alpha\beta}$ in Eq.~(\ref{eq:na_conductivity_I}).
%
%
Thus, the hump reflects the sign competition among distinct metric-derived contributions and the structure of the quantum metric for $(\bm k,{{\bm Q}})$-space for the Dirac electron.

We relate the metric-derived structure above to the parameter dependence of the Dirac Bloch eigenstates. In the honeycomb lattice, interference among inter-sublattice hopping paths makes the relative phase and mixing of the two sublattice components momentum dependent. 
A modulation of these hoppings by the optical phonon further makes the Bloch eigenstate {$|u_{n}(\bm{k};{\bm Q})\rangle$} dependent on the phonon coordinate {${\bm Q}$}.
The quantum metric characterizes this deformation of the eigenstate in the combined {$(\bm{k},{\bm Q})$} parameter space; for the Dirac Hamiltonian, it can equivalently be expressed as Eqs.~(\ref{eq:2band_qm_QQ}) and (\ref{eq:2band_qm_kQ}).
In the continuum expansion, this phonon-induced deformation of the Bloch states is encoded in the terms proportional to $gQ_\alpha$ and $\lambda k_iQ_\alpha$ in Eq.~(\ref{eq:Ham_conDirac}). 
In particular, the $\lambda k_iQ_\alpha$ terms make the phonon-induced deformation itself momentum dependent, thereby generating the nontrivial $g^{{{\bm Q}}{{\bm Q}}}$ and $g^{\bm k{{\bm Q}}}$ structures entering $S^{\rm na}$. 
The hump-like structure can therefore be viewed as a manifestation of the mixed {$(\bm{k},{\bm Q})$}-space quantum geometry of the phonon-coupled Dirac electrons.

\subsubsection{ Parameter dependence }

\begin{figure}[t]
   \includegraphics[width=1.0\hsize]{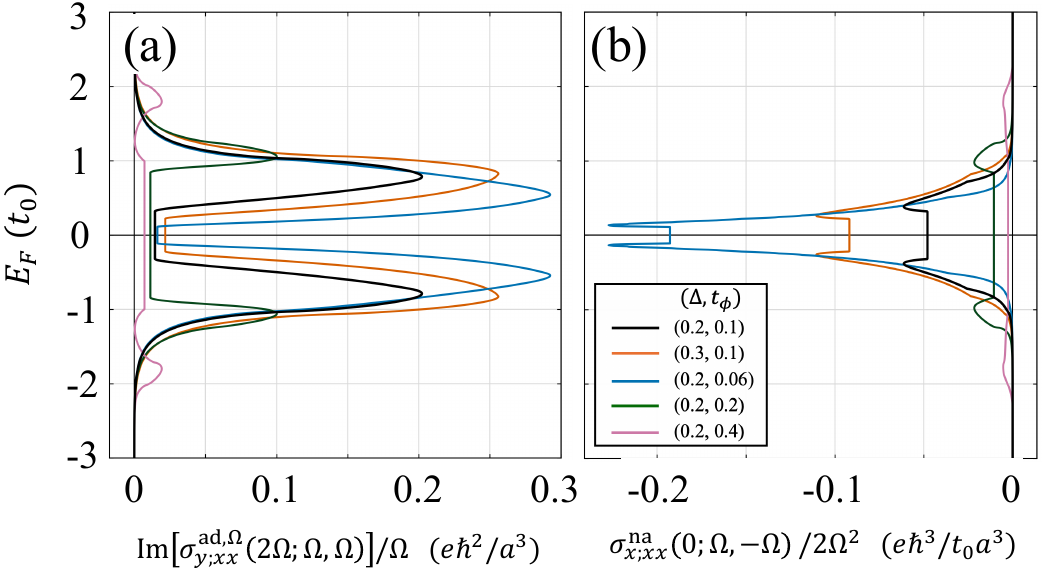}
\caption{
{Fermi-level ($E_F$) dependences of (a) the adiabatic SHG conductivity and (b) nonadiabatic rectification conductivity} for selected values of $(\Delta,t_\phi)$ on the Haldane model of Eq.~(\ref{eq:Haldane_ham0}) with $t_0=1$.
We consider $\Delta/t_0=0.3$, $0.4$, and $0.5$ at fixed $t_{\phi}/t_0=0.1$, as well as $t_{\phi}/t_0=0.04$, $0.06$, $0.08$, $0.2$, $0.3$, and $0.4$ at fixed $\Delta/t_0=0.2$. 
Among them, the results for selected values are shown on each panel.
The conductivities $(\Delta,t_\phi)=(0.2, 0.1)$ depicted by the black solid lines in each panel correspond to the result in Fig.~\ref{fig:sigma_Haldanemodel}.
}
\label{fig:sigma_Haldanemodel_para}
\end{figure}

Next, we examine 
the $\Delta$ and $t_\phi$ dependence of the conductivity in the Haldane model.
%
%
%
%
Figure~\ref{fig:sigma_Haldanemodel_para} shows 
the Fermi-level dependences of $S^{\rm ad,\Omega}_{y;xx}$ and $S^{\rm na}_{x;xx}$ for selected values of $(\Delta,t_\phi)$.
%
%
Since the results for $S^{\rm ad,\partial^2}_{y;xx}$ are almost same as $S^{\rm ad,\Omega}_{y;xx}$, we only show the $E_F$ dependences of $S^{\rm ad,\Omega}_{y;xx}$ for the adiabatic conductivity.
Except for $t_\phi/t_0=0.2$ and $0.4$ at $\Delta/t_0=0.2$, 
both $S^{\rm ad,\Omega}_{y;xx}$ and $S^{\rm na}_{x;xx}$ retain the qualitative $E_F$ dependences of the original parameter, which is depicted by the black solid line in Fig.~\ref{fig:sigma_Haldanemodel_para}.
These results can be understood by focusing on the gap structure of $K$ and $K'$ points as follows.
Near the $K$ and $K'$ valleys, the Haldane model reduces to a massive continuum Dirac model with the valley-dependent mass, $m_\xi=\Delta-3\sqrt{3}\xi t_\phi$.
In the considered parameter range, increasing $\Delta$ narrows the gap at $K$ but widens that at $K'$ on the band structure, and increasing $t_\phi$ widens the gaps at both valleys, in agreement with Fig.~\ref{fig:sigma_Haldanemodel_para}. 
%
%
%
This also means the conductivities are expected to remain qualitatively unchanged when varying the parameter, as long as the continuum Dirac model description holds.
%
%
%
%
%

In some exceptional cases, however, neither $S^{\rm ad,\Omega}_{y;xx}$ nor $S^{\rm na}_{x;xx}$ exhibits qualitatively the same $E_F$ dependence as the black solid curve shows.
The conductivities exhibit finite plateaus with the gap determined by the energy of the $M$-point in the BZ, peaks just outside this gap, and a subsequent decrease with increasing $|E_F|$, which are depicted by the magenta solid line in Fig.~\ref{fig:sigma_Haldanemodel_para}. 
This is because $|m_\xi|$ becomes comparable to or larger than the $M$-point energy, $E_F \approx1.0t_0$, and therefore the description by the continuum Dirac model no longer holds.
%
In such cases, the global gap edge is instead determined by the energy of the $M$-point, and gives rise to a qualitatively different behavior. 
%

\subsection{Kane-Mele model}\label{sec:KaneMele}


\subsubsection{Hamiltonian}

Next, we extend our analysis from the spinless Haldane model to a spinful lattice model relevant to experimentally realizable spin-orbit-coupled materials, such as silicene, germanene, and monolayer jacutingaite. 
We consider the following extended Kane-Mele model.
\begin{align}\label{eq:KanemMele_full_Ham}
    \hat{H} = {\hat H}^{(0)} + \hat{H}_V+ \delta \hat{H}({{\bm Q}}(t)).
\end{align}
{where ${{\bm Q}}=(Q_x,Q_y)$ are two in-plane optical modes orthonormal each other.}
%

The first term of Eq.~(\ref{eq:KanemMele_full_Ham}), ${\hat H}^{(0)}$, is the Kane-Mele Hamiltonians~\cite{KaneMele2005_1, KaneMele2005_2} and read
\begin{align}\label{eq:KaneMele_Ham0}
        {\hat H}^{(0)}
        &=
         -t_0
         \sum_{\bm R}\sum_{l=1}^3
         \left(
               {\hat c}^{\dagger}_{A, \bm R + \bm \tau_l}{\hat c}_{B,\bm R}
               +
               h.c.
         \right)
         \nonumber\\
         &\ \ \ 
        + 
        i\lambda_{\rm SOC} \sum_{\braket{\braket{ij}}} 
         \left( \frac{2}{\sqrt{3}} \hat{\bm d}^{ij}_1 \times \hat{\bm d}^{ij}_2 \right)_z
          {\hat c}^{\dagger}_{i,s} (\hat{\sigma}_z )_{ss'} {\hat c}_{j,s'},
\end{align}
where we define the spinor ${\hat c}_{A/B,\bm R} = ( {\hat c}_{A/B,\bm R, \uparrow}, {\hat c}_{A/B,\bm R, \downarrow} )^{\rm T}$ with ${\hat c}_{A/B,\bm R, s}$ being the creation operator of electron on $A(B)$-sublattice of unit cell $\bm R$ with the spin $s$.
$\hat{\sigma}_\mu$ is the Pauli matrices in the spin space.
$\hat{\bm d}^{ij}_{1/2}$ are the two normalized nearest-neighbor hopping vectors from the site $i$ to $j$.
The original Kane-Mele model is centrosymmetric, and therefore none of the adiabatic and nonadiabatic conductivities appear as same as the Haldane model.


We also consider possible mass term, $\hat{H}_{ V}$, in Eq.~(\ref{eq:KanemMele_full_Ham}). 
In the pristine Kane-Mele model, the phonon-induced current vanishes due to the inversion symmetry.
Depending on how it couples to
the spin and sublattice degrees of freedom, various mass terms can be introduced into the Kane-Mele model. 
Table~I summarizes these mass terms and their symmetry properties under inversion and time-reversal operation. 
We consider the Semenoff mass $\Delta \hat{\tau}_z \hat{\sigma}_0$, 
the Zeeman-type mass $\delta_z \hat{\tau}_0 \hat{\sigma}_z$, 
and their composite, i.e., the spin-dependent sublattice mass $\delta_s \hat{\tau}_z \hat{\sigma}_z$.
Because the Kane-Mele spin-orbit coupling preserves both time-reversal and inversion symmetries, the nonadiabatic conductivity vanishes when only the Semenoff mass exists, in contrast to the Haldane model. 
In the following analysis, we consider the mass pair $\hat{H}^{\rm mass} = \Delta \hat{\tau}_z\hat{\sigma}_0 + \delta_s \hat{\tau}_z\hat{\sigma}_z$.
The Semenoff mass $\Delta \sigma_0\tau_z$ may originate from a heteronuclear honeycomb lattice or from substrate-induced sublattice asymmetry. 
Meanwhile, the spin and sublattice dependent mass $\delta_s\sigma_z\tau_z$ can be interpreted as a staggered exchange field induced by, e.g., antiferromagnetic proximity coupling.
As summarized in Table I, this combination of mass terms breaks all of inversion symmetry, time-reversal symmetry, and $\mathcal{PT}$ symmetry. 
The resulting Hamiltonian therefore provides a minimal model in which both the adiabatic and nonadiabatic conductivities can be finite.

The third term of Eq.~(\ref{eq:KanemMele_full_Ham}) represents the electron-phonon coupling. 
We consider this coupling up to the first order in the phonon coordinate in order to investigate current generation induced by the phonons.
The phonon-induced modulation of the NN hopping is treated in the same manner as in the Haldane model. 
For the NNN hopping, however, we must additionally take into account the phonon-induced change in the length of the bond vector $\bm d^{ij}_{1/2}$. 
We find that, for the $\Gamma$-point optical phonon, 
the resulting electron-phonon coupling Hamiltonian is given by (detailed deviation in Appendix.~C),
\begin{align}\label{eq:KaneMele_ephHam}
\delta {\hat H}({{\bm Q}}(t))
        &=
        -t_0 \sum_{\bm R}\sum_{l=1}^3 
            \left( \frac{\beta}{a^2}  (\bm\tau_l \cdot {{\bm Q}}(t)) \; 
                   {\hat c}^\dagger_{A\bm R+ \bm \tau_l} {\hat c}_{B\bm R} + h.c. 
            \right)
        \nonumber\\
        &\ \ \ 
        +\frac{2i\lambda_{\rm SOC}}{ \sqrt{3}}
         \sum_{\braket{\braket{ij}}} 
         \Biggl(
               \Bigl(\hat{\bm d}^{ij}_1 + \hat{\bm d}^{ij}_2 \Bigr) \times {{\bm Q}}(t)
         \Biggr)_z
         {\hat c}^{\dagger}_{i} \hat{\sigma}_z  {\hat c}_{j}.
\end{align}
By performing the Fourier transformation, the Bloch Hamiltonian for the full Hamiltonian $\hat{H}= \hat{H}^{(0)}+\delta \hat{H}$ is obtained as follows, 
      \begin{align}\label{eq:KaneMele_fullHam_Bloch_diag}
        \hat{\mathcal H} (\bm k, {{\bm Q}})
        =
        \begin{pmatrix}
          \bm R^{\uparrow}(\bm k, {{\bm Q}}) \cdot \hat{\bm \tau} & 0  \\
          0 & \bm R^{\downarrow}(\bm k, {{\bm Q}}) \cdot \hat{\bm \tau} 
        \end{pmatrix},
      \end{align}
where $\hat{\tau}_\mu$ with $\mu=x,y,z$ denotes the Pauli matrices in the sublattice spaces. 
We use the basis such that 
$({\hat c}_{A,\bm k,\uparrow}, {\hat c}_{B,\bm k,\uparrow}, {\hat c}_{A,\bm k,\downarrow}, {\hat c}_{B,\bm k,\downarrow})^{\rm T}$
to separate the Bloch Hamiltonian into each spin sector.
The vector $\bm R^s (\bm k, {{\bm Q}})$ is defined as follows, 
      \begin{align}\label{eq:KaneMele_fullHam_Bloch_R}
        &R^{\uparrow}_x
        = R^{\downarrow}_x
        = 
        -t_0  \sum_{l=1}^3 
              \left( 1 + \frac{\beta }{a^2} ({{\bm Q}}(t) \cdot \bm \tau_l) \right)
              {\rm cos}(\bm k\cdot\bm \tau_l),
        \\
        &R^{\uparrow}_y
        = R^{\downarrow}_y
        = 
        -t_0  \sum_{l=1}^3 
              \left( 1 + \frac{\beta }{a^2} ({{\bm Q}}(t) \cdot \bm \tau_l) \right)
              {\rm sin}(\bm k\cdot\bm \tau_l),
        \\
        &R^{\uparrow/\downarrow}_z
        =
        \mp 2\lambda_{\rm SOC} \sum_{n=1}^3
         \Biggl(
               1 + \frac{2}{\sqrt{3}a^2} \Bigl({{\bm Q}}(t) \times \bm a_n \Bigr)_z
         \Biggr)
        {\rm sin}(\bm k\cdot \bm a_n).
      \end{align}

\begin{table}[t]\label{tab:symmetry}
    \centering
\setlength{\tabcolsep}{14.pt}
\renewcommand{\arraystretch}{1.25}
\begin{tabular}{cccc}
\hline\hline
mass 
& ${\mathcal P}$
& ${\mathcal T}$
& ${\mathcal PT}$ \\
\hline
${\rm sin}(\bm k\cdot\bm a_n)\; \lambda_{\rm SOC}\; \hat{\tau}_z \hat{\sigma}_z$
& $\checkmark$
& $\checkmark$
& $\checkmark$ \\

$\Delta\; \hat{\tau}_z \hat{\sigma}_0$
& $\times$
& $\checkmark$
& $\times$ \\

$\delta_z\; \hat{\tau}_0 \hat{\sigma}_z$
& $\checkmark$
& $\times$
& $\times$ \\

$\delta_s\; \hat{\tau}_z \hat{\sigma}_z$
& $\times$
& $\times$
& $\checkmark$ \\

\hline\hline
\end{tabular}
\caption{
Symmetry properties of mass terms. 
$\lambda_{\rm SOC}$ denotes the Kane-Mele spin-orbit coupling presented in Eq.~(\ref{eq:KaneMele_Ham0}).
$\Delta$ is the Semenoff mass, i.e., the sublattice dependent mass, while $\delta_Z$ is the Zeeman type mass, i.e., the spin dependent mass.
$\delta_s$ is the composite of the sublattice and spin dependent mass.
}
\end{table}

\begin{figure*}[t]
   \includegraphics[width=1.0\hsize]{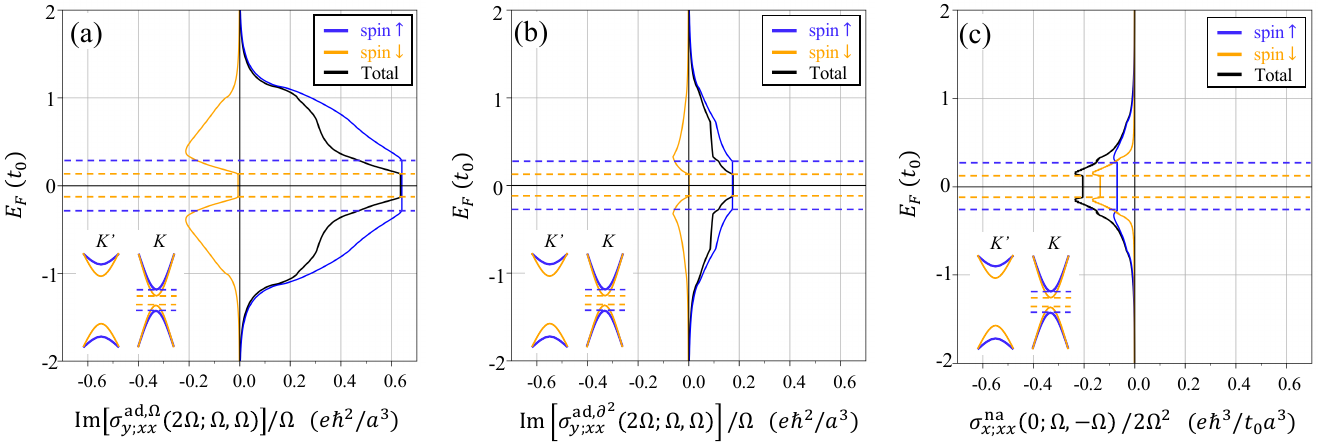}
\caption{
{Fermi-level ($E_F$) dependences of 
the adiabatic SHG conductivities [(a) and (b)] 
and 
the nonadiabatic rectification conductivity (c) 
for the extended Kane-Mele model in Eq.~(\ref{eq:KanemMele_full_Ham}) with the $\Delta \hat{\tau}_z \hat{\sigma}_0$-mass and the $\delta_s \hat{\tau}_z \hat{\sigma}_z$-mass are shown by the black solid line in each panel. 
The nonadiabatic SHG conductivity is related to the rectification conductivity by $\sigma^{\rm na}_{x;xx,\xi}(0;\Omega,-\Omega)=-2\sigma^{\rm na}_{x;xx,\xi}(2\Omega;\Omega,\Omega)$.}
The blue (orange) line represents the contribution from electrons with up~(down) spin. 
The model parameters are set as $a=1, t_0=1, \lambda_{\rm SOC}=0.04 t_0, \Delta=0.2 t_0,$ and $\delta_s= 0.3 t_0$.
On this parameter setting, 
by denoting the mass term on the valley $\xi$ with the spin $s$ as $m_{\xi,s}$, 
the mass on each spin and valley reads $m_{K,\uparrow} = 0.275$, $m_{K',\uparrow} = 0.725$, $m_{K,\downarrow} = 0.125$, and $m_{K',\downarrow} = -0.325$.
The band structure around the $K$ and $K'$ valleys are schematically depicted as inset in each panel.
The blue (orange) dotted line depicts the band edges of the $K$ valley for electrons with up~(down) spin, i.e., $\pm|m_{K,\uparrow}|$~($\pm|m_{K,\downarrow}|$). 
}
\label{fig:sigma_KaneMelemodel}
\end{figure*}

\subsubsection{Conductivity}\label{sec:KaneMele_conductivity}


The nonlinear conductivity for the Hamiltonian of Eq.~(\ref{eq:KanemMele_full_Ham}) 
can be calculated using the same formulas for the Haldane model.
As shown in Eq.~(\ref{eq:KaneMele_fullHam_Bloch_diag}), 
due to the $S_z$ conservation, the $4\times4$ extended Kane-Mele model decomposes into two separate spin sectors, each of which is a $2\times2$ Hamiltonian equivalent to the Haldane model. 
Consequently, the formulas in Sec.~IV, Eqs.~(\ref{eq:ad_conductivity_Omega}), (\ref{eq:ad_conductivity_partial}), and (\ref{eq:na_conductivity}), can be applied to the two spin
sectors of the extended Kane-Mele model. 
We replace $R_{\xi,\mu}$ in Eq.~(\ref{eq:Ham_conDirac}) with $R^{\uparrow/\downarrow}_{\mu}$ in Eq.~(\ref{eq:KaneMele_fullHam_Bloch_R}), evaluate the conductivity in each spin sector, and then sum those two contributions.
For the numerical calculations, we used 
$a=1, t_0=1, \lambda_{\rm SOC}=0.04 t_0, \Delta=0.2 t_0,$ and $\delta_s= 0.3 t_0$.



The black solid lines in Fig.~\ref{fig:sigma_KaneMelemodel} show the $E_F$ dependences of the adiabatic and nonadiabatic conductivities.
We also plot the band edge of $K$ valley for up-spin~(down) electrons as a blue~(orange) dotted line.
In Fig.~\ref{fig:sigma_KaneMelemodel}, we show only the $(y;xx)$-component for the adiabatic conductivity and the $(x;xx)$-component for the nonadiabatic conductivity for the same reason explained in the Haldane model section.
%
%
%
Figure~\ref{fig:sigma_KaneMelemodel}(a) shows the adiabatic conductivity originating from the Berry curvature, $S^{\rm ad,\Omega}_{y;xx}$. 
Within the gap common to both the $K$ and $K'$ valleys and to both spin sectors, indicated by the orange dotted lines, the conductivity reaches its maximum magnitude and forms a plateau.
Outside the common gap, the conductivity decreases approximately monotonically with increasing $|E_F|$, apart from a weak hump near $|E_F|=1.0$ corresponding to the energy of the $M$ point in BZ.
The overall $E_F$ dependence of $S^{\rm ad,\partial^2}_{y;xx}$ shown in Fig.~\ref{fig:sigma_KaneMelemodel}(b) is qualitatively identical to that of $S^{\rm ad,\Omega}_{y;xx}$. 
Compared with $S^{\rm ad,\Omega}_{y;xx}$, the curve departs more abruptly from the plateau inside the common gap as $|E_F|$ increases, resulting in a more pronounced kink at the edge of the plateau.

Figure~\ref{fig:sigma_KaneMelemodel}(c) shows the nonadiabatic conductivity, $S^{\rm na}_{x;xx}$. Its overall $E_F$ dependence is similar to that of the adiabatic conductivities described above. 
It forms a plateau inside the common gap and decreases with increasing $|E_F|$ outside the gap. 
However, as in the continuum Dirac and Haldane models, the conductivity exhibits a hump at the band edge of the $K$ valley for down-spin electrons, which is indicated by the orange dotted lines.
%
%
With increasing $|E_F|$, the conductivity again exhibits a weak hump at the band-edge energy of the $K$-valley for up-spin electrons, indicated by the blue dotted line, and then decreases monotonically.


The $E_F$ dependence of the black solid curves can be understood by decomposing the total conductivity into contributions from the two spin sectors.
In each panel of Fig.~\ref{fig:sigma_KaneMelemodel}, the blue and orange solid curves represent the up-spin and down-spin contributions, respectively. 
Each spin-resolved contribution can be understood in terms of the corresponding Haldane model, and therefore, in terms of the continuum Dirac model discussed in Sec.~\ref{sec:conDirac}.
Indeed, upon expanding the Kane-Mele Hamiltonian around the $K$ and $K'$ valleys, each valley and spin sector is described by a massive Dirac Hamiltonian. 
Denoting the valley and spin indices by 
$\xi=+1$~($K$), $\xi=-1$~($K'$), $s=+1$~(up), and $s=-1$~(down) respectively, 
the corresponding Dirac mass is given by
\begin{align}\label{eq:KaneMele_mass}
    m_{\xi,s}
    =
    \Delta+s\delta_s-\xi s\, 3\sqrt{3}\lambda_{\rm SO}.
\end{align}
For the present parameter set, the masses in the up-spin sector are positive at both valleys, 
\begin{align}\label{eq:KaneMele_mass_upspin}
    m_{K,\uparrow}
    &=
    \Delta+\delta_s- \, 3\sqrt{3}\lambda_{\rm SO} = 0.275\; > 0,
    \\
    m_{K',\uparrow}
    &=
    \Delta+\delta_s+ \, 3\sqrt{3}\lambda_{\rm SO} = 0.725 \; > 0.
\end{align}
Consequently, in the up-spin sector, the conductivities from each valley contribute in the same sign direction because of Eqs.~(44) and (47), and therefore the blue solid lines in Figs.~\ref{fig:sigma_KaneMelemodel} (a) and (b) are formed.
On the other hand, the masses in the down-spin sector have the ooposite sign at each valley, 
\begin{align}\label{eq:KaneMele_mass_downspin}
    m_{K,\downarrow}
    &=
    \Delta-\delta_s+ \, 3\sqrt{3}\lambda_{\rm SO} = 0.125\; > 0,
    \\
    m_{K',\downarrow}
    &=
    \Delta-\delta_s- \, 3\sqrt{3}\lambda_{\rm SO} = -0.325 \; < 0.
\end{align}
The situation for this down-spin sector is the same as that for the previous continuum Dirac model in Sec.~\ref{sec:conDirac}, and therefore the conductivities from each valley contribute in the opposite sign direction. 
Consequently, the orange solid lines in Figs.~\ref{fig:sigma_KaneMelemodel} (a) and (b) are formed similarly to Figs.~\ref{fig:sigma_conDiracmodel} (a) and (b).
The total conductivity behavior can therefore be understood as arising from the sum of the conductivities generated in the individual spin sectors, each governed by the continuum Dirac model with mass $m_{\xi,s}$.

%
%
%

For the nonadiabatic conductivity, the sign of the Dirac mass is irrelevant, and the conductivities from the two valleys always have opposite signs within each spin sector.
This behavior follows from Eq.~(\ref{eq:na_conductivity_an}) where the valley index $\xi=\pm$ determines the sign of the conductivities from each valley. 
Consequently, the conductivities in each spin sector exhibit the $E_F$ dependences similar to Fig.~\ref{fig:sigma_conDiracmodel}(c), which is depicted by the blue and orange solid lines in Fig.~\ref{fig:sigma_KaneMelemodel}(c). 
For the present parameter set, $|m_{K,s}| < |m_{K',s}|$ holds as shown in the insets of Fig.~\ref{fig:sigma_KaneMelemodel}. 
The corresponding plateau regions and hump have different widths. 
Their sum therefore makes the two separate humps of the black solid line in Fig.~\ref{fig:sigma_KaneMelemodel}(c).




\begin{figure}[t]
   \includegraphics[width=1.0\hsize]{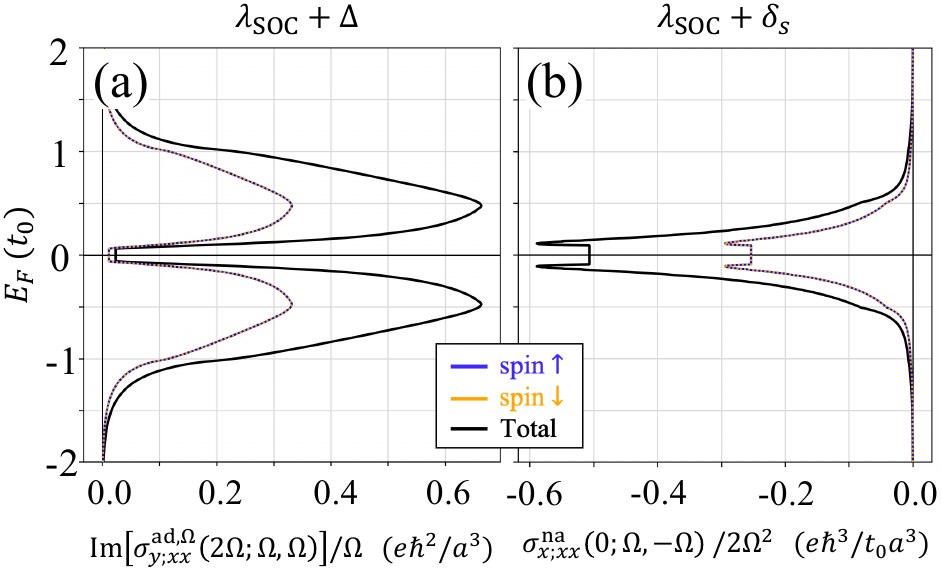}
\caption{
Fermi-level ($E_F$) dependence of the conductivities for different mass terms on the extended Kane-Mele model in Eq.~(\ref{eq:KanemMele_full_Ham}).
(a) {the adiabatic SHG conductivity} for $\Delta \hat{\tau}_z\hat{\sigma}_0$ ($\Delta=0.2t_0$);
(b) {the nonadiabatic rectification conductivity} for $\delta_s \hat{\tau}_z\hat{\sigma}_z$ ($\delta_s=0.3t_0$);
Other parameters are $a=t_0=1$ and $\lambda_{\rm SOC}=0.04t_0$.
Black curves show the total conductivity, and blue (orange) curves the up-spin (down-spin) contribution.
}
\label{fig:sigma_KaneMelemodel_othermass1}
\end{figure}

\subsubsection{Other mass pair}

Finally, we comment on other combinations of mass terms in the extended Kane-Mele model.
In addition to the mass pair considered above, we have calculated the conductivities for the following four patterns, namely,
$\hat{H}^{\rm mass} = \Delta \hat{\tau}_z\hat{\sigma}_0 $, 
$\delta_s \hat{\tau}_z\hat{\sigma}_z$,
$\delta_Z \hat{\tau}_0\hat{\sigma}_z$, and
$\Delta \hat{\tau}_z\hat{\sigma}_0+\delta_Z \hat{\tau}_0\hat{\sigma}_z$.
Regarding the pattern of the mass term $\hat{H}^{\rm mass} = \delta_Z \hat{\tau}_0\hat{\sigma}_z$, none of the adiabatic and nonadiabatic conductivity appears since the corresponding model breaks only the time-reversal symmetry, but not the inversion symmetry.
%
%
%
%
%
%
%
%
%
%
%
Figure~\ref{fig:sigma_KaneMelemodel_othermass1}(a) shows the $E_F$ dependence of $S^{\rm ad,\Omega}_{y;xx}$ for the mass term $\hat{H}^{\rm mass} = \Delta \hat{\tau}_z\hat{\sigma}_0 $ which allows only the adiabatic conductivity to appear.
Because time-reversal symmetry is preserved, the bands remain spin degenerate, and the two spin sectors contribute identically to the conductivity. 
The resulting behavior is qualitatively similar to the down-spin contribution shown in Fig.~\ref{fig:sigma_KaneMelemodel}(a). 
We omit $S^{\rm ad,\partial^2}_{y;xx}$, since its Fermi-energy dependence is nearly identical to that of $S^{\rm ad,\Omega}_{y;xx}$.
Figure~\ref{fig:sigma_KaneMelemodel_othermass1}(b) shows the $E_F$ dependence of $S^{\rm na}_{x;xx}$ for the mass term $\hat{H}^{\rm mass} = \delta_s \hat{\tau}_z\hat{\sigma}_z$ where only the nonadiabatic conductivity is allowed.  
The conductivity exhibits a finite plateau inside the gap, hump near the band edges, and a monotonic decrease away from the gap, reproducing the characteristic behavior of the black solid curve in Fig.~\ref{fig:sigma_KaneMelemodel}(c).
%
%

\begin{figure}[t]
   \includegraphics[width=1.0\hsize]{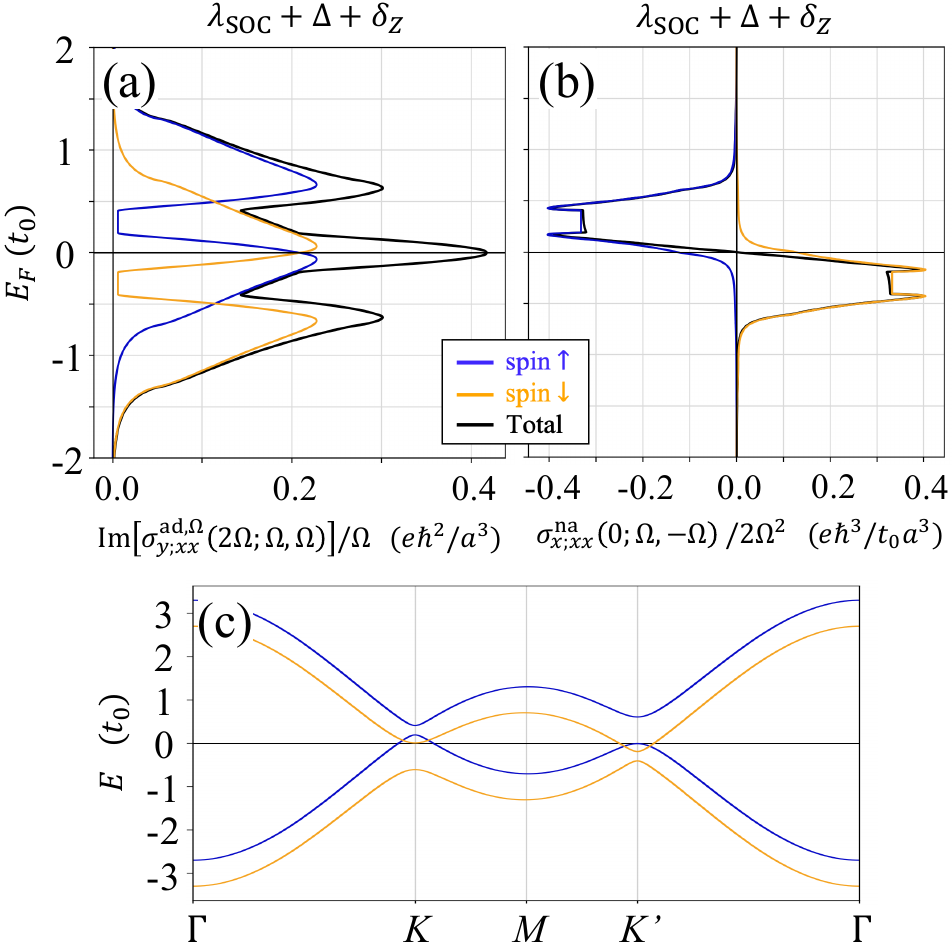}
\caption{
The current generation on the extended Kane-Mele model with the mass term $\Delta \hat{\tau}_z\hat{\sigma}_0+\delta_Z \hat{\tau}_0\hat{\sigma}_z$.
The Fermi-level dependences of 
(a)~{the adiabatic SHG conductivity} and 
(b)~{the nonadiabatic rectification conductivity}, which are depicted by the black solid lines.
The blue~(orange) solid line corresponds to the contribution of the up~(down)-spin electron.
(c) The band structure along the high symmetry lines.
The model parameters are set as $a=1, t_0=1,$ $\lambda_{\rm SOC}=0.04t_0, $ $\Delta=0.1t_0$, and $\delta_Z=0.3t_0$.
}
\label{fig:sigma_KaneMelemodel_othermass2}
\end{figure}

However, for the mass term $\hat{H}^{\rm mass} = \Delta \hat{\tau}_z\hat{\sigma}_0+\delta_Z \hat{\tau}_0\hat{\sigma}_z$, 
the behavior of the conductivities apparently differ from the results of Fig.~\ref{fig:sigma_KaneMelemodel}.
Figures~\ref{fig:sigma_KaneMelemodel_othermass2}(a) and (b) show the $E_F$ dependences of $S^{\rm ad,\Omega}_{y;xx}$ and $S^{\rm na}_{x;xx}$.
We omit the result for $S^{\rm ad,\partial^2}_{y;xx}$ because its $E_F$ dependence closely follows that of $S^{\rm ad,\Omega}_{y;xx}$.
$S^{\rm ad,\Omega}_{y;xx}$ does not exhibit a finite plateau but instead shows three peaks
, whereas the two distinct copies of the behavior shown in Fig.~\ref{fig:sigma_KaneMelemodel}(c) appear on the $S^{\rm na}_{x;xx}$.
Their behavior can be understood by examining the spin-resolved contributions together with the band structure shown in Fig.~\ref{fig:sigma_KaneMelemodel_othermass2}(c). 
The Zeeman term shifts the two spin sectors in opposite directions in energy, causing their bands to overlap near $E_F=0$ and therefore the above $E_F$ dependences.
Indeed, for both $S^{\rm ad,\Omega}_{y;xx}$ and $S^{\rm na}_{x;xx}$, each spin-resolved contribution retains essentially the same qualitative behavior as in Fig.~\ref{fig:sigma_KaneMelemodel}. 
For all these choices, the $E_F$ dependences of the resulting adiabatic and nonadiabatic conductivities can be understood again within the continuum Dirac description, which described in Sec.~\ref{sec:KaneMele_conductivity}.
%
%

\section{Discussion and Conclusion}\label{sec:discuss&concl}

{
In this work, we developed a semiclassical wave-packet formalism for electron-phonon-coupled systems, incorporating nonadiabatic inter-band mixing induced by coherent lattice vibrations.
From the resulting equations of motion, 
we derived general expressions for the second-order phonon-dynamics-driven current, including rectification and second-harmonic generation.
We showed that these responses contain adiabatic and nonadiabatic contributions, characterized by distinct quantum-geometric structures in the ${\bm Q}$-space and the $(\bm k,{{\bm Q}})$-mixed space. 
The adiabatic contribution is linear in the phonon frequency and governed by Berry-curvature-related structure, 
whereas the leading nonadiabatic contribution is quadratic in frequency and originates from the quantum metrics and the Christoffel symbols.
}
By applying the theory to massive Dirac, Haldane, and Kane-Mele models, 
we found that both the adiabatic and nonadiabatic currents induce a finite current even when the Fermi level lies inside the gap. 
In addition, the nonadiabatic current shows a characteristic hump in the chemical-potential dependence near the band edge, arising from the electron-phonon quantum geometry generated by the interplay between Dirac electrons and phonons.

%
Quantum geometry has mainly been explored through momentum-dependent electronic Bloch states \cite{Nagaosa2010, Bandyopadhyay2024,Liu2025, AGaoNN2025}.
Previous studies have extended this viewpoint to additional electronic parameters, in mainly magnetic systems.
Mixed $(\bm k,\bm m)$-space quantum geometry has been suggested to govern current-induced spin torques and its inverse responses \cite{Freimuth2014,Freimuth2015,Hanke2017,BXiong2018,CXiao2021,CXiao2022_2ndSOT,CXiao2023,JTang2024,Manchon2024,KHattori2024,XFeng2025,LXiang2025gyro,YRen2025,KHattori2025,Meguro2025,LXiang2025spinQGT,KHattori2026,Meguro_PRB2026}.
%
%
More recent works also showed intriguing geometric structures, such as quantum metrics and Christoffel symbols, naturally appearing in general extended parameter spaces~\cite{YOnishi_PRB2026,YRen_PRB2026,HYoshida_PRL2026}. 
Our work demonstrates that the phonon coordinates provide another physical realization of such an extended parameter space, where quantum geometry plays an interesting role.

Related to phonons, a recent work also connected the electron-phonon quantum geometry to phonon-mediated optical responses \cite{JHu_PRL2025}.
That theory addresses resonant optical processes, including phonon-mediated shift-current physics.
Our approach instead starts from the phonon dynamics itself near the adiabatic limit and systematically captures the leading corrections generated by the phonon-dynamics-induced interband mixing.
In particular, our formulation reveals quantum-metric and Christoffel-symbol contributions directly in the electron equations of motion.
The two approaches therefore probe complementary regimes of the electron-phonon quantum geometry.

We anticipate that the proposed responses will be evident in Dirac materials with strong electron-phonon coupling.
The nonadiabatic current requires broken inversion and time-reversal symmetries, suggesting engineered graphene-based heterostructures.
Aligned graphene/hBN can provide inversion breaking, while magnetic proximity can break time-reversal symmetry \cite{EWang_NatPhys2016, PWei_NatMat2016}.
Gate tuning of the Fermi level would enable direct tests of the predicted plateau and band-edge hump.
Because the nonadiabatic response originates from phonon-induced interband mixing, it is sensitive to interband dephasing.
High-quality samples with weak disorder and reduced scattering are therefore preferable, and 
hBN-based graphene heterostructures are promising in this respect \cite{CDean_NatTech2010}.
Transition-metal dichalcogenides provide another promising platform for phonon-dynamics-driven nonlinear transport. 
Their low-energy states near the $K$ and $K'$ valleys are described by massive Dirac electrons~\cite{DXiao_TMD2012}. 
Coherent phonons have been directly observed in several TMDs, together with strong coupling to electronic and excitonic states \cite{TJeong_ACS2016, CJin_NatPhys2017, CWang_PRL2023, SKusaba_APL2024}. 
Magnetic proximity provides an established route to break time-reversal symmetry in TMDs. 
Large exchange-induced valley splittings have been observed in several WSe$_2$, WS$_2$, and MoSe$_2$-based heterostructure~\cite{CZhao_NatTech2017, LSeyler_Nano2018, TNorden_NatCom2019, DZhong_NatTech2020}
Magnetically proximitized TMDs may therefore provide a promising setting for the nonadiabatic current generation predicted here.

The Kane-Mele model also provides a connection to realistic spin-orbit-coupled honeycomb materials. 
Germanene has experimentally demonstrated quantum-spin-Hall behavior, while jacutingaite is also described by Kane-Mele physics at low energies \cite{PBampoulis2023,PFachin2025}. 
Additional inversion and time-reversal symmetry breakings may be introduced through substrates or magnetic proximity similar to those of transition-metal dichalcogenides as mentioned above. 
The Kane-Mele model studied in Sec.~\ref{sec:KaneMele} corresponds to such cases.
By decomposing the currents into contributions from the $K$ and $K'$ valleys, we found that their principal features can be well understood in terms of the continuum Dirac model. 
Such heterostructure engineering and gate tuning of the Fermi level will provide Kane-Mele-based material platforms for the phonon-driven currents we discussed.

Finally, we address broader implications and possible extensions of our theory.
One natural extension is to explore responses beyond electric currents.
Phonon dynamics can also induce spin and magnetic responses through effective fields or the angular-momentum transfer \cite{Nova2017,JFransson2023}.
These phenomena have mainly been investigated in the adiabatic limit.
Our results suggest an additional geometric route to such responses. 
It is naturally expected that 
quantum-metric and Christoffel-symbols-related terms also contribute to phonon-driven spin and magnetic observables. 
More broadly, this possibility reflects a different perspective on electron-phonon coupling.
In conventional transport theory, phonons often enter primarily through carrier scattering and relaxation.
Here, coherent lattice dynamics instead modifies the electronic Bloch wave functions themselves.
This deformation generates geometric driving terms directly in the electron equations of motion.  
From this viewpoint, the construction is not inherently restricted to phonons.
The coordinates $Q_\alpha(t)$ can be replaced by generic multi-component dynamical parameters $X_a(t)$.
Together with crystal momentum, these components form an enlarged parameter space 
and 
form mixed geometric tensors that can enter the corresponding nonlinear electronic responses. 
Our results therefore provide a route for studying driven electron dynamics through quantum geometry in enlarged parameter spaces.

\section*{Acknowledgments}

The authors thank Takashi Oka for fruitful discussions.
This work was supported by JST SPRING, Grant No. JPMJSP2136 (T.M.), 
JSPS KAKENHI, Grant No. JP23K03275 (H.I.), JP25H00841 (H.I.), JP25H01250 (K.N.), JP26K00652 (K.N.)
and JST PRESTO, Grant No. JPMJPR2452 (H.I.).

%

\newpage

\section*{Appendix}\label{sec:Appendix}
\appendix
{\section{Derivation of Eq.~(\ref{eq:Cn_solution})}}

In order to determine the wave-packet coefficient, we here require the wave-packet follows the time-dependent Schr{\"o}dinger equation,
      \begin{align}\label{aeq:TDSE}
        i\hbar \frac{\partial}{\partial t}\ket{\Psi_{n,\bm x_c, \bm k_c}} 
        &=
        {\hat H}^{\rm tot}\ket{\Psi_{\bm x_c, \bm k_c}}.
      \end{align}
Substituting Eq.~(\ref{eq:psi_wave-packet}) into Eq.~(\ref{aeq:TDSE}) and 
multiplying $\bra{u_{m}(\bm k;{{\bm Q}})}e^{-i\bm k\cdot \hat{\bm x}}$ from the left side, the differential equation for $M_{ln}$ is obtained as follows
      \begin{align}\label{eq:diffeq_Cl}
        i\hbar \frac{\partial}{\partial t} C_{n\bm k}(l;t)
        &=
        \varepsilon_{l\bm k}({{\bm Q}}) C_{n\bm k}(l;t)
        %
       +\hbar \sum_{m} 
              ({\dot{{\bm Q}}}\cdot\bm a^{{{\bm Q}}}_{lm}) C_{n\bm k}(m;t),
       \end{align}
      for all $l$.
Here, we denote $\bm a^{{{\bm Q}}}_{lm}$ as the ${{\bm Q}}$-space Berry connection including both of the intra-band and the inter-band one; 
$\bm a^{{{\bm Q}}}_{lm} = \bm A^{{{\bm Q}}}_{l}$ for $l=m$ and $\bm a^{{{\bm Q}}}_{lm}$ for $l\ne m$.
To make its structure more transparent, we rewrite Eq.~(\ref{eq:diffeq_Cl}) explicitly in matrix form.

If the system has $D$ bands in total, this matrix equation reads,
\begin{widetext}
\footnotesize
      \begin{align}\label{eq:diffEq_Cl_mat}
        i\hbar \frac{\partial}{\partial t}
        \begin{pmatrix}
           C_{n\bm k}(1;t)  \\
           C_{n\bm k}(2;t)  \\
           \vdots              \\
           C_{n\bm k}(D;t)  \\
        \end{pmatrix}
        &=
        \begin{pmatrix}
           \varepsilon_{1\bm k}({{\bm Q}}) C_{n\bm k}(1;t)  \\
           \varepsilon_{2\bm k}({{\bm Q}}) C_{n\bm k}(2;t)  \\
           \vdots              \\
           \varepsilon_{D\bm k}({{\bm Q}}) C_{n\bm k}(D;t)  \\
        \end{pmatrix}
       +\hbar \dot{Q}_{\alpha}
        \begin{pmatrix}
           A^{{{\bm Q}}}_{1,\alpha} C_{n\bm k}(1;t) + a^{{{\bm Q}}}_{12,\alpha} C_{n\bm k}(2;t) +a^{{{\bm Q}}}_{13,\alpha} C_{n\bm k}(3;t) +\cdots a^{{{\bm Q}}}_{1D,\alpha} C_{n\bm k}(D;t)  \\
           a^{{{\bm Q}}}_{21,\alpha} C_{n\bm k}(1;t) + A^{{{\bm Q}}}_{2,\alpha} C_{n\bm k}(2;t) +a^{{{\bm Q}}}_{23,\alpha} C_{n\bm k}(3;t)+\cdots a^{{{\bm Q}}}_{2D,\alpha} C_{n\bm k}(D;t)  \\
           \vdots              \\
           a^{{{\bm Q}}}_{D1,\alpha} C_{n\bm k}(1;t) + a^{{{\bm Q}}}_{D2,\alpha} C_{n\bm k}(2;t) + a^{{{\bm Q}}}_{D3,\alpha} C_{n\bm k}(3;t)\cdots  + A^{{{\bm Q}}}_{D,\alpha} C_{n\bm k}(D;t) \\
        \end{pmatrix}.
        \end{align}
\normalsize
%
With defining matrices 
\footnotesize
      \begin{align}\label{aeq:L_R_mat}
        \hat{L}(t)
        &=
        \begin{pmatrix} 
          \varepsilon_{1\bm k}({{\bm Q}})+\hbar \dot{Q}_{\alpha} A^{{{\bm Q}}}_{1,\alpha} & 0 & \dots  & 0 \\
          0 & \varepsilon_{2\bm k}({{\bm Q}})+\hbar \dot{Q}_{\alpha} A^{{{\bm Q}}}_{2,\alpha} & \dots  & 0 \\
          \vdots & \vdots & \ddots & \vdots \\
          0 & 0 & \dots  & \varepsilon_{D\bm k}({{\bm Q}})+\hbar \dot{Q}_{\alpha} a^{{{\bm Q}}}_{D,\alpha}
       \end{pmatrix}, 
       \hat{R}(t)
       =\hbar \dot{Q}_{\alpha}
       \begin{pmatrix} 
          0 & a^{{{\bm Q}}}_{12,\alpha} & \dots  & a^{{{\bm Q}}}_{1D,\alpha} \\
          a^{{{\bm Q}}}_{21,\alpha} & 0 & \dots  & a^{{{\bm Q}}}_{2D,\alpha} \\
          \vdots & \vdots & \ddots & \vdots \\
          a^{{{\bm Q}}}_{D1,\alpha} & a^{{{\bm Q}}}_{D2,\alpha} & \dots  & 0
       \end{pmatrix},
      \end{align}
\normalsize
\end{widetext}
Equation (\ref{eq:diffEq_Cl_mat}) can be expressed in the same form as ordinary Schr{\"o}dinger equation,
\begin{align}\label{aeq:C_diffeq_LR}
  i\hbar \frac{\partial}{\partial t} {\bm C}_{n\bm k}(t) &= \Bigl( \hat{L}(t) + \hat{R}(t) \Bigr){\bm C}_{n\bm k}(t),
\end{align}
where ${\bm C}_{n\bm k}(t) = \Bigl( C_{n\bm k}(1;t), C_{n\bm k}(2;t), \cdots, C_{n\bm k}(D;t) \Bigr)^{\rm T}$.
In Eq.~(\ref{aeq:L_R_mat}), the matrix $\hat{L}(t)$ represents the dynamical phase and geometric phase accumulated by the intra-band Berry connection during the adiabatic evolution of the phonon coordinate.
On the other hand, the matrix $\hat{R}(t)$ stands for the inter-band mixing effect; the nonadiabatic effect.

By analogy with the way of solving Schr{\"o}dinger equation in the interaction picture,
we solve Eq.~(\ref{aeq:C_diffeq_LR}) up to the second order of the nonadiabatic contribution taking $t_0$ as the initial time,
      \begin{widetext}
      \begin{align}\label{aeq:Cl_generalSol}
        {\bm C}_{n\bm k}(t) 
        &=
        {\mathcal T}e^{-\frac{i}{\hbar} \int_{t_0}^t d\tau \hat{L}(\tau)}{\bm C}_{n\bm k}(t_0)
       -
        \frac{i}{\hbar}\int_{t_0}^t d\tau
        \left( {\mathcal T}e^{-\frac{i}{\hbar} \int_{\tau}^t d\tau' \hat{L}(\tau')} \right)
        \hat{R}(\tau)
        \left( {\mathcal T}e^{-\frac{i}{\hbar} \int_{t_0}^\tau d\tau' \hat{L}(\tau')} \right){\bm C}_{n\bm k}(t_0)
        \notag \\
        &\ \ \ \ \ 
       -
        \frac{1}{\hbar^2} \int_{t_0}^t d\tau_1 \int_{t_0}^{\tau_1} d\tau_2
        \left( {\mathcal T}e^{-\frac{i}{\hbar} \int_{\tau_1}^t d\tau' \hat{L}(\tau')} \right)
        \hat{R}(\tau_1)
        \left( {\mathcal T}e^{-\frac{i}{\hbar} \int_{\tau_2}^{\tau_1} d\tau' \hat{L}(\tau')} \right)
        \hat{R}(\tau_2)
        \left( {\mathcal T}e^{-\frac{i}{\hbar} \int_{t_0}^{\tau_2} d\tau' \hat{L}(\tau')} \right)  {\bm C}_{n\bm k}(t_0),
      \end{align}
\end{widetext}
where ${\mathcal T}$ is the time-ordering product.
      Because $\hat L$ is the diagonal matrix, its exponentiation is also diagonal.
      Hence, the exponential of the first term in the right hand side of Eq.~(\ref{aeq:Cl_generalSol}) become the matrix whose diagonal component consists of the dynamical and the intra-band geometric phase, and off-diagonal components are all zero; 
      %
      %
      %
      %
      %
      %
      %
      \begin{align}
        \left( {\mathcal T}e^{-\frac{i}{\hbar} \int_{t_0}^t d\tau \hat{L}(\tau)} \right)_{lm}
        &=
        e^{-i\gamma_l(t,t_0)}~\delta_{lm},
      \end{align}
      where $\gamma_l$ is the dynamical and geometric phase with respect to $l$-th band and defined as follows, 
      \begin{align}
        \gamma_l(t,t_0) 
        = 
        \frac{1}{\hbar}\int_{t_0}^t d\tau
        \Bigl(
             \varepsilon_{l\bm k}({{\bm Q}})+\hbar \dot{Q}_{\alpha} A^{{{\bm Q}}}_{l,\alpha}
        \Bigr).
      \end{align}
      With the above expression, one can find that the second term in the right hand side of Eq.(\ref{aeq:Cl_generalSol}) is just the product of the diagonal matrix and a matrix $\hat R$ which has only the off-diagonal components.
      Hence, the resulting matrix has only the off-diagonal components, and one can find the following expression after some straightforward calculations, 
      %
      %
      %
      %
      %
      %
      \begin{align}
        &\left(
               \left( {\mathcal T}e^{-\frac{i}{\hbar} \int_{\tau}^t d\tau' \hat{L}(\tau')} \right)
               \hat{R}(\tau)
               \left( {\mathcal T}e^{-\frac{i}{\hbar} \int_{t_0}^\tau d\tau' \hat{L}(\tau')} \right)
        \right)_{lm}
        \nonumber\\
        &\ \ \ \ \ \ 
        = \hbar\dot{Q}_\alpha(\tau)
        e^{-i\gamma_l(t,\tau)} a^{{{\bm Q}}}_{lm,\alpha}(\tau) e^{-i\gamma_m(\tau,t_0)} (1-\delta_{lm}).
      \end{align}
      The second order correction is little bit complex but can be calculated systematically, and the expression is as follows (diagonal component is also finite), 
      \begin{widetext}
      \begin{align}
       &\Biggl(
        \left( {\mathcal T}e^{-\frac{i}{\hbar} \int_{\tau_1}^t d\tau' \hat{L}(\tau')} \right)
        \hat{R}(\tau_1)
        \left( {\mathcal T}e^{-\frac{i}{\hbar} \int_{\tau_2}^{\tau_1} d\tau' \hat{L}(\tau')} \right)
        \hat{R}(\tau_2)
        \left( {\mathcal T}e^{-\frac{i}{\hbar} \int_{t_0}^{\tau_2} d\tau' \hat{L}(\tau')} \right)
        \Biggr)_{lm}
        \notag \\
       &\ \ \ \ \ \ \ \ \ \ \ \ \ \ \ \ \ \ \ \ 
        = \hbar^2 \dot{Q}_\alpha(\tau_1) \dot{Q}_\beta(\tau_2)
        e^{-i\gamma_l(t,\tau_1)}
        \left(
               \sum_{s=1}^D  a^{{{\bm Q}}}_{ls,\alpha}(\tau_1)  e^{-i\gamma_s(\tau_1,\tau_2)}  a^{{{\bm Q}}}_{sm,\beta}(\tau_2)
               (1-\delta_{ls})(1-\delta_{sm})
        \right)
        e^{-i\gamma_m(\tau_2,t_0)}.
      \end{align}

      %
      We can now write down the components of Eq.(\ref{aeq:Cl_generalSol}) by classifying the terms in order of the nonadiabatic effect as 
      $C_{n\bm k}(l;t) = C_{n\bm k}^{(0)}(l;t) + C_{n\bm k}^{(1)}(l;t) + C_{n\bm k}^{(2)}(l;t)$ where
%
      \begin{align} 
        C_{n\bm k}^{(0)}(l;t)
        &=
        e^{-i\gamma_l(t,t_0)} C_{n\bm k}(l;t_0),
        \\
        C_{n\bm k}^{(1)}(l;t)
        &=
        -i \int_{t_0}^t d\tau \;\dot{Q}_\alpha(\tau)
        \sum_{\substack{m=1\\ (m\ne l)}}^{D} 
          e^{-i\gamma_l(t,\tau)} 
           a^{{{\bm Q}}}_{lm,\alpha} (\tau) 
          e^{-i\gamma_m(\tau,t_0)}
          C_{n\bm k}(m;t_0),
        \\
        C_{n\bm k}^{(2)}(l;t)
        &=
        -\int_{t_0}^t d\tau\;\dot{Q}_\alpha(\tau)  \int_{t_0}^\tau d\tau'\;\dot{Q}_\beta(\tau')
        \sum_{m=1}^{D} 
          e^{-i\gamma_l({t,\tau})} 
          \notag\\
          &\ \ \ \ \ \ \ \ \ 
          \times
          \left(
                \sum_{s=1}^D
                a^{{{\bm Q}}}_{ls,\alpha}(\tau)  
                 e^{-i\gamma_s({\tau,\tau'})}
                a^{{{\bm Q}}}_{sm,\beta}(\tau')
                (1-\delta_{ls})(1-\delta_{sm})
          \right)
          e^{-i\gamma_m({\tau',t_0}})
          C_{n\bm k}(m;t_0).
      \end{align}
%
%
%
%
For the initial condition $C_{n\bm k}(l;t_0) = \delta_{ln}$ which means the wave-packet is in the focused band $n$ at the initial time, 
the intra-band coefficient reads
            \begin{align}
            \label{aeq:Cnn_sol}
              C_{n\bm k}(n;t)
             &=
              e^{-i\gamma_n({t,t_0})}
              \left\{
                    1
                   -\int_{t_0}^t d\tau\; \dot{Q}_\alpha(\tau) \int_{t_0}^\tau d\tau'\; \dot{Q}_\beta(\tau')
                    e^{-i\gamma_n({t_0,\tau})} 
                    \left(
                          \sum_{l (\ne n)}
                          a^{{{\bm Q}}}_{nl,\alpha}(\tau)  
                           e^{-i\gamma_l({\tau,\tau'})}
                          a^{{{\bm Q}}}_{ln,\beta}(\tau')
                    \right)
                    e^{-i\gamma_n({\tau',t_0})}
              \right\},
            \end{align}
      while
      the inter-band coefficient reads ($l\ne n$), 
            \begin{align}
            \label{aeq:Cnl_sol}
              C_{n\bm k}(l;t)
              &=
              -i\int_{t_0}^t d\tau\; \dot{Q}_\alpha(\tau)
                e^{-i\gamma_l(t,\tau)} a^{{{\bm Q}}}_{ln,\alpha} (\tau) e^{-i\gamma_n (\tau,t_0)}
              \notag \\
              &\ \ \ 
              -\int_{t_0}^t d\tau\; \dot{Q}_\alpha(\tau) \int_{t_0}^\tau d\tau'\; \dot{Q}_\beta(\tau')
               e^{-i\gamma_l({t,\tau})}
               \left(
                     \sum_{m=1}^D
                     a^{{{\bm Q}}}_{lm}(\tau)  e^{-i\gamma_m(\tau,\tau')}  a^{{{\bm Q}}}_{mn}(\tau')
                     (1-\delta_{lm})(1-\delta_{mn})
               \right)
               e^{-i\gamma_n(\tau',t_0)}.
           \end{align}

      One can write $C_{n\bm k}(l;t)$ in Eq.~(\ref{aeq:Cnl_sol}) in terms of $C_{n\bm k}(n;t)$ in Eq.~(\ref{aeq:Cnn_sol}) as follows.
      Firstly, we rearrange Eq.~(\ref{aeq:Cnn_sol}) to solve for $e^{-i\gamma_n (\tau,t_0)} $, 
      \begin{align}
            \label{aeq:Cnn_sol_inverse}
              e^{-i\gamma_n({t,t_0})}
             &=
              C_{n\bm k}(n;t)
              \left\{
                    1
                   -\int_{t_0}^t d\tau\; \dot{Q}_\alpha(\tau) \int_{t_0}^\tau d\tau'\; \dot{Q}_\beta(\tau')
                    e^{-i\gamma_n({t_0,\tau})} 
                    \left(
                          \sum_{l (\ne n)}
                          a^{{{\bm Q}}}_{nl,\alpha}(\tau)  
                           e^{-i\gamma_l({\tau,\tau'})}
                          a^{{{\bm Q}}}_{ln,\beta}(\tau')
                    \right)
                    e^{-i\gamma_n({\tau',t_0})}
              \right\}^{-1}
              \notag\\
              &\approx
              C_{n\bm k}(n;t) + \mathcal{O}(Q^2),
            \end{align}
        where $Q=|{{\bm Q}}(t)|$ is the amplitude of the phonon coordinate.
        Then, using a relation, 
        \begin{align}
        e^{-i\gamma_n (\tau,t_0)} 
        = e^{-i\gamma_n (\tau,t)} e^{-i\gamma_n (t,t_0)}
        =
        e^{-i\gamma_n (\tau,t)}C_{n\bm k}(n;t) + \mathcal{O}(Q^2), 
        \end{align}
        and substituting this relation into Eq.~(\ref{aeq:Cnl_sol}), one can obtain the following equation up to the second order of $Q$, 
        \begin{align}
            C_{n\bm k}(l;t)
              &=
              -i\int_{t_0}^t d\tau\; \dot{Q}_\alpha(\tau)
                e^{-i\gamma_l(t,\tau)} a^{{{\bm Q}}}_{ln,\alpha} (\tau) 
                e^{-i\gamma_n (\tau,t)}C_{n\bm k}(n;t).
        \end{align}
        This is nothing but Eq.~(\ref{eq:Cn_solution}).

\end{widetext}

{\section{ Derivations of Eqs.~(\ref{eq:EOM_x}), (\ref{eq:WPvel_ad}) and (\ref{eq:WPvel_na}) }}

Firstly, we introduce the coefficients $M_{ln,\bm k}(t)$ which is defined as $ C_{n\bm k}(l;t) = M_{ln,\bm k}(t) C_{n\bm k}(n;t)$.
From the solution of $C_{n\bm k}(l;t)$ in Eq.~(\ref{aeq:Cnl_sol}), these coefficients can be written as follows up to the second order of the phonon coordinate, 
      ($l \ne n$),
      \begin{align}\label{aeq:Mln_sol}
        &M_{ln,\bm k} (t)
         =
         \notag\\
         &\ 
        -i\int_{t_0}^t d\tau\; \dot{Q}_\alpha(\tau)
           e^{-i\gamma_l({t,\tau}) + i\gamma_n({t,\tau})}
           a^{{{\bm Q}}}_{ln,\alpha} (\tau)
           \nonumber\\
           &\ 
        -\int_{t_0}^t d\tau\; \dot{Q}_\alpha(\tau) \int_{t_0}^\tau d\tau'\; \dot{Q}_\beta(\tau')
               e^{-i\gamma_l({t,\tau}) + i\gamma_n({t,\tau'}) }
                \notag\\
                &\ \ \ \times
                     \sum_{m=1}^D
                     a^{{{\bm Q}}}_{lm}(\tau)  e^{-i\gamma_m({\tau,\tau'})}  a^{{{\bm Q}}}_{mn}(\tau')
                     (1-\delta_{lm})(1-\delta_{mn}).
      \end{align}
In the following, we denote the first and second term of the right hand side of Eq.~(\ref{aeq:Mln_sol}) as $M_{ln,\bm k}^{(1)} (t)$ $M_{ln,\bm k}^{(2)} (t)$, respectively.

The full equation of motion including the nonadiabatic effect is given in Eq.~(\ref{eq:EOM_x}).
We here present derivations of Eqs.~(\ref{eq:WPvel_ad}) and (\ref{eq:WPvel_na}) from Eq.~(\ref{eq:EOM_x}) by expanding it in the order of the wave-packet coefficient $M_{ln,\bm k}$.
Substituting $M_{ln,\bm k} = M_{ln,\bm k}^{(1)}+M_{ln,\bm k}^{(2)}$ into the full equation of motion in Eq.~(\ref{eq:EOM_x}) and 
defining $\dot{\bm x}^{(N)}_c$ such that it includes $M_{ln}$ in the $N$-th order, 
we obtain
      \begin{align}\label{aeq:vel_0_1_2}
        \dot{\bm x}_c
        &=
        \dot{\bm x}^{(0)}_c + \dot{\bm x}^{(1)}_c + \dot{\bm x}^{(2)}_c.
      \end{align}

\subsection{\underbar{Adiabatic velocity}}

In Eq.~(\ref{aeq:vel_0_1_2}), 
$\dot{\bm x}^{(0)}_c$ is defined such that it does not include the nonadiabatic effect, and therefore denotes the adiabatic velocity $\dot{\bm x}^{\rm ad}_c$.
The expression for $\dot{\bm x}^{(0)}_c$ is obtained by retaining only the terms that do not involve the coefficient $M_{ln,\bm k}$.
It follows simply by replacing the wave-packet energy with the instantaneous eigenenergy and $\widetilde{\Omega}\rightarrow\Omega$ in the
anomalous-velocity term in Eq.~(\ref{eq:EOM_x}).
Hence, that expression reads
\begin{align}\label{aeq:ad_vel_general}
        \dot{x}^{\rm ad}_{c,i}
        =
        \frac{\partial}{\hbar \partial k_{c,i}} 
        \varepsilon_{n\bm k_c} ({{\bm Q}})
       -
        \Omega^{\bm k {{\bm Q}}}_{n,i\alpha} \dot{Q}_{\alpha}.
\end{align}

To extract the expression in the second-order of the phonon coordinate, we expand the instantaneous eigenstates in Eq.(\ref{eq:inst_eigeneq}) with the assumption that $|{{\bm Q}} |\ll a$, 
      \begin{align}\label{aeq:unk_expand}
        \ket{ u_{n}(\bm k;{{\bm Q}}) }
        &=
        \ket{ u_{n}(\bm k;\bm 0) }
       +Q_{\alpha} 
        \left. 
             \ket{ \partial_{Q_\alpha} u_{n}(\bm k;{{\bm Q}})  } 
        \right|_{{{\bm Q}} = \bm 0}
        \nonumber\\
        &\ \ \ \ 
       +\frac{1}{2} Q_{\alpha} Q_{\beta} 
        \left. 
             \ket{ \partial_{Q_\alpha}\partial_{Q_\beta} u_{n}(\bm k;{{\bm Q}})  } 
        \right|_{{{\bm Q}} = \bm 0}
       +\cdots.
      \end{align}
      In the following, let us denote 
      $\left. \ket{ \partial_{Q_\alpha} u_{n}(\bm k;{{\bm Q}})  } \right|_{{{\bm Q}} = \bm 0}$ 
      as $\ket{ \partial_{Q_\alpha} u_{n}(\bm k;\bm 0)  }$
      for notational simplicity.

On the first term of Eq.~(\ref{aeq:ad_vel_general}), i.e., 
the group velocity, the instantaneous eigenvalue can be generally decomposed into the original electronic part and the electron-phonon coupling part; $\hat{H}({{\bm Q}})=\hat{H}^{(0)} +  Q_{\alpha} \hat{H}^{e-p}_\alpha$.
Up to the second-order of the phonon coordinate, 
this eigenenergy is expressed in terms of $ \ket{ u_{n}(\bm k;\bm 0) } $-basis as follows, 
      \begin{align}\label{aeq:inst_eigenvalue_expand}
        \varepsilon_{n\bm k_c} ({{\bm Q}})
       &=
        \varepsilon^{(0)}_{n\bm k_c} 
       +
        Q_\alpha \varepsilon^{(0)}_{n\bm k_c} 
         \; {\rm Re}
        \Bigl[
              \braket{ u_{n}(\bm k_c;\bm 0) |\partial_{Q_\alpha}u_{n}(\bm k_c;\bm 0) } 
        \Bigr]
        \nonumber\\
        &\ \ \ 
        +
        Q_{\alpha}
        \bra{ u_{n}(\bm k_c;\bm 0) } \hat{H}^{e-p}_\alpha  \ket{ u_{n}(\bm k_c;\bm 0) }
        \nonumber\\
        &\ \ \  
       +
        \frac{1}{2} Q_{\alpha} Q_{\beta} 
         \; \varepsilon^{(0)}_{n\bm k_c}
         \; {\rm Re}
        \Bigl[ 
              \braket{ u_{n}(\bm k_c;\bm 0) |\partial_{Q_\alpha}  \partial_{Q_\beta} u_{n}(\bm k_c;\bm 0)} 
        \Bigr] 
       \notag \\
       &\ \ \ 
       +Q_{\alpha} Q_{\beta} 
       \; {\rm Re}
        \Bigl[ 
              \bra{ u_{n}(\bm k_c;\bm 0) }
              \hat{H}^{e-p}_\beta
              \ket{ \partial_{Q_\alpha}u_{n}(\bm k_c;\bm 0) } 
        \Bigr].
      \end{align}
The third and forth terms in Eq.~(\ref{aeq:inst_eigenvalue_expand}) are $\varepsilon^{(2)}_{n\bm k_c}$ in the main text, 
\begin{align}
    \varepsilon^{(2)}_{n\bm k_c}
    &=
    \frac{1}{2} Q_{\alpha} Q_{\beta} 
         \; \varepsilon^{(0)}_{n\bm k_c}
         \; {\rm Re}
        \Bigl[ 
              \braket{ u_{n}(\bm k_c;\bm 0) | \partial_{Q_\alpha}  \partial_{Q_\beta} u_{n}(\bm k_c;\bm 0) }
        \Bigr] 
       \notag \\
       &\ \ \ 
       +Q_{\alpha} Q_{\beta} 
       \; {\rm Re}
        \Bigl[ 
              \bra{ u_{n}(\bm k_c;\bm 0) }
              \hat{H}^{e-p}_\beta
              \ket{ \partial_{Q_\alpha} u_{n}(\bm k_c;\bm 0) } 
        \Bigr].
\end{align}
We write $\hat{H}^{e-p}_\alpha$ as $\partial_{Q_\alpha} \hat{H} ({{\bm Q}})|_{{{\bm Q}} = \bm 0}$ in the main text.

The second term of $\dot{x}^{\rm ad}_{c,i}$ in Eq.~(\ref{aeq:ad_vel_general}) is the anomalous velocity. 
The Berry curvature is defined as 
\begin{align}
  \Omega^{\bm k {{\bm Q}}}_{n,i\alpha} \dot{Q}_{\alpha}
  &=
  \Bigl(
        \partial_{k_{c,i}}    A^{{{\bm Q}}}_{n,\alpha}(\bm k_c, {{\bm Q}})
       -\partial_{Q_{\alpha}} A^{\bm k}_{n,i}(\bm k_c, {{\bm Q}})
  \Bigr) \dot{Q}_{\alpha}
\end{align} 
The Berry connection $A^{\bm k/{{\bm Q}}}_{n,i/\alpha}(\bm k_c, {{\bm Q}})$ is defined by the instantaneous Bloch eigenstates. 
Here, the $\bm k_c$ and ${{\bm Q}}$ dependences are explicitly showed to 
make the following expansion with respect to ${{\bm Q}}$ more transparent.
%
We expand $A^{{{\bm Q}}}_{n,\alpha}(\bm k_c, {{\bm Q}})$ up to the first-order of ${{\bm Q}}$ using Eq.~(\ref{aeq:unk_expand}) as follows, 
      \begin{align}
        A^{{{\bm Q}}}_{n,\alpha}(\bm k_c, {{\bm Q}})
        &=
        -i\braket{ u_{n}(\bm k_c;{{\bm Q}}) | \partial_{Q_{\alpha}} u_{n}(\bm k_c;{{\bm Q}}) }
        \notag \\ 
        &=
        A^{{{\bm Q}}}_{n,\alpha}(\bm k_c, \bm 0)
       -
        i Q_{\beta} 
        \braket{ \partial_{Q_{\beta}}  u_{n}(\bm k_c;\bm 0) | \partial_{Q_{\alpha}} u_{n}(\bm k_c;\bm 0) }
        \nonumber\\
        &\ \ 
       -
        i Q_{\beta} 
        \braket{ u_{n}(\bm k_c;\bm 0) | \partial_{Q_{\alpha}} \partial_{Q_{\beta}} u_{n}(\bm k_c;\bm 0) }
       +
        {\mathcal O}(Q^2),
       \\
       \partial_{k_i} A^{{{\bm Q}}}_{n,\alpha}(\bm k_c, {{\bm Q}})
       &=
       \partial_{k_i} A^{{{\bm Q}}}_{n,\alpha}(\bm k_c, \bm 0)
       \notag \\
       &\ 
       -
        i Q_{\beta} 
        \Bigl(\partial_{k_i} \bra{ \partial_{Q_{\beta}}  u_{n}(\bm k_c;\bm 0) }\Bigr)
        \ket{ \partial_{Q_{\alpha}} u_{n}(\bm k_c;\bm 0) }
        \nonumber\\
        &\ 
       -
        i Q_{\beta} 
        \bra{ \partial_{Q_{\beta}}  u_{n}(\bm k_c;\bm 0) }
        \partial_{k_i}\ket{ \partial_{Q_{\alpha}} u_{n}(\bm k_c;\bm 0) }
       \notag \\
       &\ 
       -
        i Q_{\beta} 
        \braket{ \partial_{k_i}u_{n}(\bm k_c;\bm 0) | \partial_{Q_{\alpha}} \partial_{Q_{\beta}} u_{n}(\bm k_c;\bm 0) }
        \nonumber\\
        &\ 
       -
        i Q_{\beta} 
        \bra{ u_{n}(\bm k_c;\bm 0) }
        \partial_{k_i}
              \ket{ \partial_{Q_{\alpha}} \partial_{Q_{\beta}} u_{n}(\bm k_c;\bm 0) }
         +
        {\mathcal O}(Q^2).
      \end{align}

Unlike $A^{{{\bm Q}}}_{n,\alpha}(\bm k_c, {{\bm Q}})$, 
we have to expand $A^{\bm k}_{n,i}(\bm k_c, {{\bm Q}})$ up to the second-order of ${{\bm Q}}$ because of the derivative $\partial_{Q_\alpha}$, 
\begin{align}
        A^{\bm k}_{n,i}(\bm k_c, {{\bm Q}})
        &=
        -i\bra{ u_{n}(\bm k_c;{{\bm Q}}) } \partial_{k_{i}} \ket{ u_{n}(\bm k_c;{{\bm Q}}) }
        \nonumber\\
        &=
        A^{\bm k}_{n,i}(\bm k_c, \bm 0)
        -
        i Q_{\beta} 
        \braket{ \partial_{Q_{\beta}}  u_{n}(\bm k_c;\bm 0) | \partial_{k_i} u_{n}(\bm k_c;\bm 0) }
        \nonumber\\
        &\ 
       -
        i Q_{\beta} 
        \braket{ u_{n}(\bm k_c;\bm 0) | \partial_{k_i} \partial_{Q_{\beta}} u_{n}(\bm k_c;\bm 0) }
        \nonumber\\
        &\ 
        -i\Bigl(
               Q_{\beta} \bra{ \partial_{Q_{\beta}} u_{n\bm k} (\bm 0)}
          \Bigr)
          \partial_{k_i}
          \Bigl(
               Q_{\lambda} \ket{ \partial_{Q_{\lambda}} u_{n}(\bm k_c;\bm 0) }
          \Bigr)
        \nonumber\\
        &\ 
        -i\Biggl( \frac{1}{2}
               Q_{\beta}Q_{\lambda} \bra{ \partial_{Q_{\beta}} \partial_{Q_{\lambda}} u_{n}(\bm k_c;\bm 0) }
          \Biggr)
          \ket{ \partial_{k_i} u_{n}(\bm k_c;\bm 0) }
        \nonumber\\
        &\ 
        -i\bra{u_{n}(\bm k_c;\bm 0)}
          \partial_{k_i}
          \Biggl(\frac{1}{2}
               Q_{\beta}Q_{\lambda} \ket{ \partial_{Q_{\beta}} \partial_{Q_{\lambda}} u_{n}(\bm k_c;\bm 0) }
          \Biggr).
\end{align}
\begin{align}
       \partial_{Q_\alpha} A^{\bm k }_{n,i}(\bm k_c, {{\bm Q}})
       &=
       -iQ_{\lambda}\bra{ \partial_{Q_{\alpha}} u_{n}(\bm k_c;\bm 0) }
         \partial_{k_i}
         \ket{ \partial_{Q_{\lambda}} u_{n}(\bm k_c;\bm 0) }
       \nonumber\\
       &\ \ 
       -iQ_{\beta} \bra{ \partial_{Q_{\beta}} u_{n}(\bm k_c;\bm 0) }
          \partial_{k_i}
          \ket{ \partial_{Q_{\alpha}} u_{n}(\bm k_c;\bm 0) }
        \notag \\
       &\ \ 
       -i Q_{\beta} \braket{ \partial_{Q_{\beta}} \partial_{Q_{\alpha}} u_{n}(\bm k_c;\bm 0) | \partial_{k_i} u_{n}(\bm k_c;\bm 0) }
       \nonumber\\
       &\ \ 
       -i Q_{\beta} \bra{u_{n}(\bm k_c;\bm 0)}
         \partial_{k_i}
         \ket{ \partial_{Q_{\beta}} \partial_{Q_{\alpha}} u_{n}(\bm k_c;\bm 0) }
       \nonumber\\
       &\ \ 
       + \mathcal{O}(Q^2)
      \end{align}
Using these expressions, 
the second order correction of the mixed Berry curvature, says $\Omega^{\bm k{{\bm Q}} (2)}_{n,i\alpha}$, {reads} 
      \begin{align}
        \Omega^{\bm k{{\bm Q}} (2)}_{n,i\alpha}
        &=
        2{\rm Im}
        \braket{ \partial_{Q_{\alpha}} u_{n}(\bm k_c;\bm 0) | \partial_{k_i}\partial_{Q_{\beta}} u_{n}(\bm k_c;\bm 0) } Q_\beta
        \nonumber\\
        &\ \ 
        -2{\rm Im}
        \bra{ \partial_{k_i}u_{n}(\bm k_c;\bm 0) }
        \ket{ \partial_{Q_{\alpha}} \partial_{Q_{\beta}} u_{n}(\bm k_c;\bm 0) } Q_\beta
        \notag \\
        &=
        -2{\rm Im}
        \partial_{Q_{\beta}}
        \Bigl(
        \braket{ \partial_{k_i}u_{n}(\bm k_c;\bm 0) | \partial_{Q_{\alpha}}  u_{n}(\bm k_c;\bm 0) } 
        \Bigr)
        Q_\beta.
      \end{align}
      %
To progress the calculation of $\Omega^{\bm k{{\bm Q}} (2)}_{n,i\alpha}, $
we need to calculate the $\partial_{Q_{\beta}}$-derivative of {mixed quantum geometric tensor}, 
${\mathcal Q}^{\bm k{{\bm Q}}}_{n,i\alpha}(\bm k, {{\bm Q}}) =\braket{ \partial_{k_i}u_{n}(\bm k;{{\bm Q}}) | \partial_{Q_{\alpha}}  u_{n}(\bm k;{{\bm Q}}) } = \sum_{l(\ne n)} a^{\bm k}_{nl,i} a^{{{\bm Q}}}_{ln,\alpha} $ at ${{\bm Q}}=\bm 0$.
Now let us introduce a following geometric quantity; $ {\mathcal T}^{\bm k{{\bm Q}}}_{n,i\alpha\beta} = \partial_{Q_{\beta}}{\mathcal Q}^{\bm k{{\bm Q}}}_{n,i\alpha}$.
      \begin{align}\label{aeq:T_def}
        {\mathcal T}^{\bm k{{\bm Q}}}_{n,i\alpha\beta} (\bm k, {{\bm Q}})
        &= 
        \partial_{Q_{\beta}}{\mathcal Q}^{\bm k{{\bm Q}}}_{n,i\alpha}(\bm k, {{\bm Q}})
        \notag \\
        &=
        \sum_{l(\ne n)}\Bigl\{
                       (\partial_{Q_{\beta}}a^{\bm k}_{nl,i}) a^{{{\bm Q}}}_{ln,\alpha} 
                       + a^{\bm k}_{nl,i} (\partial_{Q_{\beta}}a^{{{\bm Q}}}_{ln,\alpha})
                       \Bigr\}
        \notag \\
        &=
        \sum_{l(\ne n)}\Bigl\{
                       ({\mathcal D}_{Q_{\beta}}a^{\bm k}_{nl,i}) a^{{{\bm Q}}}_{ln,\alpha} 
                       + a^{\bm k}_{nl,i} ({\mathcal D}_{Q_{\beta}}a^{{{\bm Q}}}_{ln,\alpha})
                       \Bigr\}.
      \end{align}
Here, we introduce the covariant derivative which acts the band-resolved quantity $f_{nl}$,
      \begin{align}
        {\mathcal D}_{Q_{\beta}}  f_{nl}
        &=
        \partial_{Q_{\beta}} f_{nl}
        + i ( A^{{{\bm Q}}}_{n,\beta}-A^{{{\bm Q}}}_{l,\beta} )f_{nl},
      \end{align}
to make the gauge invariance of ${\mathcal T}^{\bm k{{\bm Q}}}_{n,i\alpha\beta} $ explicit.
In fact, its gauge invariance can be confirmed as follows. 
Let us transform the instantaneous eigenstates as 
$\ket{ u_{n}(\bm k;{{\bm Q}}) } \rightarrow \ket{ {\tilde u}_{n}(\bm k;{{\bm Q}}) } = e^{i\chi_n(\bm k,{{\bm Q}})}\ket{ u_{n}(\bm k;{{\bm Q}}) }$, 
and then the first term of RHS of Eq.~(\ref{aeq:T_def}) transforms as 
      \begin{align}\label{aeq:T_gauge_confirm}
        ({\mathcal D}_{Q_{\beta}}a^{\bm k}_{nl,i})a^{{{\bm Q}}}_{ln,\alpha}
        &\rightarrow
        ({\mathcal {\tilde D}}_{Q_{\beta}} {\tilde a}^{\bm k}_{nl,i}) {\tilde a}^{{{\bm Q}}}_{ln,\alpha}
        \notag \\
        &=
        \partial_{Q_{\beta}}\Bigl( e^{-i(\chi_n-\chi_l)} a^{\bm k}_{nl,i} \Bigr)
                                   e^{i(\chi_n-\chi_l)}a^{{{\bm Q}}}_{ln,\alpha}
        \nonumber\\
        &\ \ 
        +i
        \Bigl( A^{{{\bm Q}}}_{n,\beta} + \partial_{Q_\beta}\chi_n -A^{{{\bm Q}}}_{l,\beta} - \partial_{Q_\beta}\chi_l \Bigr)
        a^{\bm k}_{nl,i} a^{{{\bm Q}}}_{ln,\alpha}
        \notag \\
        &=
        ({\mathcal D}_{Q_{\beta}}a^{\bm k}_{nl,i})a^{{{\bm Q}}}_{ln,\alpha}.
      \end{align}
The $Q_\beta$-derivative of the phase term $e^{-i(\chi_n-\chi_l)}$ in the first term of RHS of Eq.~(\ref{aeq:T_gauge_confirm}) cancels 
the $Q_\beta$-derivative of $\chi_n$ and $\chi_l$ in the second term.
Hence, the gauge invariance of $({\mathcal D}_{Q_{\beta}}a^{\bm k}_{nl,i})a^{{{\bm Q}}}_{ln,\alpha}$ is confirmed.
The gauge invariance of $a^{\bm k}_{nl,i} ({\mathcal D}_{Q_{\beta}}a^{{{\bm Q}}}_{ln,\alpha})$ in Eq.~(\ref{aeq:T_def}) can also be confirmed in the same way as Eq.~(\ref{aeq:T_gauge_confirm}), 
and therefore ${\mathcal T}^{\bm k{{\bm Q}}}_{n,i\alpha\beta} $ is gauge invariant {at least in the nondegenerate systems}.

Let us express ${\mathcal T}_{n,i\alpha\beta}$ in terms of the matrix elements of the velocity operator and the electron-phonon coupling Hamiltonian for numerical calculations.
With these matrix elements, i.e. 
$ v^i_{nl} = \bra{u_{n}(\bm k;{{\bm Q}})} \partial_{k_i}\hat{\mathcal H}_{\bm k}({{\bm Q}}) \ket{u_{l}(\bm k;{{\bm Q}})}$ and 
$ \lambda^\alpha_{nl} = \bra{u_{n}(\bm k;{{\bm Q}})} \partial_{Q_\alpha}\hat{\mathcal H}_{\bm k}({{\bm Q}}) \ket{u_{l}(\bm k;{{\bm Q}})}$, 
the inter-band Berry connections are given as 
\begin{align}\label{aeq:akQ_HellFeyn}
    a^{\bm k}_{nl,i}
    &=
    \frac{i v_{nl}^i}{ \varepsilon_{nl,\bm k}({{\bm Q}}) },
    \ \ \ \ \ \ 
    a^{{{\bm Q}}}_{nl,i}
    =
    \frac{i \lambda_{nl}^i}{ \varepsilon_{nl,\bm k}({{\bm Q}}) }.
\end{align}
%
%
%
With Eq.(\ref{aeq:akQ_HellFeyn}), ${\mathcal T}^{\bm k{{\bm Q}}}_{n,i\alpha\beta}$ reads
\begin{align}\label{aeq:Tiab_numerical_1}
  {\mathcal T}_{n,i\alpha\beta}^{\bm k{{\bm Q}}}
  &=
  \sum_{l(\ne n)} 
  \frac{ ( {\mathcal D}_{Q_{\beta}}v^i_{nl}) \lambda^\alpha_{ln} 
        +v^i_{nl} ( {\mathcal D}_{Q_{\beta}} \lambda^\alpha_{ln} )
       }
       { (\varepsilon_{nl,\bm k}({{\bm Q}}))^2 }
   \nonumber\\
   &\ \ \ \ \ 
  -2\sum_{l(\ne n)} 
  \frac{ v^i_{nl} \lambda^\alpha_{ln} }
       { (\varepsilon_{nl,\bm k}({{\bm Q}}))^3 }
  \Bigl(\partial_{Q_\beta} \varepsilon_{nl,\bm k}({{\bm Q}}) \Bigr).
\end{align}
The covariant derivative of the velocity operator can be calculated by its definition as follows, 
\begin{widetext}
                \begin{align}
\label{aeq:DQ_vnl}
                  {\mathcal D}_{Q_{\beta}} v^i_{nl}
                  &=
                  \partial_{Q_{\beta}}v^i_{nl} + i ( A^{{{\bm Q}}}_{n,\beta}-A^{{{\bm Q}}}_{l,\beta} )v^i_{nl},
                  \\
\label{aeq:parQ_vnl}
                  \partial_{Q_{\beta}}v^i_{nl} 
                  &=
                  \bra{\partial_{Q_{\beta}}u_{n}(\bm k;{{\bm Q}})} 
                      \partial_{k_i}\hat{\mathcal H}_{\bm k}({{\bm Q}}) 
                  \ket{u_{l}(\bm k;{{\bm Q}})}
                 +
                  \bra{u_{n}(\bm k;{{\bm Q}})} 
                      \partial_{Q_{\beta}}\partial_{k_i}\hat{\mathcal H}_{\bm k}({{\bm Q}}) 
                  \ket{u_{l}(\bm k;{{\bm Q}})}
                 +
                  \bra{u_{n}(\bm k;{{\bm Q}})} 
                      \partial_{k_i}\hat{\mathcal H}_{\bm k}({{\bm Q}}) 
                  \ket{\partial_{Q_{\beta}}u_{l}(\bm k;{{\bm Q}})}
                  %
                  \notag \\
                  &=
                  v^{i;\beta}_{nl} 
                 -i ( A^{{{\bm Q}}}_{n,\beta} - A^{{{\bm Q}}}_{l,\beta})v^i_{nl} 
                 -i a^{{{\bm Q}}}_{nl,\beta} v^i_{ll} -i \sum_{m(\ne n,l)} a^{{{\bm Q}}}_{nm,\beta} v^i_{ml} 
                 +i  v^i_{nn}a^{{{\bm Q}}}_{nl,\beta} +i \sum_{m(\ne n,l)} v^i_{nm}a^{{{\bm Q}}}_{ml,\beta}. 
                \end{align}
\end{widetext}
Here, the diagonal elements of the velocity operator is the derivative of instantaneous eigenenergy, $v^i_{nn} = \partial_{k_i}\varepsilon_{n\bm k}({{\bm Q}})$. Substituting Eq.~(\ref{aeq:parQ_vnl}) into Eq.~(\ref{aeq:DQ_vnl}) and using Eq.~(\ref{aeq:akQ_HellFeyn}), we reach the following expression, 
                \begin{align}\label{aeq:D_vnl_numerical}
                  {\mathcal D}_{Q_{\beta}} v^i_{nl}
                  &=
                  v^{i;\beta}_{nl} 
                 -\frac{\partial_{k_i}\varepsilon_{nl,\bm k}({{\bm Q}})}{ \varepsilon_{nl,\bm k}({{\bm Q}}) } \lambda^\beta_{nl} 
                 \nonumber\\
                 &\ \ 
                 +\sum_{m(\ne n,l)}
                  \left(
                       \frac{ \lambda^\beta_{nm} v^i_{ml} }{ \varepsilon_{nm,\bm k}({{\bm Q}}) }
                      +\frac{ v^i_{nm} \lambda^\beta_{ml}  }{ \varepsilon_{lm,\bm k}({{\bm Q}}) }
                  \right),
                \end{align}
where  
$ v^{i;\beta}_{nl} = \bra{u_{n}(\bm k;{{\bm Q}})} \partial_{Q_\beta}\partial_{k_i}\hat{\mathcal H}_{\bm k}({{\bm Q}}) \ket{u_{n}(\bm k;{{\bm Q}})} $.
The covariant derivative of $\lambda_{ln}^\alpha$ can be obtained by the similar calculation to that described above, 
                \begin{align}\label{aeq:D_lamnl_numerical}
                  {\mathcal D}_{Q_{\beta}} \lambda^\alpha_{ln} 
                  &=
                  \lambda^{\alpha\beta}_{ln} 
                 -\frac{\partial_{Q_\alpha}\varepsilon_{nl,\bm k}({{\bm Q}})}{ \varepsilon_{nl,\bm k}({{\bm Q}}) } \lambda^\beta_{ln} 
                 \nonumber\\
                 &\ \ 
                 +\sum_{m(\ne n,l)}
                  \left(
                       \frac{ \lambda^\alpha_{lm} \lambda^\beta_{mn}  }{ \varepsilon_{nm,\bm k}({{\bm Q}}) }
                      +\frac{ \lambda^\beta_{lm}  \lambda^\alpha_{mn} }{ \varepsilon_{lm,\bm k}({{\bm Q}}) }
                  \right),
                \end{align}
$\lambda^{\alpha\beta}_{ln} = \bra{u_{n}(\bm k;{{\bm Q}})} \partial_{Q_\beta}\partial_{Q_\alpha}\hat{\mathcal H}_{\bm k}({{\bm Q}}) \ket{u_{n}(\bm k;{{\bm Q}})}$.
Substituting Eqs.~(\ref{aeq:D_vnl_numerical}) and (\ref{aeq:D_lamnl_numerical}) into Eq.~(\ref{aeq:Tiab_numerical_1}), 
{
${\mathcal T}_{n,i\alpha\beta}$ reads
}
\begin{widetext}
      \begin{align}\label{aeq:Tiab_numerical}
        {\mathcal T}^{\bm k{{\bm Q}}}_{n,i\alpha\beta}
        &=
        \sum_{l(\ne n)}
        \frac{ v^{i;\beta}_{nl} \lambda^\alpha_{ln} + v^{i}_{nl} \lambda^{\alpha\beta}_{ln} }
             { (\varepsilon_{nl,\bm k}({{\bm Q}}))^2 }
       -\sum_{l(\ne n)}
       \frac{  
              (\partial_{k_i} \varepsilon_{nl})      \lambda^\beta_{nl}\lambda^\alpha_{ln} 
             +(\partial_{u_\alpha} \varepsilon_{nl}) v^i_{nl} \lambda^\beta_{ln}
             +2(\partial_{u_\beta} \varepsilon_{nl}) v^i_{nl} \lambda^\alpha_{ln}
            }
            { (\varepsilon_{nl,\bm k}({{\bm Q}}))^3 }
       \notag \\
       &\ \ \ \ \ 
       +\sum_{m(\ne n,l)} 
       \Biggl\{
              \frac{\lambda^\beta_{nm} v^i_{ml} \lambda^\alpha_{ln} + v^i_{nl} \lambda^\alpha_{lm} \lambda^\beta_{mn}}
                   { (\varepsilon_{nl,\bm k}({{\bm Q}}))^2 \varepsilon_{nm,\bm k}({{\bm Q}}) }
             +\frac{v^i_{nm} \lambda^\beta_{ml} \lambda^\alpha_{ln} + v^i_{nl} \lambda^\beta_{lm} \lambda^\alpha_{mn}}
                   { (\varepsilon_{nl,\bm k}({{\bm Q}}))^2 \varepsilon_{lm,\bm k}({{\bm Q}}) }
       \Biggr\},
      \end{align}
\end{widetext}

\subsection{\underbar{Nonadiabatic velocity}}

$\dot{\bm x}^{(1)}_c$ and $\dot{\bm x}^{(2)}_c$ in Eq.~(\ref{aeq:vel_0_1_2}) represents the nonadiabatic effect for the wave-packet velocity.
Expanding the $M_{ln,\bm k}$ of the dynamical Berry curvature in Eq.~(\ref{eq:EOM_x}) with Eq.~(\ref{aeq:Mln_sol}) gives $\dot{\bm x}^{(1)}_c$, 
\footnotesize
      \begin{align}\label{aeq:x(1)_1}
        &\dot{x}^{(1)}_{c,i}
        =
        -2{\rm Re}\sum_{l(\ne n)}
        \Biggl\{
         \partial_{k_{i}} 
                  \Bigl(a^{{{\bm Q}}}_{nl,\alpha} M^{(1)}_{ln,\bm k_c}(t) \Bigr)
       - \partial_{Q_\alpha} 
                  \Bigl(a^{\bm k}_{nl,i} M^{(1)}_{ln,\bm k_c}(t) \Bigr)
         \Biggl\}\dot{Q}_{\alpha},
      \end{align}
\normalsize
whereas the left terms {such that the wave-packet energy in Eq.~(\ref{eq:EOM_x})} give $\dot{\bm x}^{(2)}_c$
%
      \begin{align}\label{aeq:x(2)_1}
        &\dot{x}^{(2)}_{c,i}
        =
        \frac{1}{\hbar}\partial_{k_{c,i}}
        \sum_{l(\ne n)}
        \varepsilon_{l \bm k_c}({{\bm Q}}) \Bigl|M^{(1)}_{ln,\bm k_c}(t)\Bigr|^2
        \notag\\
        &\ \ \ \ \ \ 
        +\frac{d}{dt} {\rm Im}
        \Biggl[ \sum_{l(\ne n)} M^{(1)*}_{ln,\bm k_c}(t) \frac{\partial M^{(1)}_{ln,\bm k_c}(t)}{\partial k_{c,i}} \Biggr]
        \nonumber\\
        &\ \ \ \ \ \ 
       +
        {\rm Im}\left[
                \sum_{l(\ne n)} \frac{\partial}{\partial k_{c,i}}
                \left(
                M^{(1)*}_{ln,\bm k_c}(t)
                \frac{d M^{(1)}_{ln,\bm k_c}(t)}{dt}\right) \right].
      \end{align}

We perform {\it{asymptotic expansion}} to $M^{(1)}_{ln,\bm k_c}(t)$. 
Let us consider the following integral, 
\begin{align*}
    \int^t_{t_0}d\tau~f(\tau) 
                e^{i\omega \int^\tau_{t} d{\tilde \tau} g({\tilde \tau})}. 
\end{align*}
Perform partial integration repeatedly, one can find that this integral takes the following form, 
              \begin{align}\label{aeq:asympt_exp_general}
                &\int^t_{t_0}d\tau~f(\tau) 
                e^{i\omega \int^\tau_{t} d{\tilde \tau} g({\tilde \tau})} 
               \nonumber\\
                &\ \ 
                =
                \frac{1}{i\omega} 
                \left.\Biggl(\frac{f(\tau)}{g(\tau)} 
                             e^{i\omega \int^\tau_{t} d{\tilde \tau} g({\tilde \tau})} 
                \Biggr)\right|^{t}_{t_0}
               -
                \frac{1}{i\omega} 
                \int^t_{t_0}d\tau~ 
                \Biggl( 
                       \frac{d}{d\tau} \frac{f(\tau)}{g(\tau)}
                \Biggr)e^{i\omega \int^\tau_{t} d{\tilde \tau} g({\tilde \tau})}
                \notag \\
                %
                %
                \notag \\
                &\ \ 
                = \sum_{N=1}^{\infty} 
                   \left(\frac{1}{i\omega g(t)} \right)^N 
                   \mathcal{F}_N(t)
              \end{align}
{where $\mathcal{F}_N(t)$ is a function of time; since its form is irrelevant, it is omitted here.}
For $M^{(1)}_{ln,\bm k}$, $g(t)$ corresponds to the dynamical and geometric phase, and hence Eq.~(\ref{aeq:asympt_exp_general}) is the expansion with respect to the power of 
$\hbar\Bigl( \varepsilon_{nl,\bm k}({{\bm Q}}) + \hbar(A^{{{\bm Q}}}_{n,\alpha}-A^{{{\bm Q}}}_{l,\alpha}) \dot{Q}_\alpha(t) \Bigr)^{-1}$.
In the situation where the phonon dynamics is much slower than the time scale of the electrons;  $T \gg \hbar/\varepsilon_{nl}$, the leading term of Eq.~(\ref{aeq:asympt_exp_general}) is the term of $N=1$.

Performing the derivatives in Eq.~(\ref{aeq:x(1)_1}) carefully (substitute Eq.~(\ref{aeq:Mln_sol}) into Eq.~(\ref{aeq:x(1)_1}) ), one can obtain following equations, 
\footnotesize
\begin{widetext}
\begin{align}\label{aeq:x(1)_2}
        \dot{x}_{c,i}^{(1)}
                &
                =
                2{\rm Im}
                \Biggl[
                      \sum_{l(\ne n)}
                      \Biggl(
                             \frac{\partial a^{{{\bm Q}}}_{nl,\alpha}(t)}{\partial k_{c,i}}
                            -\frac{\partial a^{\bm k}_{nl,i}(t)}{\partial Q_\alpha}
                      \Biggr)
                     \int^t_{t_0} d\tau~\dot{Q}_\beta (\tau) a^{{{\bm Q}}}_{ln,\beta}(\tau)
                      e^{ i\int_{\tau}^t d\tau' 
                         \Bigl( 
                               \frac{\varepsilon_{nl}(\tau')}{\hbar} 
                             + ( A^{{{\bm Q}}}_{n,\gamma}(\tau')-A^{{{\bm Q}}}_{l,\gamma}(\tau') )\dot{Q}_\gamma(\tau') 
                         \Bigr) }
                \Biggr]\dot{Q}_\alpha (t)
                \notag \\
                &\ \ 
                +
                2{\rm Im}
                \Biggl[
                      \int^t_{t_0} d\tau
                      \sum_{l(\ne n)}
                      \Biggl(
                             a^{{{\bm Q}}}_{nl,\alpha}(t)\frac{\partial a^{{{\bm Q}}}_{ln,\beta}(\tau)}{\partial k_{c,i}}
                            -a^{\bm k}_{nl,i}(t)\frac{\partial a^{{{\bm Q}}}_{ln,\beta}(\tau)}{\partial Q_\alpha}
                      \Biggr)
                      \dot{Q}_\beta (\tau) 
                      e^{ i\int_{\tau}^t d\tau' 
                         \Bigl( 
                               \frac{\varepsilon_{nl}(\tau')}{\hbar} 
                             + ( A^{{{\bm Q}}}_{n,\gamma}(\tau')-A^{{{\bm Q}}}_{l,\gamma}(\tau') )\dot{Q}_\gamma(\tau') 
                         \Bigr) }
                \Biggr]\dot{Q}_\alpha (t)
                \notag \\
                &\ \ 
                +
                2{\rm Re}
                \Biggl[
                      \int^t_{t_0} d\tau
                      \sum_{l(\ne n)}\dot{Q}_\beta (\tau) a^{{{\bm Q}}}_{ln,\beta}(\tau)
                      \int^t_{\tau} d\tau'
                      \Biggl\{
                          a^{{{\bm Q}}}_{nl,\alpha}(t)
                          \Bigl(
                               \frac{1}{\hbar}\frac{\partial \varepsilon_{nl}(\tau')}{\partial k_{c,i}}
                              +\frac{\partial (A^{{{\bm Q}}}_{n,\gamma}(\tau')-A^{{{\bm Q}}}_{l,\gamma}(\tau'))}{\partial k_{c,i}} \dot{Q}_\gamma (\tau')
                          \Bigr)
                          \notag \\
                          &\ \ \ \ \ \ \ \ \ \ \ \ \ \ \ \ \ \ \ \ \ \ \ \ \ \ \ \ \ \ 
                         -
                          a^{\bm k}_{nl,i}(t)
                          \Bigl(
                               \frac{1}{\hbar}\frac{\partial \varepsilon_{nl}(\tau')}{\partial Q_\alpha}
                              +\frac{\partial (A^{{{\bm Q}}}_{n,\gamma}(\tau')-A^{{{\bm Q}}}_{l,\gamma}(\tau'))}{\partial Q_\alpha} \dot{Q}_\gamma (\tau')
                          \Bigr)
                      \Biggr\}
                      e^{ i\int_{\tau}^t d\tau' 
                         \Bigl( 
                               \frac{\varepsilon_{nl}(\tau')}{\hbar} 
                             + ( A^{t}_n(\tau')-A^{t}_l(\tau') ) 
                         \Bigr) }
                \Biggr]\dot{Q}_\alpha (t), 
              \end{align}
\normalsize
where $\varepsilon_{nl}(t),  A^{\bm k/{{\bm Q}}}_{n,i/\alpha}(t), $and $ a^{\bm k/{{\bm Q}}}_{nl,i/\alpha} (t)$ are $\varepsilon_{nl,\bm k_c(t)}({{\bm Q}}(t)), A^{\bm k_c/{{\bm Q}}}_{n,i/\alpha}(\bm k_c(t), {{\bm Q}}(t)), $and $a^{\bm k_c/{{\bm Q}}}_{nl,i/\alpha} (\bm k_c(t), {{\bm Q}}(t))$, i.e., the values made by the instantaneous eigenvalues/stats respectably.
Applying the asymptotic expansion to the time-integral of Eq.~(\ref{aeq:x(1)_2}) and retaining $N=1$ term of Eq.~(\ref{aeq:asympt_exp_general}), the final line of Eq.~(\ref{aeq:x(1)_2}) eventually takes the following form, 
\begin{align}
  \dot{x}_{c,i}^{(1)}
  &=
   2\hbar~{\rm Re}
   \left[
         \sum_{l(\ne n)}
         \frac{ \frac{\partial}{\partial k_{c,i}} \Bigl( a^{{{\bm Q}}}_{nl,\alpha}(t) a^{{{\bm Q}}}_{ln,\beta}(t) \Bigr)
               -\frac{\partial}{\partial Q_\alpha} \Bigl( a^{\bm k}_{nl,i}(t) a^{{{\bm Q}}}_{ln,\beta}(t) \Bigr)
              }
              { \varepsilon_{nl}(t) + \hbar \Bigl( A^{{{\bm Q}}}_{n,\gamma}(t) - A^{{{\bm Q}}}_{l,\gamma}(t) \Bigr)\dot{Q}_\gamma (t) }
   \right] \dot{Q}_\alpha (t) \dot{Q}_\beta (t).
\end{align}
%
%
%
%

The calculations for $\dot{x}_{c,i}^{(2)}$ are more complicated, but the procedure is same as that for $\dot{x}_{c,i}^{(2)}$ and straightforward; 
preforming $\partial_{k_i}$/$\partial_{Q_\alpha}$-derivatives to $M^{(1)}_{ln,\bm k}(t)$ and applying the asymptotic expansion.
On the frist term of Eq.~(\ref{aeq:x(2)_1}), we simply apply the asymptotic expansion to $M^{(1)}_{ln,\bm k}(t)$ and obtain the following expression as a result, 
\begin{align}\label{aeq:x(2)1_2}
    \frac{1}{\hbar}\partial_{k_{i}}
        \sum_{l(\ne n)}
        \varepsilon_{l \bm k_c}({{\bm Q}}) \Bigl|M^{(1)}_{ln,\bm k_c}(t)\Bigr|^2
    &=
    \hbar \sum_{l(\ne n)}
          \frac{\partial}{\partial k_{c,i}}
          \left(
                \frac{a^{{\bm Q}}_{nl,\alpha}(t) 
                      \varepsilon_{l,\bm k_c}(t) 
                      a^{{\bm Q}}_{ln,\beta}(t) }
                     { \Bigl( \varepsilon_{nl,\bm k_c}(t) 
                      + \hbar(A^{{{\bm Q}}}_{n,\gamma}(t) - A^{{{\bm Q}}}_{l,\gamma}(t)){\dot Q}_\gamma (t)  \Bigr)^2 }
          \right) 
          \dot{Q}_\alpha (t) \dot{Q}_\beta (t)
\end{align}

Next, in the second term of Eq.~(\ref{aeq:x(2)_1}), 
the derivative with respect to $k_{c,i}$ reads, 
               \begin{align}
                 \frac{\partial M^{(1)}_{ln,\bm k_c}(t)}{\partial k_{c,i}}
                 &=
                 -i\int_{t_0}^t d\tau
                 \left\{
                        -i\frac{\partial ( \gamma_l(t,\tau)-\gamma_n(t,\tau) )}
                               {\partial k_{c,i}}
                          a^{{\bm Q}}_{ln,\alpha}(\tau)
                        +\frac{\partial a^{{\bm Q}}_{ln,\alpha}(\tau)}{\partial k_{c,i}} 
                 \right\} \dot{Q}_\alpha (\tau) 
                 e^{ -i\gamma_l(t,\tau) +i \gamma_n(t,\tau) }, 
               \end{align}
and hence the second term of RHS in Eq.~(\ref{aeq:x(2)_1}) reads, 
\begin{align}
&
\frac{d}{dt}
\left[
M^{(1)*}_{ln,\bm k_c}(t) 
\frac{\partial M^{(1)}_{ln,\bm k_c}(t)}
     {\partial k_{c,i}}
\right]
\notag\\
&
=
\frac{d}{dt}
\int_{t_0}^t d\tau \int_{t_0}^t d\tau'
\left\{
-i\frac{\partial ( \gamma_l(t,\tau')-\gamma_n(t,\tau') )}
                    {\partial k_{c,i}}
                    a^{{\bm Q}}_{ln,\alpha}(\tau')
                    +\frac{\partial a^{{\bm Q}}_{ln,\alpha}(\tau')}{\partial k_{c,i}} 
\right\}
e^{ -i\gamma_l(\tau,\tau') +i \gamma_n(\tau,\tau') }
                 ( a^{{\bm Q}}_{ln,\beta}(\tau) )^*
\dot{Q}_\alpha (\tau') \dot{Q}_\beta (\tau).
\end{align}
To make the equation transparent, 
let us define some functions, 
                \begin{align}
                  f(t,\tau')
                  &=
                  -i\frac{\partial ( \gamma_l(t,\tau')-\gamma_n(t,\tau') )}
                               {\partial k_{c,i}}
                          a^{{\bm Q}}_{ln,\alpha}(\tau')\dot{Q}_\alpha (\tau'),
                   \\
                  h(\tau') 
                  &= 
                  \frac{\partial a^{{\bm Q}}_{ln,\alpha}(\tau')}{\partial k_{c,i}}\dot{Q}_\alpha (\tau'),
                   \\
                  g(\tau,\tau')
                  &=
                  e^{ -i\gamma_l(\tau,\tau') +i \gamma_n(\tau,\tau') }
                 ( a^{{\bm Q}}_{ln,\beta}(\tau) )^*
                 \dot{Q}_\beta (\tau)
                \end{align}
                Note that $f(t,t)=0$. 
          Then, the above expression can read
                \begin{align}\label{eq:dt_MdM_1}
                  \frac{d}{dt}
\left[
M^{(1)*}_{ln,\bm k_c}(t) 
\frac{\partial M^{(1)}_{ln,\bm k_c}(t)}
     {\partial k_{c,i}}
\right]
                  &=
                  \frac{d}{dt}
                  \int_{t_0}^t d\tau \int_{t_0}^t d\tau'
                  \Bigl\{
                         f(t,\tau')+h(\tau') 
                  \Bigr\}g(\tau,\tau')
                  \notag \\
                  &
                  =
                   \int_{t_0}^{t} d\tau \; f(t,t)\, g(\tau,t)
                 + \int_{t_0}^{t} d\tau' \; f(t,\tau')\, g(t,\tau')
                 + \int_{t_0}^{t} d\tau \int_{t_0}^{t} d\tau' \; 
                    (\partial_t f(t,\tau'))\, g(\tau,\tau')
                  \notag \\
                  &\ \ \ \ 
                 + \int_{t_0}^{t} d\tau'\; h(\tau')\,g(t,\tau')
                   \;+\;
                   h(t)\int_{t_0}^{t} d\tau\; g(\tau,t).
                \end{align}
                Here, the first term in the second equality vanishes since the function $f(t,t)$ appears.
                The second term also vanishes when considering asymptotic expansion of $g(t,\tau')$~(replacing $\tau'$ to $t$).
%
%
By applying the asymptotic expansion to the function $g$, 
%
%
%
and performing the partial derivative of the function $f(t,\tau')$ with time $t$, 
%
the third term in Eq. (\ref{eq:dt_MdM_1}) reads, 
               \begin{align}
                 \int_{t_0}^{t} d\tau \int_{t_0}^{t} d\tau' \; 
                                (\partial_t f(t,\tau'))\, g(\tau,\tau')
                 &=
                  -i \hbar 
\frac{\partial \varepsilon_{ln}(t) }
     {\partial k_{c,i}}
\frac{ a^{{\bm Q}}_{nl,\alpha}(t) a^{{\bm Q}}_{ln,\beta}(t) }
     { \varepsilon^2_{nl}(t) }
\dot{Q}_\alpha (t)
\dot{Q}_\beta (t)
+{\mathcal{O}}( {\dot Q}^3 ).
            \end{align}
Therefore, the second term of Eq.~(\ref{aeq:x(2)_1}) eventually takes the following form, 
\begin{align}\label{aeq:dt_MdM}
        \frac{d}{dt} {\rm Im}
        \Biggl[ 
               \sum_{l(\ne n)} M^*_{ln}(\bm k_c, {{\bm Q}}) \frac{\partial M_{ln}(\bm k_c, {{\bm Q}})}{\partial k_{c,i}} 
        \Biggr]
        =
        -\hbar \sum_{l(\ne n)}
         \frac{\partial \varepsilon_{ln}(t) }{\partial k_{c,i}}
         \frac{ {\rm Re}[a^{{\bm Q}}_{nl,\alpha}(t) a^{{\bm Q}}_{ln,\beta}(t)] }
     { \varepsilon^2_{nl}(t) }
\dot{Q}_\alpha (t)
\dot{Q}_\beta (t)
\end{align}

Finally, let us consider the third term of Eq.~(\ref{aeq:x(2)_1}).
The time derivative of $M_{ln,\bm k}(t)$ takes the form like 
               $ \frac{d}{dt} \int^t_{t_0} d\tau f(t,\tau)g(\tau) = f(t,t)g(t) + \int^t_{t_0} d\tau (\partial_t f(t,\tau)) g(\tau) $.
Hence, by applying asymptotic expansion
               \begin{align}
                 \frac{d M^{(1)}_{ln,\bm k_c}(t)}{dt}
                 &=
                 -i \frac{d}{dt} 
                    \int_{t_0}^t d\tau~
                    e^{-i\gamma_l({t,\tau}) + i\gamma_n({t,\tau})}
                    a^{{\bm Q}}_{ln} (\tau)\dot{Q}_\alpha (\tau)
                 \notag \\
                 &\sim
                 -i a^{{\bm Q}}_{ln} (t)\dot{Q}_\alpha (t)
                 -i
                   \left(
                         \frac{\varepsilon_{ln}(t)}{\hbar}
                         +(A^{{\bm Q}}_{l,\alpha}(t) - A^{{\bm Q}}_{n,\alpha}(t))\dot{Q}_\alpha (t)
                   \right)
                   \frac{\hbar}{i}
                   \frac{ a^{{\bm Q}}_{ln} (t) \dot{Q}_\alpha (t) }
                        { \varepsilon_{ln}(t) + \hbar (A^{{\bm Q}}_{l,\beta}(t) - A^{{\bm Q}}_{n,\beta}(t))\dot{Q}_\alpha (t) }
                \notag \\
                &=
                -i a^{{\bm Q}}_{ln} (t)\dot{Q}_\alpha (t) - a^{{\bm Q}}_{ln} (t)\dot{Q}_\alpha (t).
               \end{align}
Then, one can obtain the results, 
              \begin{align}
                {\rm Im}\left[
                \sum_{l(\ne n)} \frac{\partial}{\partial k_{c,i}}
                \left(
                M^{(1)*}_{ln,\bm k_c}(t)
                \frac{d M^{(1)}_{ln,\bm k_c}(t)}{dt}\right) \right] 
                &=
                \hbar
                \sum_{l(\ne n)}
                \partial_{k_{c,i}}
                \left(
                \frac{ {\rm Re} [a^{{\bm Q}}_{nl,\alpha} (t)  a^{{\bm Q}}_{ln,\beta} (t)] }
                     { \varepsilon_{ln}(t)  }
                \right)
                \dot{Q}_\alpha (t)
                \dot{Q}_\beta (t).
              \end{align}

Let us define $\dot{x}^{\rm na}_{c,i} = \dot{x}^{(1)}_{c,i} + \dot{x}^{(2)}_{c,i}$.
Combining Eqs.~(B27), (B28), and (B39), we obtain the expression of $\dot{x}^{\rm na}_{c,i}$ up to the second order of ${{\bm Q}}$ as follows, 
\begin{align}\label{aeq:x(2)_2}
    \dot{x}^{\rm na}_{c,i}
    &=
        \hbar \sum_{l(\ne n)}
               \frac{  \partial_{k_{c,i}}{\rm Re}[a^{{{\bm Q}}}_{nl,\alpha}(0) a^{{{\bm Q}}}_{ln,\beta}(0)]
                    -2\partial_{Q_\alpha} {\rm Re}[a^{\bm k}_{nl,   i  }(0) a^{{{\bm Q}}}_{ln,\beta}(0)]
                    }
                    { \varepsilon_{nl}(0)}
        \dot{Q}_\alpha (t)\dot{Q}_\beta (t)
        \notag \\
       &\ \ \ 
        +
          \hbar \sum_{l(\ne n)}
          \frac{\partial}{\partial k_{c,i}}
          \left(
                \frac{ {\rm Re}[a^{{{\bm Q}}}_{nl,\alpha}(0) \varepsilon_{l\bm k_c}(0) a^{{{\bm Q}}}_{ln,\beta}(0)] }
                     { \varepsilon^2_{nl,\bm k_c}(0)  }
          \right) {\dot Q}_{\alpha}(t) {\dot Q}_{\beta}(t)
       -2\hbar \sum_{l(\ne n)}
         \frac{\partial \varepsilon_{ln,\bm k_c}(0) }{\partial k_{c,i}}
         \frac{ {\rm Re}[a^{{{\bm Q}}}_{nl,\alpha}(0) a^{{{\bm Q}}}_{ln,\beta}(0)] }
              { \varepsilon^2_{nl,\bm k_c}(0)  }
         {\dot Q}_{\alpha}(t) {\dot Q}_{\beta}(t)
\end{align}
The first term of Eq.~(\ref{aeq:x(2)_2}) can be transformed as follows, 
      \begin{align}
        &\hbar \sum_{l(\ne n)}
               \frac{  \partial_{k_{c,i}}{\rm Re}[a^{{{\bm Q}}}_{nl,\alpha}(t) a^{{{\bm Q}}}_{ln,\beta}(t)]
                    -2\partial_{Q_\alpha} {\rm Re}[a^{\bm k}_{nl,   i  }(t) a^{{{\bm Q}}}_{ln,\beta}(t)]
                    }
                    { \varepsilon_{nl}(t)}
        \dot{Q}_\alpha (t)\dot{Q}_\beta (t)
        \notag \\
        &=
        \hbar \sum_{l(\ne n)}
               \frac{  \partial_{k_{c,i}}{\rm Re}[a^{{{\bm Q}}}_{nl,\alpha}(t) a^{{{\bm Q}}}_{ln,\beta}(t)]
                    -\partial_{Q_\alpha} {\rm Re}[a^{\bm k}_{nl,   i  }(t) a^{{{\bm Q}}}_{ln,\beta}(t)]
                    -\partial_{Q_\beta} {\rm Re}[a^{\bm k}_{nl,   i  }(t) a^{{{\bm Q}}}_{ln,\alpha}(t)]
                    }
                    { \varepsilon_{nl}(t)}
        \dot{Q}_\alpha (t)\dot{Q}_\beta (t)
        \notag \\
        &=
         -2\hbar \sum_{l(\ne n)}
               \frac{  (\Gamma^{\bm k {{\bm Q}} {{\bm Q}}}_{i, \alpha \beta} (t))_{nl}
                    }
                    { \varepsilon_{nl}(t)}
        \dot{Q}_\alpha (t)\dot{Q}_\beta (t).
      \end{align}
The second and third terms of Eq.~(\ref{aeq:x(2)_2}) can be summed up by $\partial_{k_{c,i}}$-derivative, and to sum up, we obtain the expression of $\dot{x}^{\rm na}_{c,i}$ in the form of Eq.~(\ref{eq:WPvel_na}) in the main text.

\end{widetext}


{\section{ Electron-phonon coupling on lattice models }}

\normalsize

{
We first describe the optical phonon modes considered in the honeycomb lattice models. 
The lattice consists of two sublattices, $A$ and $B$, each with an in-plane displacement vector $\Delta\bm R_{A/B}(\bm R,t)$. 
At $\bm q=\bm 0$, the two in-plane acoustic modes correspond to rigid translations of the lattice, for which $\Delta\bm R_A(t)=\Delta\bm R_B(t)$. 
Since these displacements do not modify the relative atomic positions, they do not affect the hopping amplitudes.
On the other hand, the two in-plane optical modes involve opposite displacements of the sublattices, $\Delta\bm R_A(t)=-\Delta\bm R_B(t)$, and are degenerate in the honeycomb lattice considered here. 
Since these modes are orthogonal as described in the main text, we choose their polarization directions along the $x$ and $y$ axes. 
Absorbing the normalization factors of the normal modes into their amplitudes, we define the phonon coordinates by the relative displacement
\begin{align}
{\bm Q}(t)
&\equiv \Delta\bm R_B(t)-\Delta\bm R_A(t)
=Q_x(t)\hat{\bm x}+Q_y(t)\hat{\bm y}.
\end{align}
Here, $Q_x(t)$ and $Q_y(t)$ represent the time-dependent amplitudes of the two optical modes. 
In contrast to the general formulation, 
where the phonon coordinates label arbitrary optical modes, 
${\bm Q}(t)$ in the following lattice models represents the in-plane relative displacement of the two sublattices.
}

{
We next derive the electron-phonon coupling in the Haldane model through the modulation of the NN hopping amplitudes. 
Let $\bm\tau_l$ ($l=1,2,3$) denote the NN bond vectors connecting the $B$ and $A$ sublattices, as defined in the main text. 
In the presence of lattice displacements, the atomic positions are given by
\begin{align}
\bm r_A(t)
&=\bm R+\bm\tau_l+\Delta\bm R_A(t),
\ \ 
\bm r_B(t)
=\bm R+\Delta\bm R_B(t).
\end{align}
Consequently, the bond vector changes according to
\begin{align}
\bm\tau_l
\longrightarrow
\bm r_A(t)-\bm r_B(t)
=\bm\tau_l-{\bm Q}(t).
\end{align}
This deformation modifies the overlap between electronic orbitals and hence the hopping amplitude. 
Expanding the NN hopping to first order in the lattice displacement, we obtain
\begin{align}
t_l({\bm Q})
&=t_0+\delta t_l({\bm Q}),
\ \ \ \ \ 
\delta t_l({\bm Q})
=\frac{\beta t_0}{a^2}
\bm\tau_l\cdot{\bm Q},
\end{align}
where $\beta$ is a dimensionless effective electron-phonon coupling parameter.
The resulting electron-phonon coupling Hamiltonian reads
\begin{align}
\delta\hat H_{\rm NN}({\bm Q})
=-t_0\sum_{\bm R}\sum_{l=1}^{3}
\left[
\frac{\beta}{a^2}
(\bm\tau_l\cdot{\bm Q})
\hat c^\dagger_{A,\bm R+\bm\tau_l}
\hat c_{B,\bm R}
+{h.c.}
\right].
\end{align}
which is presented as Eq.~(\ref{eq:Haldane_HamQ}) in the main text.
Here, the NNN hopping associated with the Haldane mass is assumed to remain unchanged, since the optical phonons at the $\Gamma$-point do not modify the relative positions of atoms belonging to the same sublattice. 
}

{
We finally consider the electron-phonon coupling in the Kane-Mele model. 
As in the Haldane model, the phonon-induced modulation of the NN hopping is described by $\delta\hat H_{\rm NN}({\bm Q})$ derived above. 
In addition, lattice vibrations modify the intrinsic SOC associated with NNN hopping.
In the Kane-Mele model, the SOC hopping depends on the cross product of two consecutive NN bond vectors, $\bm d_1^{ij}$ and $\bm d_2^{ij}$, connecting the NNN sites $i$ and $j$ through an intermediate site. 
For the optical phonon at the $\Gamma$-point, these two vectors acquire opposite corrections due to the relative displacement of the sublattices. 
With the bond-vector convention used in the model, their deformation is given by
\begin{align}
\bm d_1^{ij}&\longrightarrow\bm d_1^{ij}-{\bm Q},\ \ \ \ 
\bm d_2^{ij}\longrightarrow\bm d_2^{ij}+{\bm Q}.
\end{align}
Accordingly, the cross product changes as 
\begin{align}
\left(\bm d_1^{ij}\times\bm d_2^{ij}\right)_z
\longrightarrow
\left(\bm d_1^{ij}\times\bm d_2^{ij}\right)_z
\nonumber\
+
\left[
\left(\bm d_1^{ij}+\bm d_2^{ij}\right)
\times{\bm Q}
\right]_z.
\end{align}
The correction quadratic in ${\bm Q}$ vanishes because it is proportional to ${\bm Q}\times{\bm Q}=\bm0$.
Then, the resulting electron-phonon coupling is expressed as 
\begin{align}
\delta\hat H({\bm Q})
&=
\delta\hat H_{\rm NN}({\bm Q})
+
\delta\hat H_{\rm SOC}({\bm Q}),\\
\delta\hat H_{\rm SOC}({\bm Q})
&=
\frac{2i\lambda_{\rm SOC}}{\sqrt3}
\sum_{\langle\langle ij\rangle\rangle}
\left[
\left(\hat{\bm d}_1^{ij}+\hat{\bm d}_2^{ij}\right)
\times{\bm Q}
\right]_z
\hat c_i^\dagger\hat\sigma_z\hat c_j.
\end{align}
which is presented as Eq.~(\ref{eq:KaneMele_ephHam}) in the main text.
}


\section{ Constraints under the ${\mathcal M}_x {\mathcal T}$ symmetry }
\normalsize

Here, we derive the constraints for the conductivity under the magnetic mirror symmetry ${\mathcal M}_x {\mathcal T}$.
For the Haldane model and the $S_z$-preserved Kane-Mele model discussed in the main text, 
we set the time-reversal operator by the complex conjugation $\mathcal{K}$.
Meanwhile, we define the mirror operator acting on the electron-creation/anihiration operators such that $ \hat{{\mathcal M}}_x \hat{c}_{A/B, \bm R} \hat{\mathcal M}_x^{-1} = \hat{c}_{A/B, {\mathbb{M}}_x \bm R}$, where ${\mathbb{M}}_x = {\rm diag}(-1,1)$.
We note that, for the lattice vector defined in the main text, sublattice degrees of freedom do not exchange each other under this mirror operation.
Hence, the anti-unitary operator $\hat{\Theta}_{{\mathcal M}_x {\mathcal T}}$ corresponding to the ${\mathcal M}_x {\mathcal T}$ operation is defined by the composite operation, $\hat{\Theta}_{{\mathcal M}_x {\mathcal T}} = \hat{{\mathcal M}}_x\mathcal{K}$.

The Haldane model meets the relation 
$\hat{\Theta}_{{\mathcal M}_x {\mathcal T}} \hat{H}({{\bm Q}}) \hat{\Theta}_{{\mathcal M}_x {\mathcal T}}^{-1} = \hat{H}({\mathbb{M}}_x {{\bm Q}})$.
This relation can be proved by calculating the action of the magnetic mirror operation on the Haldane model.
The pristine Haldane model is composed of the nearest neighbor hopping $\hat{H}_{\rm NN}$, sublattice-dependent potential $\hat{H}_{\Delta}$, and the Haldane term $\hat{H}_\phi$.
$\hat{H}_{\rm NN}$ is transformed as follows, 
\begin{align}
    &\hat{\Theta}_{{\mathcal M}_x {\mathcal T}} \hat{H}_{\rm NN} \hat{\Theta}_{{\mathcal M}_x {\mathcal T}}^{-1}
    \nonumber\\
    &\ \ =
    \hat{\Theta}_{{\mathcal M}_x {\mathcal T}} 
    \Biggl( -t_0 \sum_{\bm R}\sum_{l=1}^3 
             \Bigl( \hat{c}^\dagger_{A,\bm R+\bm \tau_l} \hat{c}_{B,\bm R} + {\rm h.c.} \Bigr) 
    \Biggr) 
    \hat{\Theta}_{{\mathcal M}_x {\mathcal T}}^{-1} 
    \nonumber \\
    &\ \ =
    -t_0 \sum_{\bm R}\sum_{l=1}^3 
     \Bigl( \hat{c}^\dagger_{A,{\mathbb{M}}_x\bm R+ {\mathbb{M}}_x \bm\tau_l} \hat{c}_{B,{\mathbb{M}}_x\bm R} + {\rm h.c.} \Bigr) 
    \nonumber\\
    &\ \ =
    -t_0 \sum_{\bm R'}\sum_{l'=1}^3 
     \Bigl( \hat{c}^\dagger_{A,\bm R'+ \bm\tau_{l'}} \hat{c}_{B,\bm R'} + {\rm h.c.} \Bigr) \ 
    =
    \hat{H}_{\rm NN},
\end{align}
where $\bm R' = {\mathbb{M}}_x\bm R$.
In the third equality, we use the fact that the lattice vectors $\bm \tau_l$ is transformed into the themselves; 
${\mathbb{M}}_x \bm \tau_1 = \bm \tau_1$ and ${\mathbb{M}}_x \bm \tau_{2(3)} = \bm \tau_{3(2)}$, 
and therefore it is sufficient to rewrite their label. We suppose $l'$ such that $l'=2(3)$ corresponds $l=3(2)$.
For $\hat{H}_{\Delta}$, we obtain 
$ \hat{\Theta}_{{\mathcal M}_x {\mathcal T}} \hat{H}_{\Delta} \hat{\Theta}_{{\mathcal M}_x {\mathcal T}}^{-1} = \hat{H}_{\Delta}$
by similar way to the above calculation. 
The condition for $\hat{H}_\phi$ is the same as the above ones, 
$ \hat{\Theta}_{{\mathcal M}_x {\mathcal T}} \hat{H}_{\phi} \hat{\Theta}_{{\mathcal M}_x {\mathcal T}}^{-1} = \hat{H}_{\phi}$,
and this can be proved as follows, 
\begin{align}
    \hat{\Theta}_{{\mathcal M}_x {\mathcal T}} \hat{H}_{\phi} \hat{\Theta}_{{\mathcal M}_x {\mathcal T}}^{-1}
    &=
    \hat{\Theta}_{{\mathcal M}_x {\mathcal T}} 
    \Biggl( 
           t_\phi \sum_{\braket{\braket{ij}}} e^{i\nu_{ij}\phi} \hat{c}^\dagger_{i} \hat{c}_{j}
    \Biggr) 
    \hat{\Theta}_{{\mathcal M}_x {\mathcal T}}^{-1} 
    \nonumber \\
    &=
    t_\phi \sum_{\braket{\braket{ij}}} e^{-i\nu_{ij}\phi} \hat{c}^\dagger_{i'} \hat{c}_{j'}
    \nonumber\\
    &=
    t_\phi \sum_{\braket{\braket{i'j'}}} e^{i\nu_{i'j'}\phi} \hat{c}^\dagger_{i'} \hat{c}_{j'} \ 
    =\hat{H}_{\phi},
\end{align}
where $i'$ is the spatial coordinate transformed by the ${\mathbb{M}}_x$-operation.
In the second line, we use the anti-unitary property of $\hat{\Theta}_{{\mathcal M}_x {\mathcal T}}$, $\hat{\Theta}_{{\mathcal M}_x {\mathcal T}} z \hat{\Theta}_{{\mathcal M}_x {\mathcal T}}^{-1} = z^* \hat{\Theta}_{{\mathcal M}_x {\mathcal T}} \hat{\Theta}_{{\mathcal M}_x {\mathcal T}}^{-1}$ when $z$ is a complex number.
In the third line, we replace the factor $\nu_{ij}$ with $-\nu_{i'j'}$.
The additional minus sign comes from the fact that the clockwise hopping is transformed to anti-clockwise one by the${\mathcal M}_x {\mathcal T}$-operation.
In addition to the above pristine Haldane model, in the main text, we consider the electron-phonon coupling, $\delta\hat{H}({{\bm Q}})$.
This term is transformed in the following way, 
\begin{align}
    &\hat{\Theta}_{{\mathcal M}_x {\mathcal T}} \delta\hat{H}({{\bm Q}}) \hat{\Theta}_{{\mathcal M}_x {\mathcal T}}^{-1}
    \nonumber\\
    &\ =
    \hat{\Theta}_{{\mathcal M}_x {\mathcal T}} 
    \Biggl( -t_0 \sum_{\bm R,l}
             \Bigl( \frac{\beta}{a^2}( {{\bm Q}}\cdot \bm \tau_l ) \hat{c}^\dagger_{A,\bm R+\bm \tau_l} \hat{c}_{B,\bm R} + {\rm h.c.} \Bigr) 
    \Biggr) 
    \hat{\Theta}_{{\mathcal M}_x {\mathcal T}}^{-1} 
    \nonumber \\
    &\ =
    -t_0 \sum_{\bm R,l}
    \Bigl( 
          \frac{\beta}{a^2}( {{\bm Q}}\cdot \bm \tau_l ) 
          \hat{c}^\dagger_{A,{\mathbb{M}}_x\bm R+{\mathbb{M}}_x\bm \tau_l} \hat{c}_{B,{\mathbb{M}}_x\bm R} + {\rm h.c.} 
    \Bigr)
    \nonumber\\
    &\ =
    -t_0 \sum_{\bm R',l'}
    \Bigl( 
          \frac{\beta}{a^2}( {{\bm Q}}\cdot {\mathbb{M}}_x\bm \tau_{l'} ) 
          \hat{c}^\dagger_{A,\bm R'+\bm \tau_{l'}} \hat{c}_{B,\bm R'} + {\rm h.c.} 
    \Bigr)
    \nonumber\\
    &\ =
    \delta\hat{H}({\mathbb{M}}_x{{\bm Q}}).
\end{align}
In the forth equality, we use the fact that ${\mathbb{M}}_x$ is the orthogonal matrix and therefore 
${{\bm Q}}\cdot ({\mathbb{M}}_x\bm \tau_{l'}) = ({\mathbb{M}}_x{{\bm Q}})\cdot \bm \tau_{l'}$.
From the above considerations, we obtain the condition under the ${\mathcal M}_x {\mathcal T}$ symmetry as 
\begin{align}\label{aeq:Ham_MxT}
    \hat{\Theta}_{{\mathcal M}_x {\mathcal T}} \hat{H}({{\bm Q}}) \hat{\Theta}_{{\mathcal M}_x {\mathcal T}}^{-1} = \hat{H}({\mathbb{M}}_x {{\bm Q}}).
\end{align}

By conducting the Fourier transformation with the ${\mathcal M}_x {\mathcal T}$-operation, one can obtain the condition for the Bloch Hamiltonian.
The Fourier transformation of the electron annihilation operator is transformed as follows, 
\begin{align}
    &\hat{\Theta}_{{\mathcal M}_x {\mathcal T}} \hat{c}_{A/B,\bm k} \hat{\Theta}_{{\mathcal M}_x {\mathcal T}}^{-1}
    \nonumber\\
    &=
    \frac{1}{\sqrt{N}}\sum_{\bm R}
    \hat{\Theta}_{{\mathcal M}_x {\mathcal T}}
    \Bigl(
          e^{-i\bm k\cdot\bm R}  \hat{c}_{A/B,\bm R} 
    \Bigr)
    \hat{\Theta}_{{\mathcal M}_x {\mathcal T}}^{-1}
    \nonumber\\
    &=
    \frac{1}{\sqrt{N}}\sum_{\bm R}
    e^{+i\bm k\cdot\bm R}  \hat{c}_{A/B,{\mathbb{M}}_x\bm R} 
=
    \frac{1}{\sqrt{N}}\sum_{\bm R'}
    e^{-i(-{\mathbb{M}}_x\bm k)\cdot\bm R}\hat{c}_{A/B,\bm R'} 
    \nonumber\\
    &= 
    \hat{c}_{A/B,-{\mathbb{M}}_x\bm k}.
\end{align}
Accordingly, the condition for the Bloch Hamiltonian is obtained as follows. 
Defining the operator $\hat{\Psi}_{\bm k} = ( \hat{c}_{A,\bm k}, \hat{c}_{B,\bm k} )^{\rm T}$ and the roman indices $a,b=A,B$,
\begin{align}\label{aeq:BlochHam_MxT_1}
    &\hat{\Theta}_{{\mathcal M}_x {\mathcal T}} \hat{H}({{\bm Q}}) \hat{\Theta}_{{\mathcal M}_x {\mathcal T}}^{-1}
    \nonumber\\
    &=
    \hat{\Theta}_{{\mathcal M}_x {\mathcal T}}
    \left(
    \sum_{\bm k,\alpha,\beta} \hat{\Psi}_{\bm k,\alpha}^{\dagger} {\mathcal{H}}_{\alpha\beta}(\bm k,{{\bm Q}})\hat{\Psi}_{\bm k,\beta}
    \right)
    \hat{\Theta}_{{\mathcal M}_x {\mathcal T}}^{-1}
    \nonumber\\
    &=
    \sum_{\bm k,\alpha,\beta} {\mathcal{H}}^*_{\alpha\beta}(\bm k,{{\bm Q}}) 
    \hat{\Theta}_{{\mathcal M}_x {\mathcal T}} 
    \left(\hat{\Psi}_{\bm k,\alpha}^{\dagger} \hat{\Psi}_{\bm k,\beta} \right)
    \hat{\Theta}_{{\mathcal M}_x {\mathcal T}}^{-1}
    \nonumber\\
    &=
    \sum_{\bm k,\alpha,\beta} {\mathcal{H}}^*_{\alpha\beta}(\bm k,{{\bm Q}}) 
    \hat{\Psi}_{-{\mathbb{M}}_x\bm k,\alpha}^{\dagger}  \hat{\Psi}_{-{\mathbb{M}}_x\bm k,\beta}
    \nonumber\\
    &=
    \sum_{\bm k,\alpha,\beta} 
    \hat{\Psi}_{\bm k',\alpha}^{\dagger}
    {\mathcal{H}}^*_{\alpha\beta}(-{\mathbb{M}}_x\bm k',{{\bm Q}}) 
    \hat{\Psi}_{\bm k',\beta}
    \nonumber\\
    &=
    \sum_{\bm k,\alpha,\beta} \hat{\Psi}_{\bm k,\alpha}^{\dagger} {\mathcal{H}}_{\alpha\beta}(\bm k,{\mathbb{M}}_x{{\bm Q}})\hat{\Psi}_{\bm k,\beta},
\end{align}
where we define $\bm k'={\mathbb{M}}_x\bm k$ in the fourth equality.
In the fifth equality, we use the condition of Eq.~(\ref{aeq:Ham_MxT}).
$-{\mathbb{M}}_x\bm k=(k_x,-k_y)^{\rm T}$ and ${\mathbb{M}}_x{{\bm Q}}=(-Q_x,Q_y)^{\rm T}$, and hence,
with Eq.~(\ref{aeq:BlochHam_MxT_1}), we obtain the following condition, 
\begin{align}\label{aeq:BlochHam_MxT_2}
    \hat{\mathcal{H}}(\bm k,{{\bm Q}})
    &=
    \hat{\mathcal{H}}^*(-{\mathbb{M}}_x\bm k, {\mathbb{M}}_x{{\bm Q}})
    \nonumber\\
    &=
    \hat{\mathcal{H}}^*(k_x,-k_y,-Q_x,Q_y),
\end{align}
under the ${{\mathcal M}_x {\mathcal T}}$ symmetry.
In the following, we introduce the sign $s^{\bm k}_{x,y}$ and $s^{{{\bm Q}}}_{x,y}$ as $s^{\bm k}_{x}=+1, s^{\bm k}_{y}=-1, s^{{{\bm Q}}}_{x}=-1, $ and $s^{{{\bm Q}}}_{y}=+1$ 
for notational transparency.

Using Eq.~(\ref{aeq:BlochHam_MxT_2}), we can also derive the condition for the matrix elements of the velocity operator, $v^{nm}_{i}(\bm k,{{\bm Q}})$.
We decompose the momentum derivative in the velocity operator, 
\begin{widetext}
\begin{align}\label{aeq:vel_MxT_calc}
    v^{nm}_{i}(\bm k,{{\bm Q}})
    &=
    \bra{ u_{n\bm k} ({{\bm Q}})} \frac{ \partial \hat{\mathcal{H}}(\bm k,{{\bm Q}}) }{\partial k_i} \ket{ u_{m\bm k} ({{\bm Q}})}
    =
    \bra{ u_{n\bm k} ({{\bm Q}})} 
        \frac{ \hat{\mathcal{H}}(\bm k+\Delta k \hat{\bm i},{{\bm Q}}) - \hat{\mathcal{H}}(\bm k,{{\bm Q}}) }
        {\Delta k} 
    \ket{ u_{m\bm k} ({{\bm Q}})}
    \nonumber\\
    &=
    \bra{ u_{n\bm k} ({{\bm Q}})} 
        \frac{ \hat{\mathcal{H}}^*(-{\mathbb{M}}_x\bm k-\Delta k {\mathbb{M}}_x \hat{\bm i},{\mathbb{M}}_x{{\bm Q}}) - \hat{\mathcal{H}}^*(-{\mathbb{M}}_x\bm k,{\mathbb{M}}_x{{\bm Q}}) }
        {\Delta k} 
    \ket{ u_{m\bm k} ({{\bm Q}})}
    \nonumber\\
    &=
    \Bigl(\bra{ u_{n,-{\mathbb{M}}_x\bm k} ({\mathbb{M}}_x{{\bm Q}})} \Bigr)^* 
        \frac{ \hat{\mathcal{H}}^*(k_x+\Delta k \hat{i}_x, -k_y+\Delta k \hat{i}_y,{\mathbb{M}}_x{{\bm Q}}) - \hat{\mathcal{H}}^*(k_x, -k_y,{\mathbb{M}}_x{{\bm Q}}) }
        {\Delta k} 
    \Bigl(\ket{ u_{m,-{\mathbb{M}}_x\bm k} ({\mathbb{M}}_x{{\bm Q}})} \Bigr)^*
    \nonumber\\
    &=
    s^{\bm k}_i
    \Bigl(\bra{ u_{n,-{\mathbb{M}}_x\bm k} ({\mathbb{M}}_x{{\bm Q}})} \Bigr)^* 
        \frac{ \hat{\mathcal{H}}^*(k_x + s^{\bm k}_x \Delta k \hat{i}_x, -k_y- s^{\bm k}_y\Delta k \hat{i}_y,{\mathbb{M}}_x{{\bm Q}}) - \hat{\mathcal{H}}^*(k_x, -k_y,{\mathbb{M}}_x{{\bm Q}}) }
        {s^{\bm k}_i\Delta k} 
    \Bigl(\ket{ u_{m,-{\mathbb{M}}_x\bm k} ({\mathbb{M}}_x{{\bm Q}})} \Bigr)^* .
\end{align}
\end{widetext}
Here, we use the condition of Eq.~(\ref{aeq:BlochHam_MxT_2}) in the third equality.
In the fourth equality, we use the fact that 
$\ket{ u_{n,\bm k} ({{\bm Q}})} =\ket{ u_{n,-{\mathbb{M}}_x\bm k} ({\mathbb{M}}_x{{\bm Q}})} ^*$ holds except for the $U(1)$-gauge degrees of freedom, which can be derived by the condition of Eq.~(\ref{aeq:BlochHam_MxT_2}).
In the fifth equality, we introduce the sign factor $s^{\bm k}_i$ to make the arguments of the Bloch Hamiltonian and the Bloch eigenstates consistent.
The last line of Eq.~(\ref{aeq:vel_MxT_calc}) is nothing but the definition of an matrix element of the velocity operator, which is composed of $\ket{ u_{n,-{\mathbb{M}}_x\bm k} ({\mathbb{M}}_x{{\bm Q}})} ^*$.
Therefore, we eventually obtain the following condition, 
\begin{align}\label{aeq:v_MxT}
    v^{nm}_{i}(\bm k,{{\bm Q}})
    &=
    s^{\bm k}_i  \Bigl( v^{nm}_{i}(-{\mathbb{M}}_x\bm k,{\mathbb{M}}_x{{\bm Q}}) \Bigr)^*.
\end{align}
One can also obtain the condition for an matrix element for the electron-phonon coupling operator by the similar calculation above, and that condition reads, 
\begin{align}\label{aeq:lambda_MxT}
    \lambda^{nm}_{\alpha}(\bm k,{{\bm Q}})
    &=
    s^{{{\bm Q}}}_\alpha \Bigl( \lambda^{nm}_{\alpha}(-{\mathbb{M}}_x\bm k,{\mathbb{M}}_x{{\bm Q}}) \Bigr)^*.
\end{align}
Accordingly, by using the Hellman-Feynman relation, the Berry connections are supposed to have the following conditions under the ${{\mathcal M}_x {\mathcal T}}$ symmetry, 
\begin{align}
    \label{aeq:ak_MxT_condition}
    a^{\bm k}_{nm,i}(\bm k, {{\bm Q}}) 
    &=
    -s^{\bm k}_i  \Bigl( a^{\bm k}_{nm,i}(-{\mathbb{M}}_x\bm k,{\mathbb{M}}_x{{\bm Q}}) \Bigr)^*, 
    \\
    \label{aeq:aQ_MxT_condition}
    a^{{{\bm Q}}}_{nm,\alpha}(\bm k, {{\bm Q}}) 
    &=
    -s^{\bm k}_\alpha  \Bigl( a^{{{\bm Q}}}_{nm,\alpha}(-{\mathbb{M}}_x\bm k,{\mathbb{M}}_x{{\bm Q}}) \Bigr)^*.
\end{align}

From the condition of Eqs.~(\ref{aeq:ak_MxT_condition}) and (\ref{aeq:aQ_MxT_condition}), the quantum geometric quantities are supposed to follow the condition below, 
\begin{align}
    \label{aeq:gkQ_MxT}
    g^{\bm k {{\bm Q}}}_{i\alpha,n}(\bm k, {{\bm Q}})
    &=
    {\rm Re}\left[ \sum_{m(\ne n)} a^{\bm k}_{nm,i}(\bm k, {{\bm Q}}) a^{{{\bm Q}}}_{mn,\alpha}(\bm k, {{\bm Q}}) \right]
    \nonumber\\
    &=
    {\rm Re}
    \Biggl[
          \sum_{m(\ne n)}
          (-s^{\bm k}_i)  \Bigl( a^{\bm k}_{nm,i}(-{\mathbb{M}}_x\bm k,{\mathbb{M}}_x{{\bm Q}}) \Bigr)^*
    \nonumber\\
    &\ \ \ \ \ \ \ \ \ \ \ \ \ \ \ \ 
          (-s^{\bm k}_\alpha)  \Bigl( a^{{{\bm Q}}}_{nm,\alpha}(-{\mathbb{M}}_x\bm k,{\mathbb{M}}_x{{\bm Q}}) \Bigr)^*
    \Biggr]
    \nonumber\\
    &=
    s^{\bm k}_i s^{{{\bm Q}}}_\alpha\; g^{\bm k {{\bm Q}}}_{i\alpha,n}(-{\mathbb{M}}_x\bm k,{\mathbb{M}}_x{{\bm Q}}),
    \\
    \label{aeq:gQQ_MxT}
    g^{{{\bm Q}} {{\bm Q}}}_{\alpha\beta,n}(\bm k, {{\bm Q}})
    &=
    s^{{{\bm Q}}}_\alpha s^{{{\bm Q}}}_\beta\; g^{{{\bm Q}} {{\bm Q}}}_{\alpha\beta,n}(-{\mathbb{M}}_x\bm k,{\mathbb{M}}_x{{\bm Q}}),
    \\
    \label{aeq:OmegakQ_MxT}
    \Omega^{\bm k {{\bm Q}}}_{i\alpha,n}(\bm k, {{\bm Q}})
    &=
    -2{\rm Im}\left[ \sum_{m(\ne n)} a^{\bm k}_{nm,i}(\bm k, {{\bm Q}}) a^{{{\bm Q}}}_{mn,\alpha}(\bm k, {{\bm Q}}) \right]
    \nonumber\\
    &=
    -2{\rm Im}
    \Biggl[
          \sum_{m(\ne n)}
          (-s^{\bm k}_i)  \Bigl( a^{\bm k}_{nm,i}(-{\mathbb{M}}_x\bm k,{\mathbb{M}}_x{{\bm Q}}) \Bigr)^*
    \nonumber\\
    &\ \ \ \ \ \ \ \ \ \ \ \ \ \ \ \ 
          (-s^{\bm k}_\alpha)  \Bigl( a^{{{\bm Q}}}_{nm,\alpha}(-{\mathbb{M}}_x\bm k,{\mathbb{M}}_x{{\bm Q}}) \Bigr)^*
    \Biggr]
    \nonumber\\
    &=
    -s^{\bm k}_i s^{{{\bm Q}}}_\alpha\; \Omega^{\bm k {{\bm Q}}}_{i\alpha,n}(-{\mathbb{M}}_x\bm k,{\mathbb{M}}_x{{\bm Q}}),
    \\
    \label{aeq:OmegaQQ_MxT}
    \Omega^{{{\bm Q}} {{\bm Q}}}_{\alpha\beta,n}(\bm k, {{\bm Q}})
    &=
    -s^{{{\bm Q}}}_\alpha s^{{{\bm Q}}}_\beta\; \Omega^{{{\bm Q}} {{\bm Q}}}_{\alpha\beta,n}(-{\mathbb{M}}_x\bm k,{\mathbb{M}}_x{{\bm Q}}).
\end{align}

For the matrix elements, $v^{i;\alpha}_{nm}$ and $\lambda^{\alpha\beta}_{nm}$, which appeared in $\mathcal{T}^{\partial^2}_{n,i\alpha\beta}$, one can obtain the conditions by the similar calculations to Eq.~(\ref{aeq:vel_MxT_calc}). It reads, 
\begin{widetext}
\begin{align}\label{aeq:via_MxT_calc}
    v^{i;\alpha}_{nm}
    &=
     \bra{ u_{n\bm k} ({{\bm Q}})} \frac{ \partial^2 \hat{\mathcal{H}}(\bm k,{{\bm Q}}) }{\partial Q_\alpha \partial k_i} \ket{ u_{m\bm k} ({{\bm Q}})}
    \nonumber\\
    &=
    \bra{ u_{n\bm k} ({{\bm Q}})} 
        \frac{ \hat{\mathcal{H}}(\bm k+\Delta k \hat{\bm i},{{\bm Q}} +\Delta Q \hat{\bm j}) 
              -\hat{\mathcal{H}}(\bm k+\Delta k \hat{\bm i},{{\bm Q}})
              }
             {\Delta k \Delta Q}
       -
        \frac{ \hat{\mathcal{H}}(\bm k,{{\bm Q}} +\Delta Q \hat{\bm j}) 
              -\hat{\mathcal{H}}(\bm k,{{\bm Q}})
              }
             {\Delta k \Delta Q} 
    \ket{ u_{m\bm k} ({{\bm Q}})}
    \nonumber\\
    &=
    \Bigl(\bra{ u_{n,-{\mathbb{M}}_x\bm k} ({\mathbb{M}}_x{{\bm Q}})} \Bigr)^* 
        \frac{ \hat{\mathcal{H}}^*(-{\mathbb{M}}_x\bm k-\Delta k {\mathbb{M}}_x\hat{\bm i},{\mathbb{M}}_x{{\bm Q}} +\Delta Q {\mathbb{M}}_x\hat{\bm j}) 
              -\hat{\mathcal{H}}^*(-{\mathbb{M}}_x\bm k-\Delta k {\mathbb{M}}_x\hat{\bm i},{\mathbb{M}}_x{{\bm Q}})
              }
             {\Delta k \Delta Q}
    \Bigl(\ket{ u_{m,-{\mathbb{M}}_x\bm k} ({\mathbb{M}}_x{{\bm Q}})} \Bigr)^*
    \nonumber\\
    &\ \ \ \ \ 
    -
    \Bigl(\bra{ u_{n,-{\mathbb{M}}_x\bm k} ({\mathbb{M}}_x{{\bm Q}})} \Bigr)^* 
        \frac{ \hat{\mathcal{H}}^*(-{\mathbb{M}}_x\bm k,{\mathbb{M}}_x{{\bm Q}} +\Delta Q {\mathbb{M}}_x\hat{\bm j}) 
              -\hat{\mathcal{H}}^*(-{\mathbb{M}}_x\bm k,{\mathbb{M}}_x{{\bm Q}})
              }
             {\Delta k \Delta Q} 
    \Bigl(\ket{ u_{m,-{\mathbb{M}}_x\bm k} ({\mathbb{M}}_x{{\bm Q}})} \Bigr)^*
    \nonumber\\
    %
    %
\end{align}
\end{widetext}
Once the quantity is decomposed as above, the following constraints are obtained by appropriately inserting $s^{\bm k}_i$ and $s^{{{\bm Q}}}_\alpha$ into the arguments involving the infinitesimal quantities $\Delta k$ and $\Delta Q$, in the same manner as in Eq.~(\ref{aeq:vel_MxT_calc}); 
\begin{align}
    \label{aeq:via_MxT}
    v^{i;\alpha}_{nm} (\bm k,{{\bm Q}})
    &=
    s^{\bm k}_i s^{{{\bm Q}}}_\alpha
    \Bigl( v^{i;\alpha}_{nm} (-{\mathbb{M}}_x\bm k,{\mathbb{M}}_x{{\bm Q}}) \Bigr)^*,
    \\
    \label{aeq:lambdaab_MxT}
    \lambda^{\alpha\beta}_{nm} (\bm k,{{\bm Q}})
    &=
    s^{{{\bm Q}}}_\alpha s^{{{\bm Q}}}_\beta
    \Bigl( \lambda^{\alpha\beta}_{nm} (-{\mathbb{M}}_x\bm k,{\mathbb{M}}_x{{\bm Q}}) \Bigr)^*.
\end{align}

The conductivity involves not the quantum metric itself but its derivatives. 
By the definition of the quantum metric, 
one finds that derivatives of the Bloch eigenstates appear in addition to derivatives of the Hamiltonian. 
Applying to these derivatives the same analysis as that used in Eq.~(\ref{aeq:vel_MxT_calc}), we obtain conditions of the same form as Eq.~(\ref{aeq:gkQ_MxT}), namely,
\begin{align}
    \partial_{Q_\beta}g^{\bm k {{\bm Q}}}_{i\alpha,n} (\bm k,{{\bm Q}})
    &=
    s^{\bm k}_i s^{{{\bm Q}}}_\alpha s^{{{\bm Q}}}_\beta \; \partial_{(s^{{{\bm Q}}}_\beta Q_\beta)}g^{\bm k {{\bm Q}}}_{i\alpha,n}(-{\mathbb{M}}_x\bm k,{\mathbb{M}}_x{{\bm Q}})
    \\
    \partial_{k_i}g^{{{\bm Q}} {{\bm Q}}}_{\alpha\beta,n} (\bm k,{{\bm Q}})
    &=
    s^{\bm k}_i s^{{{\bm Q}}}_\alpha s^{{{\bm Q}}}_\beta \; \partial_{(s^{\bm k}_i k_i)}g^{{{\bm Q}} {{\bm Q}}}_{\alpha\beta,n}(-{\mathbb{M}}_x\bm k,{\mathbb{M}}_x{{\bm Q}})
\end{align}
Utilizing these conditions with Eqs.~(\ref{aeq:OmegakQ_MxT}), (\ref{aeq:OmegaQQ_MxT}), (\ref{aeq:v_MxT}), (\ref{aeq:lambda_MxT}), (\ref{aeq:via_MxT}), and (\ref{aeq:lambdaab_MxT}), 
one can eventually obtain the constraints for the conductivity as follows, 
\begin{align}
    \sigma^{\rm ad,\Omega/\partial^2}_{i;\alpha\beta} &= -s^{\bm k}_{i}s^{{{\bm Q}}}_{\alpha}s^{{{\bm Q}}}_{\beta} \sigma^{\rm ad,\Omega/\partial^2}_{i;\alpha\beta},
    \\
    \sigma^{\rm na}_{i;\alpha\beta} &= s^{\bm k}_{i}s^{{{\bm Q}}}_{\alpha}s^{{{\bm Q}}}_{\beta} \sigma^{\rm na}_{i;\alpha\beta}.
\end{align}

{\section{ Additional multipole coefficients }}

{
Here, we present the remaining coefficients of the angular multipole expansions in Eqs. (67) and (68), which are omitted from the main text. We use the polar coordinates $\bm k=(k\cos\theta,k\sin\theta)$ around each Dirac point, with $\xi=\pm$ denoting the valley index.
The integrands for the adiabatic and nonadiabatic conductivities can be decomposed into angular multipoles, as shown in Eqs. (67) and (68). 
For the Berry-curvature-related adiabatic contribution, the additional coefficients are
\begin{align}
C_{\xi,1}^{\mathrm{ad},\Omega}(k)
&=-\frac{3\xi m_\xi\lambda^2\hbar^2v^2k^3}
{4R_\xi^5},
\\
C_{\xi,2}^{\mathrm{ad},\Omega}(k)
&=-\frac{3m_\xi g\lambda\hbar^2v^2k^2}
{4R_\xi^5},
\end{align}
where $R_\xi=\sqrt{\hbar^2v^2k^2+m_\xi^2}$ and $m_\xi=\Delta-\xi\Delta_\phi$. For the other adiabatic contribution, the only remaining nonzero coefficient is
\begin{align}
C_{\xi,1}^{\mathrm{ad},\partial^2}(k)
=\frac{\xi m_\xi\lambda^2 k}{4R_\xi^3}.
\end{align}
For the nonadiabatic contribution, we write the coefficients as $C_{\xi,l}^{\mathrm{na}}(k)=A_{\xi,l}(k)/(16R_\xi^7)$. In addition to the monopole and octupole coefficients presented in Eqs. (71) and (72), we obtain
\begin{align}
A_{\xi,1}(k)
&=g^2\left(7m_\xi^2 k
+\frac{11}{4}\hbar^2v^2k^3\right)
\nonumber\\
&
+\lambda^2\left(
-2m_\xi^4k
+9m_\xi^2\hbar^2v^2k^3
+\frac52\hbar^4v^4k^5
\right),\\
A_{\xi,2}(k)
&=-\xi g\lambda
\left(m_\xi^2\hbar^2v^2k^2
+\hbar^4v^4k^4\right),\\
A_{\xi,4}(k)
&=-\frac{17}{2}\xi g\lambda\hbar^4v^4k^4,\\
A_{\xi,5}(k)
&=\frac{17}{4}\lambda^2\hbar^4v^4k^5.
\end{align}
The remaining $l=0$ and $l=3$ coefficients are given in the main text. These expressions show that the conductivity integrands generally contain higher angular multipoles even within the Dirac model presented in the main text.
}

\end{document}